\documentclass[preprint,12pt]{elsarticle}

\usepackage{amssymb}
\usepackage{amsmath}
\usepackage{siunitx}
\usepackage{multirow}
\usepackage{makecell}
\usepackage[hidelinks,colorlinks=true,linkcolor=blue,urlcolor=blue,citecolor=blue]{hyperref}
\usepackage{placeins}
\usepackage{subcaption}
\usepackage{xcolor}
\usepackage[commandnameprefix=always, markup=default,authormarkup=none, final]{changes}

\definechangesauthor[name={Ruiyang Zhang}, color=orange]{rzhang}

\usepackage{lineno}

\journal{Nuclear Instruments and Methods in Physics Research Section A: Accelerators, Spectrometers, Detectors and Associated Equipment}

\begin{document}

\begin{frontmatter}



\title{Simulation of MAPS and a MAPS-based Inner Tracker for the Super Tau-Charm Facility}


\author[label1,label2]{Ruiyang Zhang} %% Author name
\author[label1,label2]{Dongwei Xuan}
\author[label1,label2]{Jiajun Qin}
\author[label1,label2]{Lei Zhao}
\author[label3]{Le Xiao}
\author[label3]{Xiangming Sun}
\author[label1,label2]{Lailin Xu\corref{Lailin Xu}}
\cortext[Lailin Xu]{Corresponding author 1: Lailin Xu (lailinxu@ustc.edu.cn)}
\author[label1,label2]{Jianbei Liu\corref{Jianbei Liu}}
\cortext[Jianbei Liu]{Corresponding author 2: Jianbei Liu (liujianb@ustc.edu.cn)}

\affiliation[label1]{
            organization={State Key Laboratory of Particle Detection and Electronics},%Department and Organization
            addressline={University of Science and Technology of China}, 
            city={Hefei},
            postcode={230026}, 
            country={China}}
\affiliation[label2]{
            organization={Department of Modern Physics},%Department and Organization
            addressline={University of Science and Technology of China}, 
            city={Hefei},
            postcode={230026}, 
            country={China}}
\affiliation[label3]{
            organization={PLAC, Key Laboratory of Quark and Lepton Physics (MOE)},
            addressline={Central China Normal University}, 
            city={Wuhan},
            postcode={430079}, 
            country={China}}

%% Abstract
\begin{abstract}
Monolithic Active Pixel Sensors (MAPS) are a promising detector candidate for the inner tracker of the Super Tau-Charm Facility (STCF). To evaluate the performance of MAPS and the MAPS-based inner tracker, a dedicated simulation workflow has been developed, offering essential insights for detector design and optimization.

The intrinsic characteristics of MAPS, designed using several fabrication processes and pixel geometries, were investigated through a combination of Technology Computer Aided Design~(TCAD) and Monte Carlo simulations. Simulations were conducted with both minimum ionizing particles and $^{55}$Fe X-rays to assess critical parameters such as detection efficiency, cluster size, spatial resolution, and charge collection efficiency. Based on these evaluations, a MAPS featuring a strip-like pixel and a high-resistivity epitaxial layer is selected as the baseline sensor design for the STCF inner tracker due to its excellent performance.

Using this optimized MAPS design, a three-layer MAPS-based inner tracker was modeled and simulated. The simulation demonstrated an average detection efficiency exceeding 99\%, spatial resolutions of \SI{44.8}{\micro\meter} in the $z$ direction and \SI{8.2}{\micro\meter} in the $r-\phi$ direction, and an intrinsic sensor time resolution of \SI{5.9}{\nano\second} for \SI{1}{GeV/c} $\mu^-$ particles originating from the interaction point. These promising results suggest that the MAPS-based inner tracker fulfills the performance requirements of the STCF experiment.
\end{abstract}

\begin{graphicalabstract}
\end{graphicalabstract}

\begin{highlights}
\item A simulation workflow is established to evaluate the performance of various MAPS designs, guiding the selection of an optimized configuration.
\item A detailed digitization process is integrated into the simulation to produce results closely reflecting realistic detector behavior.
\item Simulation results confirm the feasibility of a MAPS-based inner tracker for the STCF detector.
\end{highlights}

%% Keywords

\end{frontmatter}

% \linenumbers


\section{Introduction}
\label{intro}
% In the past decade, the rapid development of monolithic active pixel sensors (MAPS) have facilitated their extensive application in high energy physics experiments, particularly in tracking and vertex systems~\cite{star,its2}. A key feature of MAPS is the integration of sensing unit and readout circuit design monolithically within commercial CMOS processes. This approach further leverages cutting-edge semiconductor manufacturing techniques, such as epitaxy and stitching, to develop detectors characterized by high resolution, low material budget, low noise, and large active areas~\cite{picoad,moss}. However, to meet the stringent demands of particle detection, usually dedicated modifications to standard CMOS processes are essential to achieve high detection efficiency and withstand harsh radiation environments. Numerous efforts in this direction have yielded promising results, with simulation playing a crucial role in guiding and optimizing these developments~\cite{modified,modified2,tcad_sim_65nm}.

The proposed Super Tau-Charm Facility (STCF) is a next generation high-luminosity $e^+ e^-$ collider designed to operate in the tau-charm energy region, covering 2 to \SI{7}{GeV}\cite{CDR}. It aims to achieve a peak luminosity exceeding $0.5\times10^{35}$~\si{cm^{-2}s^{-1}} at a center-of-mass energy of \SI{4}{GeV}. With an anticipated data sample approximately 100 times larger than that of the current $\tau-c$ factory, BEPCII~\cite{BESIII:2009fln}, the STCF provides a unique platform for exploring matter-antimatter asymmetry, conducting in-depth studies of the internal structure of hadrons, investigating the nature of non-perturbative strong interactions, and searching for exotic hadrons and physics beyond the Standard Model.

To fully exploit its physics potential, the STCF detector is designed to enable precise particle detection and identification in a broad kinematic range, even while operating in a high-radiation environment. As a core component of the STCF detector, the tracking system plays a crucial role in achieving high-precision trajectory measurements, especially for low-momentum final-state charged particles. To ensure high tracking efficiency and momentum resolution for charged particles within the STCF momentum range of 50\,\si{MeV/c} to 3.5\,\si{GeV/c}---in particular, achieving a tracking efficiency of over 90\% at 100\,\si{MeV/c}, over 99\% at 300\,\si{MeV/c}, and a momentum resolution better than 0.5\% at 1\,\si{GeV/c}---minimizing the material budget is a key design consideration~\cite{stcf_tracking}. Moreover, given that the expected physics event rate can reach up to \SI{400}{kHz}, a track timestamp resolution of \SI{10}{ns} is essential for correctly associating tracks with their corresponding collision events, thereby mitigating event pile-up~\footnote{{While the average interval between events at 400\,\si{kHz} is 2.5\,\si{\micro\second}, a time resolution of 10\,\si{\nano\second} is required to ensure track separation between two events with a 50\,\si{\nano\second} interval, corresponding to a 5$\sigma$ discrimination threshold. This helps maintain pile-up probability at an acceptable level.}}. In addition to this timing requirement, the detector near the interaction point must operate at high radiation levels~\cite{DONG2024169582}. At larger radii, however, the radiation environment becomes significantly less severe. Taking these factors into account, the tracking system is designed as a hybrid structure, incorporating an inner tracker with multiple layers of finely segmented detectors and an outer drift chamber.
% Combining these requirements, as well as the high radiation levels near the interaction point, the tracking system is designed as a hybrid structure, incorporating an outer drift chamber and an inner tracker composed of multiple layers of finely segmented detectors.

Monolithic Active Pixel Sensors (MAPS) are being considered for the inner tracker due to their ability to combine low power consumption, a low material budget, good temporal resolution, and excellent spatial resolution~\cite{star,its2}. The baseline design of the MAPS-based inner tracker (ITKM) consists of three layers with a material budget of approximately 0.3\%~$X_0$ each, which demands a MAPS chip with power consumption below \SI{50}{mW/cm^2}. According to the latest background simulation studies, the innermost layer of the ITKM is required to withstand a hit rate of \SI{1}{MHz/cm^2}, a total ionizing dose (TID) of \SI{1.0}{Mrad/year} and a non-ionizing energy loss (NIEL) of $1.0\times10^{11}$\,\SI{1}{MeV}\,\si{ n_{eq}/cm^2/year} (safety factor is not applied)~\cite{bkg}, with a total operation time of ten years. Besides, a key requirement for STCF MAPS is the ability to provide precise hit-level timing information, which is crucial for achieving the target track-level timing.
A hit time resolution of better than \SI{20}{ns} is desired for the conceptual design.
In contrast, the vertex reconstruction requirements at the STCF experiment are relatively moderate. A hit spatial resolution of less than \SI{100}{\micro\meter} in the $r-\phi$ direction would be sufficient. Given that MAPS technology can readily achieve sub-\SI{10}{\micro\meter} resolution in both pixel dimensions, there is considerable flexibility in optimizing the sensor and readout electronics to balance power consumption and time resolution, while still satisfying the spatial precision needs. The above requirements for the STCF MAPS-based inner tracker are summarized in Table~\ref{tab0}.

\begin{table}[htbp]
\centering
\small
\caption{Summary of the requirements for the STCF MAPS-based inner tracker.}\label{tab0}
{
\renewcommand{\arraystretch}{1.4}
\begin{tabular}{l|l}
    \hline
     & Requirements \\
    \hline
    Hit position resolution & $<$\SI{100}{\micro\meter} \\
    Hit time resolution & $\sim$\SI{20}{\nano\second} \\
    Power consumption & $\sim$\SI{50}{mW/cm^2} \\
    Material budget & $\sim$0.3\%~$X_0$ per layer \\
    Hit rate & $>$\SI{1}{MHz/cm^2} \\
    TID & $>$\SI{1.0}{Mrad/year} \\
    NIEL & $>1.0\times10^{11}$\,\SI{1}{MeV}\,\si{n_{eq}/cm^2/year} \\
    \hline
\end{tabular}
}
\end{table}
% This higher resolution offers the potential to enhance vertex reconstruction and enable additional physics studies, such as precise $\tau$ lifetime measurements~\cite{tau_lifetime}. 
% An additional consideration for STCF MAPS is the ability to measure the time of arrival and dE/dx of charged particles, which could significantly improve background suppression and particle identification. Consequently, the MAPS design targets a time resolution of less than \SI{50}{ns} and incorporates time-over-threshold techniques to record collected charge~\cite{timepix,monopix,mupix}.

To meet the unique demands of the STCF experiment, a small collection electrode MAPS is chosen as the baseline design, primarily to minimize power consumption and thereby reduce the material budget~\cite{review_pixels}. In this baseline design, both Time-of-Arrival (ToA) and Time-over-Threshold (ToT) measurement capabilities are implemented to enable precise timing~\cite{timepix,monopix,mupix}. The measured ToT provides additional dE/dx information of charged particles, which can enhance the tracking performance for low-momentum particles and also contributes to improving the spatial resolution. Multiple different pixel sizes, geometries and fabrication processes have been investigated to reach the best overall performance. 
% The MAPS prototypes are  developed using three different CMOS processes: the well-established TowerJazz \SI{180}{nm} process, and two alternative processes with smaller technology nodes: GSMC \SI{130}{nm}~\cite{GSMC} and backside-illuminated CIS~(BCIS) \SI{90}{nm}.

During the MAPS design phase, simulation plays a crucial role in evaluating the feasibility of various fabrication processes and exploring sensor geometry variations~\cite{modified,modified2,tcad_sim_65nm}. Beyond individual sensor studies, a full-system simulation of the ITKM is essential to assessing its expected performance in the STCF experiment and guiding further detector optimization. To this end, dedicated simulation of MAPS and ITKM for the STCF experiment was performed, integrating Technology Computer-Aided Design~(TCAD) and Monte Carlo~(MC) methods. This workflow enables systematic evaluation of various sensor designs and supports the selection of an optimal baseline configuration. Based on the chosen sensor design, the STCF ITKM was implemented and fully simulated within the Super Tau-Charm Facility's Offline Software framework, OSCAR~\cite{oscar}. This paper presents the detailed simulation workflow, spanning from sensor modeling to the full ITKM system. The expected performance of ITKM under optimized design, including its detection efficiency, spatial resolution, and timing resolution, is also discussed to demonstrate its feasibility for the STCF experiment.
% This paper introduces the simulated performance of the MAPS and the ITKM, providing insights into their capabilities for the STCF experiment.

\section{Simulation workflow}
\label{sim_workflow}
% The suitability of a technology for MAPS design is assessed based on a range of factors. In addition to the conventional considerations that affect circuit design and performance, as seen in traditional ASICs, the impact of the process on detection capabilities is critical for MAPS applications. Key parameters such as sensor capacitance, charge collection speed, detection efficiency, and radiation tolerance are heavily influenced by the pixel geometry and doping profile, which together define the depletion region of the MAPS sensor.

\begin{figure}[htbp]
\centering
\includegraphics[width=\textwidth]{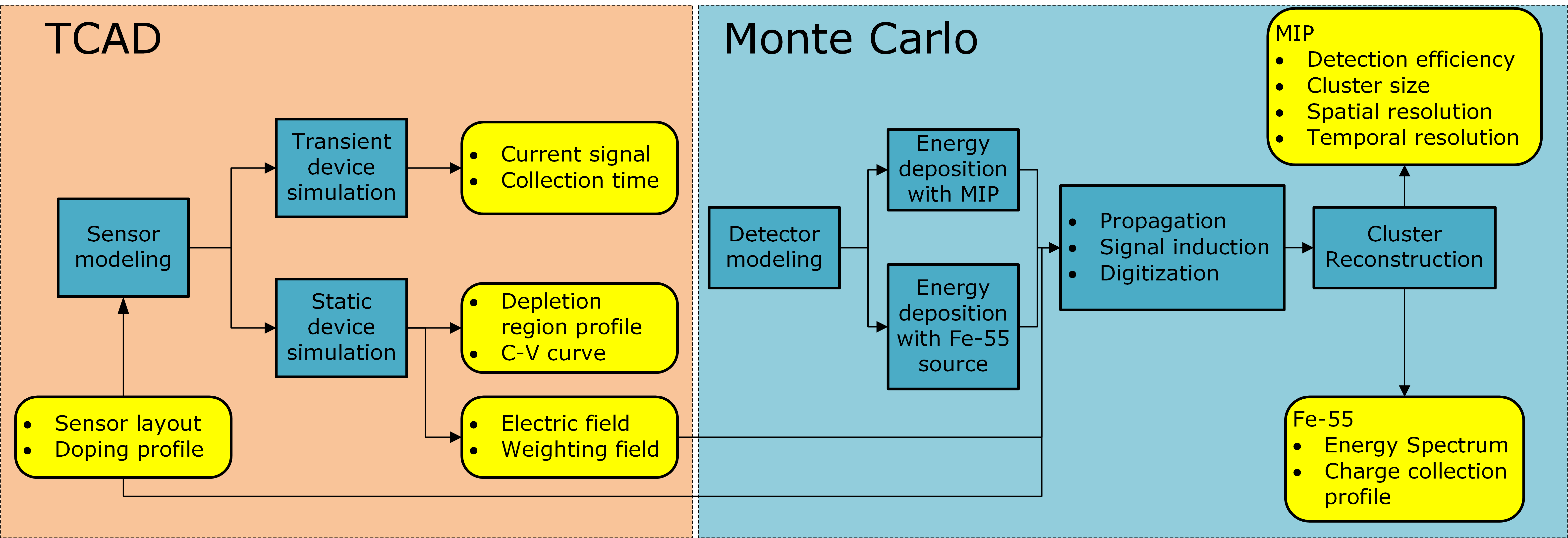}
\caption{Overview of the MAPS simulation workflow and key outcomes.}\label{fig_workflow}
\end{figure}

To characterize the performance of MAPS, both TCAD and MC simulations serve as powerful tools. These methods enable detailed studies of key sensor properties, such as capacitance, charge collection efficiency, and radiation tolerance, offering a comprehensive understanding of the sensor response prior to physical testing~\cite{trasient_simulation,mc_sim_arcadia,mc_sim_65nm}. Building upon these techniques, a flexible simulation workflow has been developed and applied to evaluate the performance of MAPS under the specific conditions and design parameters relevant to the STCF experiment. The primary aim of this simulation is to guide the choice of sensor geometry and fabrication process, while also providing results that can be compared with future laboratory and beam test data. Furthermore, the majority of the simulation should be integrated into the STCF offline software, necessitating a realistic modeling of the STCF MAPS behaviour for the future physics analysis. A flow chart illustrating the designed workflow is shown in Fig.~\ref{fig_workflow}.

For a given MAPS variant, a three-dimensional sensor model of our pixel layout is constructed within TCAD, which is going to be discussed in detail in Section~\ref{TCAD}. The key parameters that determines the sensor performance, including doping profiles and diode geometry, are defined according to the given fabrication process and the MAPS layout~\cite{simulating_workflow}. Once the sensor model is established, either a static or transient device simulation is performed, depending on the desired output. The static simulation solves the Poisson equation and continuity equations to obtain the electrostatic potential and carrier distributions. It also generates a Capacitance-Voltage (C-V) curve reflecting the sensor capacitance under increasing substrate bias through small-signal AC analysis. The transient simulation is used to estimate the time-dependent current induced on the collection electrode by drifting and diffusing charge carriers. In this case, a predefined ionization track is introduced into the sensor volume to represent a traversing particle. To emulate the behavior of a minimum ionizing particle (MIP), the ionization is modeled as a straight line of uniformly distributed electron–hole pairs, with a density of 60\,\si{e^-/\micro\meter}. This track is positioned either at the pixel center or at the corner to study spatial response variations.

Despite the above results, TCAD simulations can not account for the stochastic variations in particle incidence position and energy deposition. To evaluate the sensor performance under realistic experimental conditions, MC simulations are essential. A workflow for MAPS MC simulations has been developed within the OSCAR framework to address this need. The OSCAR framework serves as a unified platform for full simulation of the STCF detector and will also support the STCF offline data analysis. The complete MAPS simulation chain under the framework includes detector modeling, event generation, ionized particle transportation, signal induction, digitization and reconstruction.

The MAPS geometry is modeled using DD4hep~\cite{DD4hep}. For each pixel, the in-pixel electric field, doping concentration, and weighting potential are defined based on results from static TCAD simulation. Geant4~\cite{geant4} is employed to simulate the passage of incident particles through the sensor. All relevant physics processes---including ionization, multiple scattering, and $\delta$-ray production---are activated. The energy deposits are recorded along the track and converted into electron–hole pairs using an average ionization energy of 3.64\,\si{eV} in silicon. Fluctuations in energy loss are accounted for by using Geant4's intrinsic energy deposition model.
The propagation of the generated carriers is subsequently simulated using a custom carrier transport model adapted from Allpix\textsuperscript{2}~\footnote{{In this work, the ``Extended Canali model'' is used for carrier mobility simulation, while the ``Combined SRH/Auger model'' is used for recombination simulation.}}~\cite{trasient_simulation,allpix2}. 
The model takes into account drift under the local electric field, thermal diffusion, and carrier recombination based on the doping concentration profile. Electron–hole pairs are propagated separately using a time-step integration method, and the induced signal on the collection electrode is calculated via the Ramo-Shockley theorem~\cite{shockley, ramo}, using the pre-computed weighting potential. The implementation of these models within the OSCAR framework has been validated to ensure consistency with the results obtained from Allpix\textsuperscript{2}. After the signal induction, a digitization procedure can be applied to convert the signal into electronic outputs, such as signal waveforms or ADC values. Typically, a single incident particle induces signals in multiple adjacent pixels due to charge sharing, with varying amounts of charge collected per pixel. Clusters of adjacent pixels exceeding a predefined charge threshold are identified, and their positions are reconstructed using the charge-weighted center-of-gravity method. The simulation framework is capable of reproducing the detector’s response to both MIPs and $^{55}$Fe X-rays, providing complementary performance assessments. Also, it can be readily extended to full detector simulation using predefined event generators, facilitating realistic studies of physical event responses.
% \footnote{\chadded[id=rzhang]{In this work, the ``Extended Canali model'' is used for carrier mobility simulation, while the ``Combined SRH/Auger model'' is used for recombination simulation.}}

The above Monte Carlo workflow is basically the same for the simulation described in Section~\ref{maps_sim} and Section~\ref{itkm_sim}, while several distinctions exist depending on the specific objectives. For the MAPS simulation part, the primary goal is to compare the intrinsic performance of different pixel designs. To this end, the detector is simplified to a single pixel array with minimal supporting structures, and signal digitization is intentionally omitted to isolate intrinsic sensor behaviours. For the ITKM simulation part, however, a more realistic detector modeling and digitization based on the STCF MAPS design is integrated, which will be further illustrated in Section~\ref{detector_setup} and~\ref{digitization}.

\section{MAPS process and geometry variants}
\label{variants}

\begin{figure}[htbp]
\centering
    \begin{subfigure}{0.45\textwidth}
        \centering
        \includegraphics[width=\linewidth]{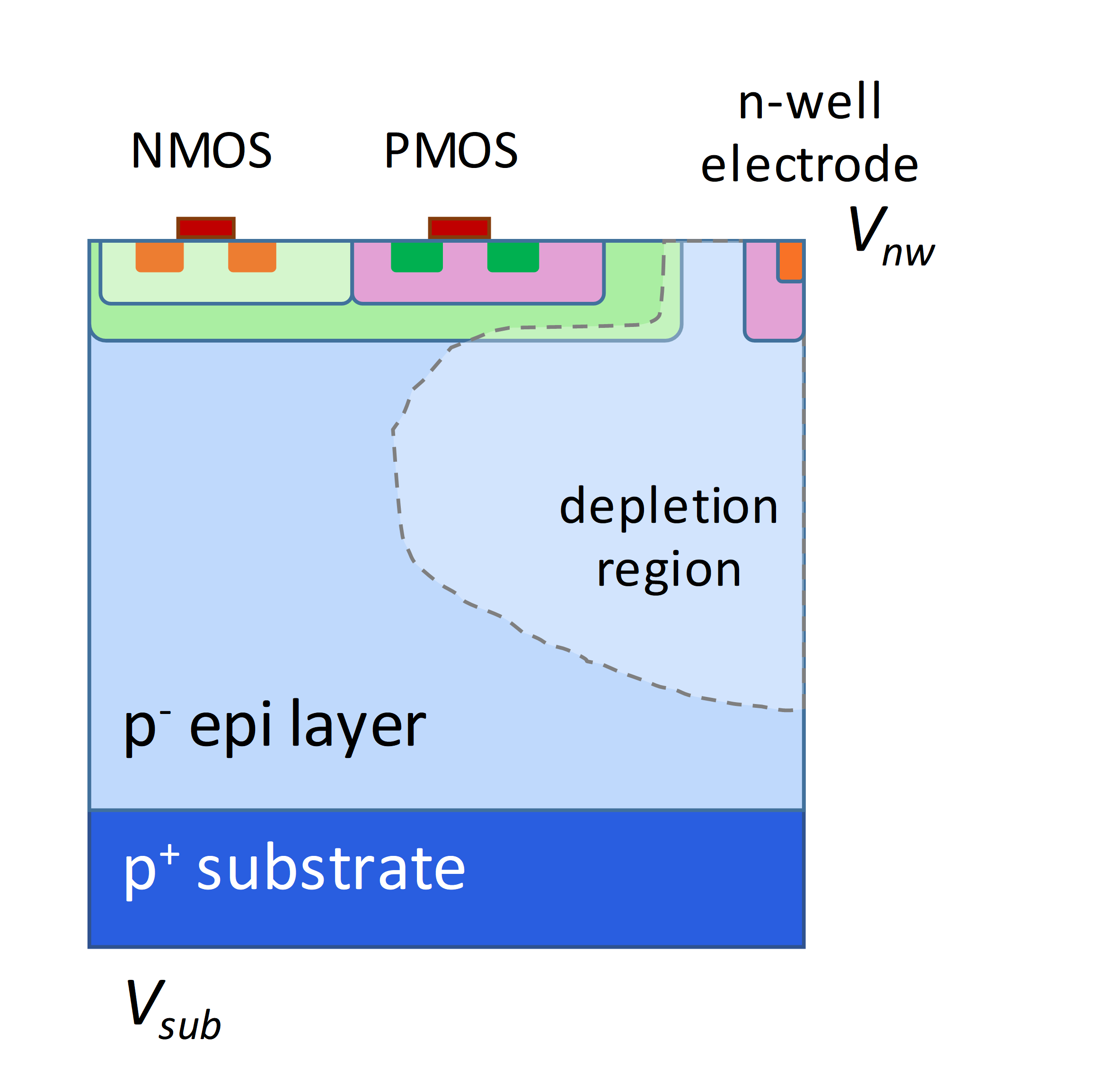}
        \caption{}
        \label{process_a}
    \end{subfigure}
    \hfill
    \begin{subfigure}{0.45\textwidth}
        \centering
        \includegraphics[width=\linewidth]{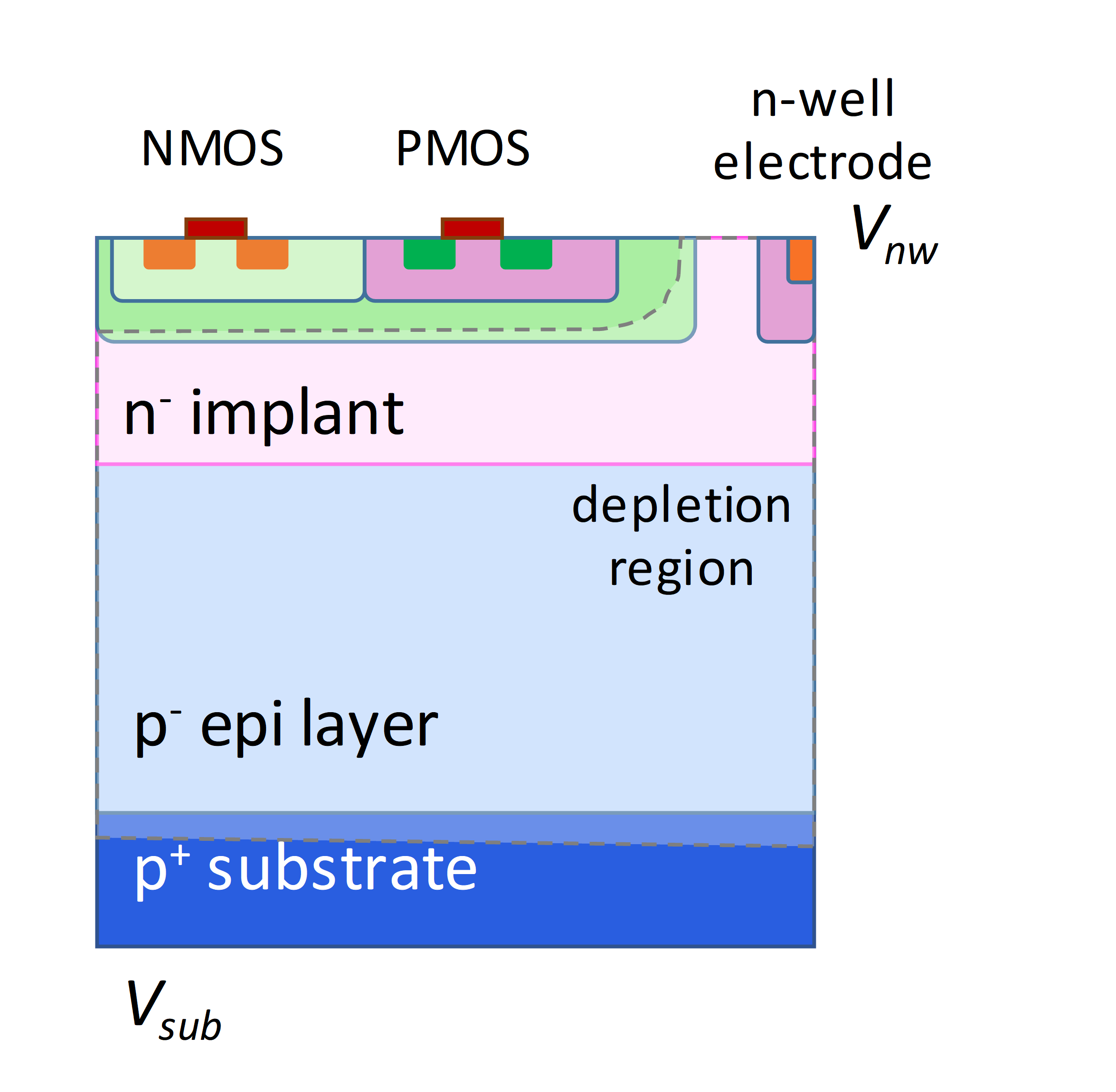}
        \caption{}
        \label{process_b}
    \end{subfigure}
    \hfill
    \begin{subfigure}{0.45\textwidth}
        \centering
        \includegraphics[width=\linewidth]{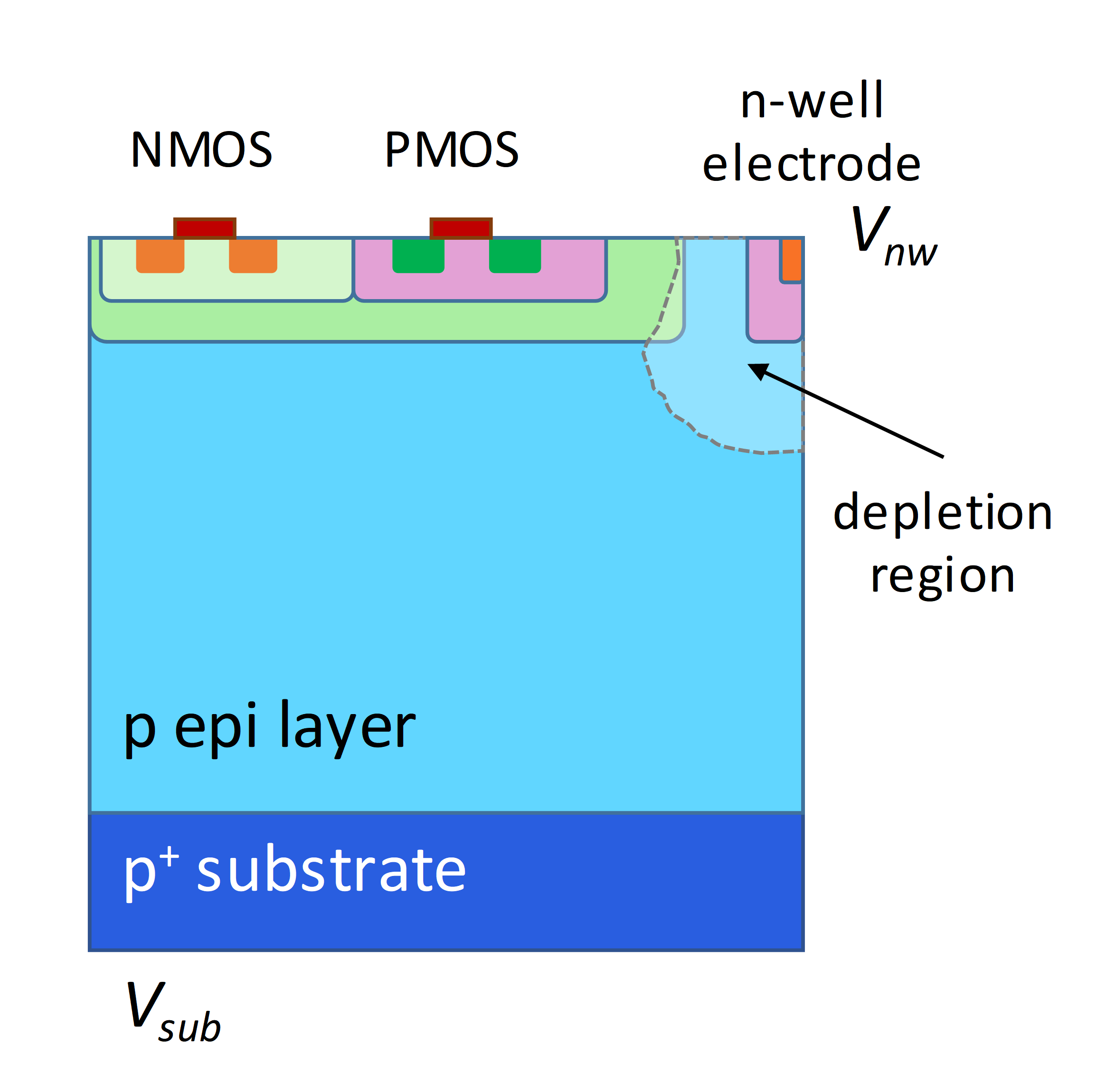}
        \caption{}
        \label{process_c}
    \end{subfigure}
    \hfill
    \begin{subfigure}{0.45\textwidth}
        \centering
        \includegraphics[width=\linewidth]{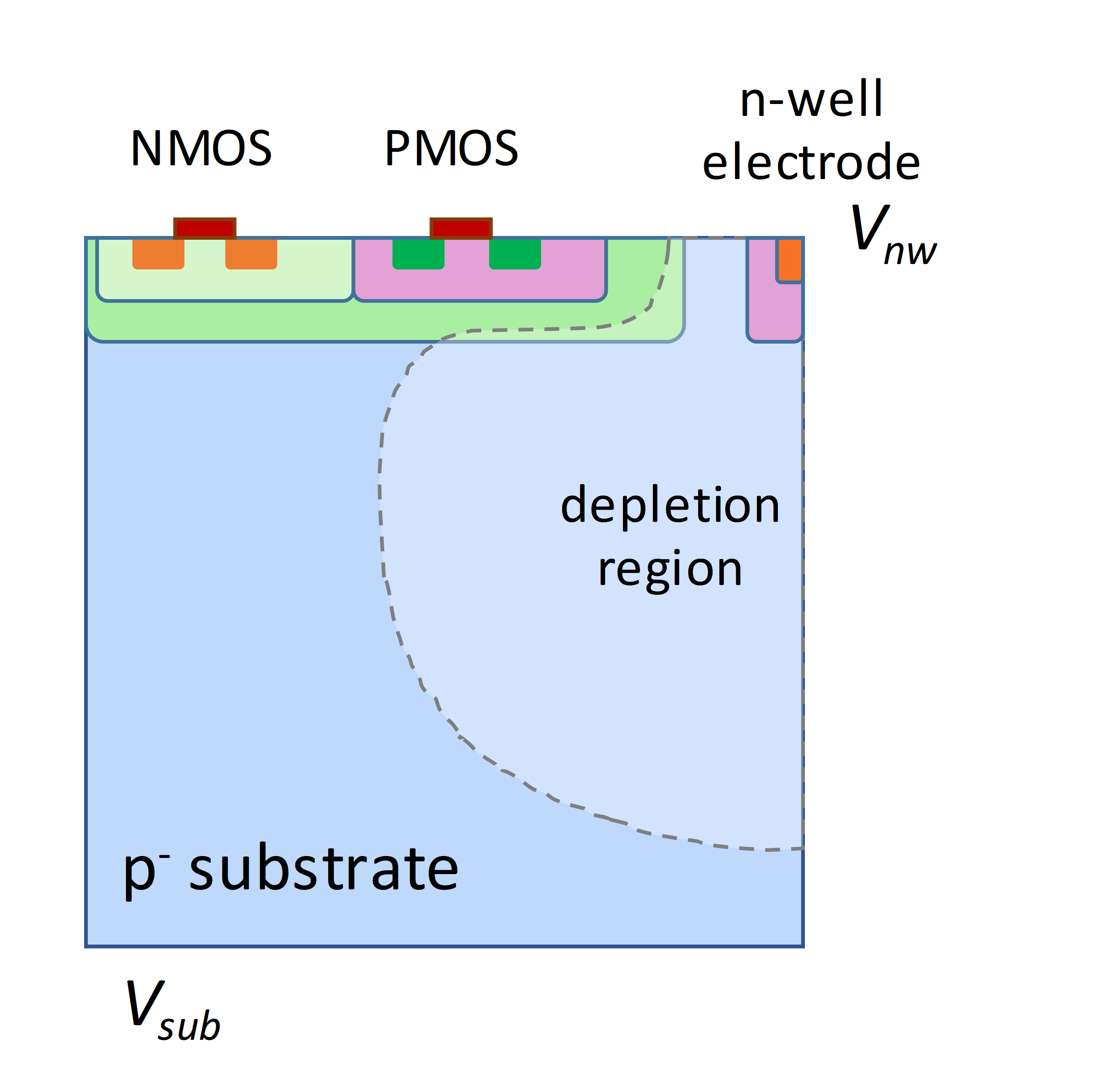}
        \caption{}
        \label{process_d}
    \end{subfigure}
    
\caption{Schematic cross section of four process variants in simulation: HR epi (a), N blanket (b), LR epi (c), and HR substrate (d)~\cite{modified}.}\label{process}
\end{figure}

For the development of the STCF MAPS, four CMOS imaging sensor processes featuring distinct MAPS-related characteristics have been considered. These process variants are described below, each identified by an alias reflecting its unique feature:

\begin{itemize}
    \item \textbf{HR (High Resistivity) epi}: This process (cf. Fig.~\ref{process_a}) is based on the standard TowerJazz \SI{180}{nm} CMOS technology. It utilizes a 20\,\si{\micro\meter} high-resistivity p-type epitaxial layer grown on a low-resistivity substrate of the same doping type. The epitaxial layer typically has a resistivity exceeding \SI{1}{k\ohm\cdot cm}, while the substrate features a doping concentration of approximately $10^{18}$\si{cm^{-3}}~\cite{ALPIDE_thesis}. Charge is collected by a small n-well electrode inside each pixel, which establishes a bulb-shaped depletion region. The pixel’s CMOS circuitry is encapsulated within a deep p-well structure to mitigate charge collection competition. This process was notably utilized in the ALPIDE chip and successfully integrated into the ALICE ITS2 detector~\cite{ALPIDE, its2}.
    \item \textbf{N blanket}: This modified TowerJazz \SI{180}{nm} process includes an additional deep, low-dose n-type implant layer on top of the epitaxial layer (cf. Fig.~\ref{process_b})~\cite{modified}. The n-type implant forms an extended depletion region within the epitaxial layer, improving the charge collection speed and efficiency. Moreover, the continuous depletion region isolates the front side from the substrate, allowing for a larger operating margin for substrate bias. A further optimized version of this process is widely adopted in the development of new generations of MAPS, such as the TJ-Monopix series~\cite{monopix,tjmp2,obelix}.  A similar process has also been realized at the \SI{65}{nm} process node and is expected to be used in the MAPS for ALICE ITS3~\cite{APTS}.
    \item \textbf{LR (Low Resistivity) epi}: This process is based on a BCIS \SI{90}{nm} technology (cf. Fig.~\ref{process_c}), featuring an epitaxial layer with low resistivity (\~{}\SI{10}{\ohm\cdot cm}) and a thickness of approximately \SI{10}{\micro\meter}. This process was primarily developed for CMOS imaging sensor and has not yet been fully optimized for MAPS, which somewhat limits its current effectiveness. Ongoing process modifications are being explored to improve its suitability for particle detection.
    \item \textbf{HR substrate}: The GSMC \SI{130}{nm} technology (cf. Fig.~\ref{process_d}) directly employs a high-resistivity wafer (\SI{>1}{k\ohm\cdot cm}) for CMOS circuitry without using an epitaxial layer~\cite{GSMC,topmetal_m}. This approach is similar to the one implemented in the MALTA chip~\cite{MALTA_Cz}, but presently lacks an n-type blanket layer. The larger low-dose p-type region compared to the ``HR epi'' process results in an expanded depletion region. Additionally, the high-resistivity substrate allows for a higher substrate bias.
    % This process is used in the development of Nupix series for the High Intensity heavy-ion Accelerator Facility (HIAF)~\cite{nupix_a1,nupix_a2}.
\end{itemize}

% \begin{figure}[htbp]
% \centering
% \includegraphics[width=\textwidth]{process variants.png}
% \caption{Schematic cross section of four process variants in simulation: HR epi (a), N blanket (b), LR epi (c), and HR substrate (d)~\cite{modified}.}\label{process}
% \end{figure}

\begin{figure}[htbp]
\centering
    \begin{subfigure}{0.4\textwidth}
        \centering
        \includegraphics[width=\linewidth]{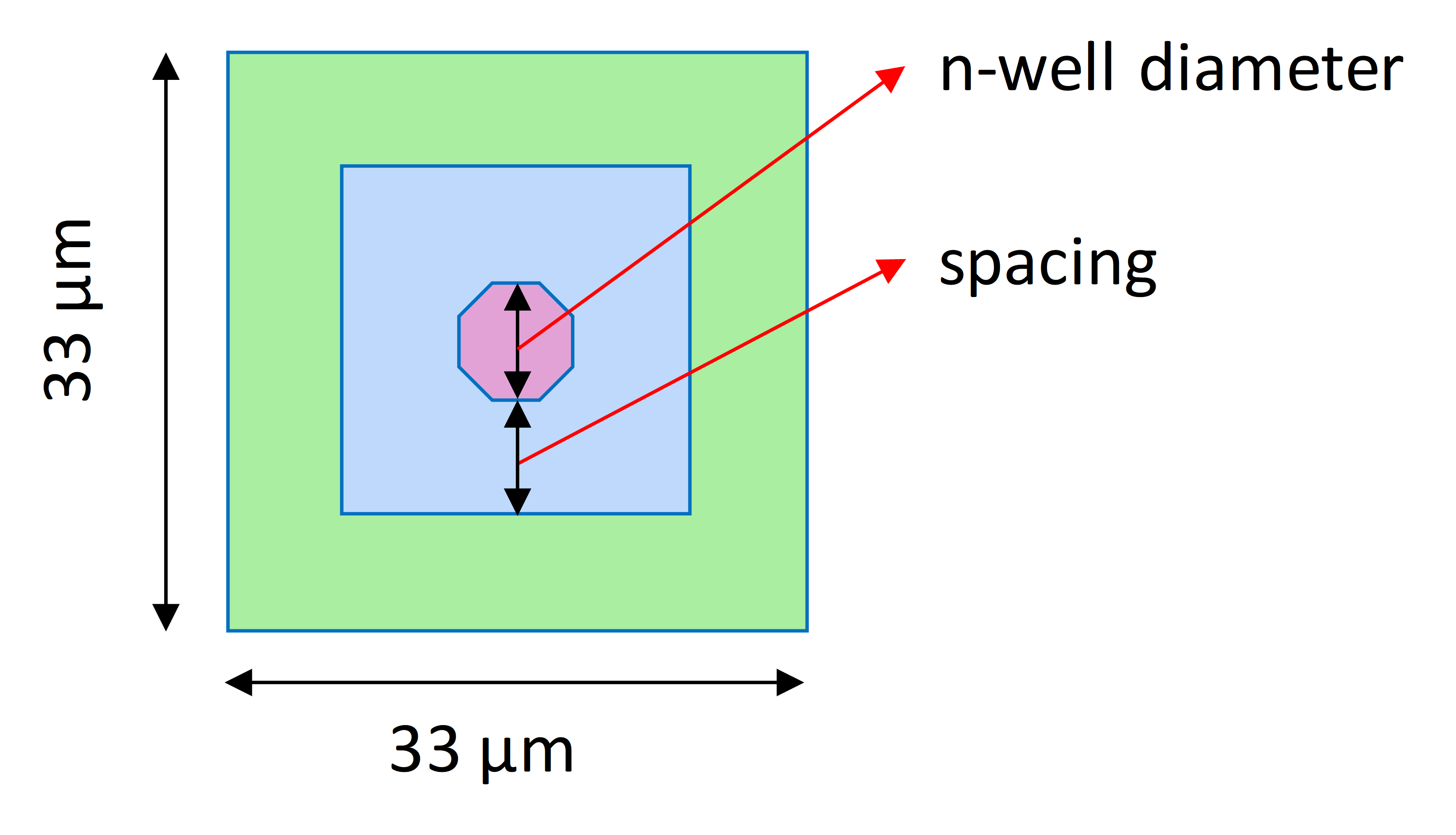}
        \caption{}
        \label{geometry_a}
    \end{subfigure}
    \hfill
    \begin{subfigure}{0.55\textwidth}
        \centering
        \includegraphics[width=\linewidth]{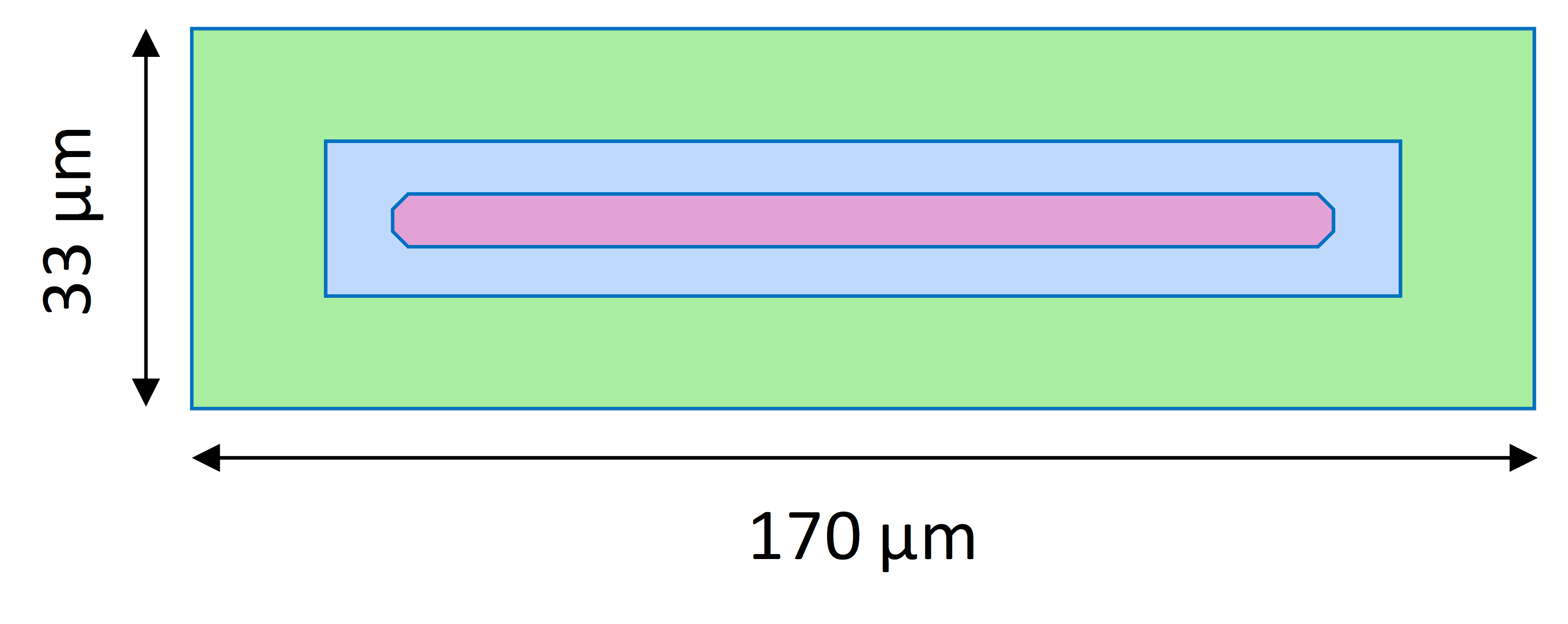}
        \caption{}
        \label{geometry_b}
    \end{subfigure}
    \hfill
    \begin{subfigure}{0.55\textwidth}
        \centering
        \includegraphics[width=\linewidth]{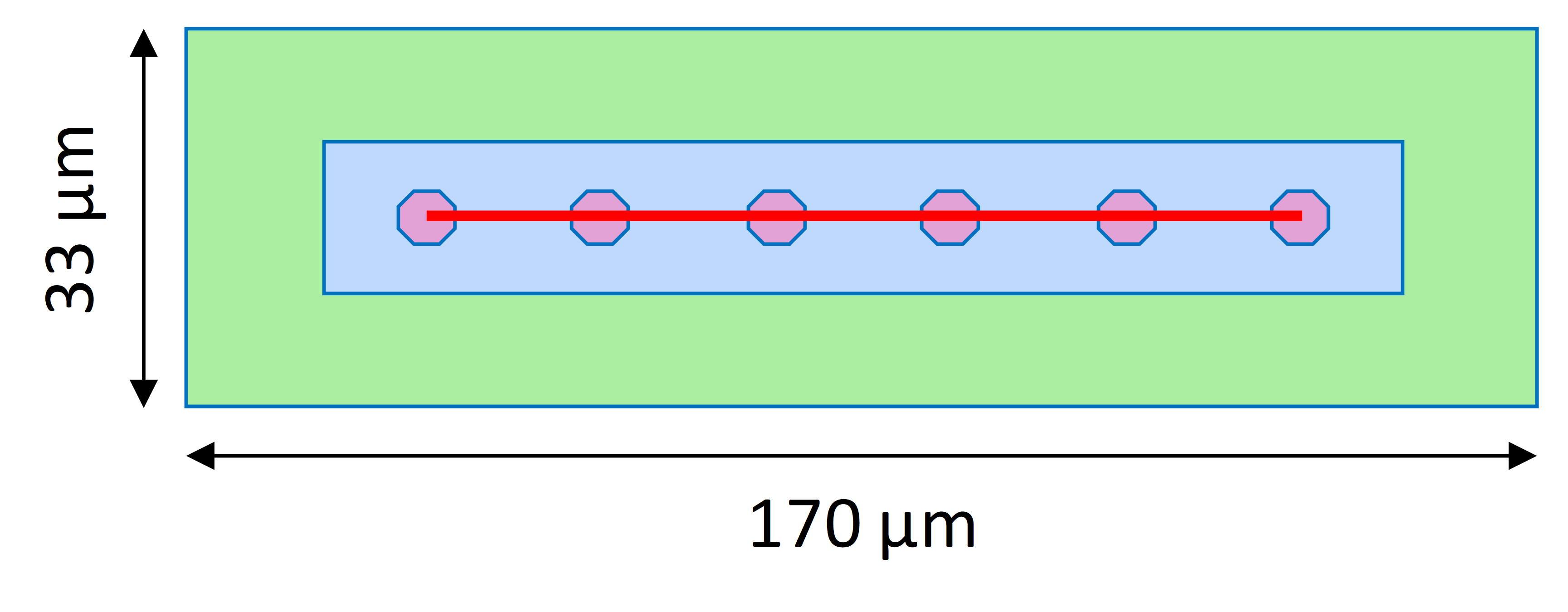}
        \caption{}
        \label{geometry_c}
    \end{subfigure}
\caption{Schematic top view of three different pixel geometry variants used in the simulation (not to scale): standard pixel (a), active-connect large pixel (b), and metal-connect large pixel (c).}\label{geometry}
\end{figure}

To investigate the pixel performance difference purely introduced by the processes, other unrelated parameters are fixed to appropriate values in the following simulations. The pixel size is set to $\SI{33}{\micro\meter}\times\SI{33}{\micro\meter}$. The n-well electrode is designed with an octagonal shape, having a diameter of \SI{2}{\micro\meter} and a \SI{2}{\micro\meter} distance from the deep p-well boundary (cf. Fig.~\ref{geometry_a}).

Given that the material budget is the primary concern in the STCF MAPS, while the requirement for spatial resolution is relatively relaxed, an strip-like pixel design has been explored as a potential solution. In this design, the pixel pitch is extended in one dimension to \SI{170}{\micro\meter}, approximately six times the size of a standard pixel, while the other dimension remains unchanged. When implemented in the STCF inner tracker, the extended dimension is aligned along the $z$-axis (the beam pipe), where a lower granularity is sufficient. This arrangement not only reserves more in-pixel space for electronics and routing but also significantly reduces the chip’s power consumption due to the smaller number of channels. Lower power consumption below \SI{50}{mW/cm^2} in turn allows for a lighter and less complex cooling system (e.g., air cooling instead of liquid cooling)~\cite{Spannagel_2017}, which is crucial for maintaining a low material budget.
Two pixel layouts featuring slightly different diode structures have been designed. The first, referred to as the ``active-connect'' type (cf. Fig.~\ref{geometry_b}), employs a single elongated n-well that spans the entire pixel length. In contrast, the ``metal-connect'' type (cf. Fig.~\ref{geometry_c}) consists of six individual n-wells interconnected by a metal line. The active-connect pixel is expected to offer a larger depletion volume, whereas the metal-connect pixel exhibits a smaller sensor capacitance, as will be discussed in the following sections.

% \begin{figure}[htbp]
% \centering
% \includegraphics[width=0.8\textwidth]{geometry variants.png}
% \caption{Schematic top view of three different pixel geometry variants used in the simulation (not to scale): standard pixel (a), active-connect large pixel (b), and metal-connect large pixel (c).}\label{geometry}
% \end{figure}

The process and geometrical parameters used in this work are summarized in Table~\ref{tab1}. It should be noted that, due to the limited available information regarding specific technologies, the process parameter values chosen here represent only typical conditions for each process. Some parameters are derived from simulation or test results of other monolithic sensors designed with the same process~\cite{simulation_ALPIDE,suljic_2018}. Systematic uncertainty arising from these parameter choices has not been considered in the present study.

% \begin{table}[htbp]
% \centering
% \small
% \caption{Process parameters of the different sensor variants.}\label{tab1}
% {
% \renewcommand{\arraystretch}{1.4}
% \begin{tabular}{l|l|l}
%     \hline
%     Process name & Layer thickness & Resistivity \\
%     \hline
%     HR epi & \SI{20}{\micro\meter} epitaxial layer & \SI{15}{k\ohm\cdot cm} \\
%     N blanket & \SI{20}{\micro\meter} epitaxial layer, \SI{6}{\micro\meter} n-blanket layer & \SI{15}{k\ohm\cdot cm} \\
%     LR epi & \SI{10}{\micro\meter} epitaxial layer & \SI{10}{\ohm\cdot cm} \\
%     HR substrate & \SI{50}{\micro\meter} substrate & \SI{1}{k\ohm\cdot cm} \\
%     \hline
% \end{tabular}
% }
% \end{table}

% \begin{table}[htbp]
% \centering
% \small
% \caption{Geometrical parameters of the different sensor variants.}\label{tab1_2}
% {
% \renewcommand{\arraystretch}{1.4}
% \begin{tabular}{l|l|l}
%     \hline
%     Geometry name & Pixel size & Common parameters\\
%     \hline
%     Standard & $\SI{33}{\micro\meter}\times\SI{33}{\micro\meter}$ & \multirow{3}{*}{\makecell[l]{N-well diameter: \SI{2}{\micro\meter}\\ Spacing: \SI{2}{\micro\meter}}} \\
%     Active-connect & $\SI{170}{\micro\meter}\times\SI{33}{\micro\meter}$ &  \\
%     Metal-connect & $\SI{170}{\micro\meter}\times\SI{33}{\micro\meter}$ &  \\
%     \hline
% \end{tabular}
% }
% \end{table}

\begin{table}[htbp]
\centering
\small
\caption{Process and geometry parameters of the different sensor variants.}
\label{tab1}
{
\renewcommand{\arraystretch}{1.4}
\begin{tabular}{l|l|l}
    \hline
    \multicolumn{3}{c}{\textbf{Process Parameters}} \\
    \hline
    Process name & Layer thickness & Resistivity \\
    \hline
    HR epi       & \SI{20}{\micro\meter} epitaxial layer & \SI{15}{k\ohm\cdot cm} \\
    N blanket    & \SI{20}{\micro\meter} epitaxial, \SI{6}{\micro\meter} n-blanket layer & \SI{15}{k\ohm\cdot cm} \\
    LR epi       & \SI{10}{\micro\meter} epitaxial layer & \SI{10}{\ohm\cdot cm} \\
    HR substrate & \SI{50}{\micro\meter} substrate & \SI{1}{k\ohm\cdot cm} \\
    \hline
    \multicolumn{3}{c}{\textbf{Geometry Parameters}} \\
    \hline
    Geometry name & Pixel size & Common parameters \\
    \hline
    Standard       & $\SI{33}{\micro\meter} \times \SI{33}{\micro\meter}$ & \multirow{3}{*}{\makecell[l]{N-well diameter: \SI{2}{\micro\meter} \\ Spacing: \SI{2}{\micro\meter}}} \\
    Active-connect & $\SI{170}{\micro\meter} \times \SI{33}{\micro\meter}$ & \\
    Metal-connect  & $\SI{170}{\micro\meter} \times \SI{33}{\micro\meter}$ & \\
    \hline
\end{tabular}
}
\end{table}

\section{MAPS characteristic simulation}
\label{maps_sim}
This section is organized into four subsections. Sections~\ref{TCAD} to \ref{xray} present detailed simulation studies based on the design variants introduced in Section~\ref{variants}. Specifically, Section~\ref{TCAD} focuses on TCAD simulation results, highlighting sensor characteristics such as capacitance and charge collection capability. Section~\ref{MIP} and Section~\ref{xray} then describe Monte Carlo simulations using MIPs and \SI{5.9}{keV} $^{55}$Fe X-rays, respectively, to evaluate the detector response under different irradiation scenarios. Finally, the simulation outcomes from both approaches are synthesized and discussed in Section~\ref{MAPS_discussion}, providing insights into the optimal MAPS configuration for STCF.
% The following sections present the results of TCAD and Monte Carlo simulations, comparing different MAPS variants. The TCAD simulation results are introduced in Section~\ref{TCAD}, while the Monte Carlo simulations of MIP and $^{55}$Fe X-ray are introduced in Section~\ref{MIP} and~\ref{xray}, respectively. The above simulation results are together discussed in Section~\ref{MAPS_discussion}.

\subsection{TCAD simulation results}
\label{TCAD}
\begin{figure}[htbp]
\centering
    \begin{subfigure}{0.45\textwidth}
        \centering
        \includegraphics[width=\linewidth]{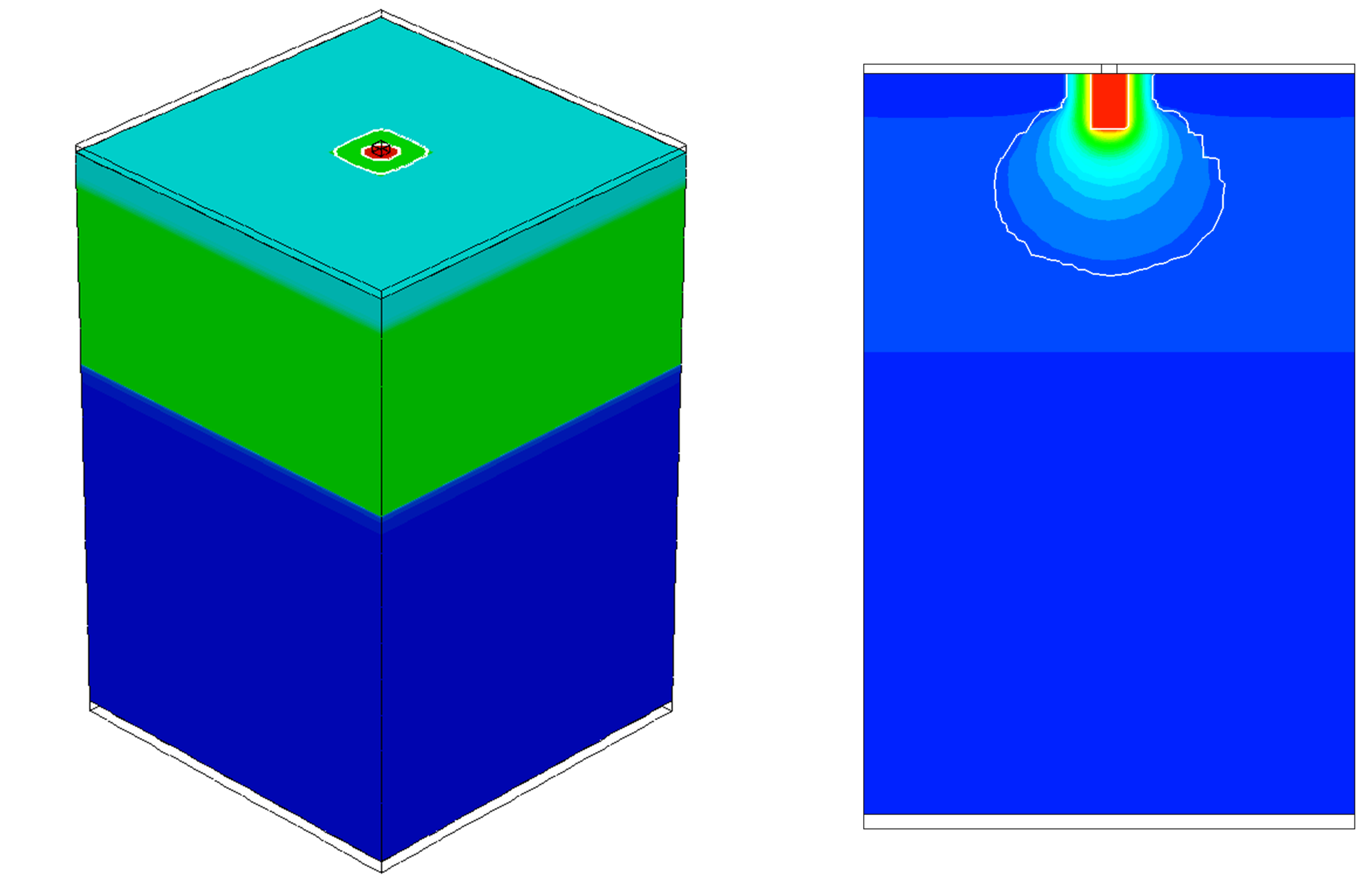}
        \caption{}
    \end{subfigure}
    \hfill
    \begin{subfigure}{0.45\textwidth}
        \centering
        \includegraphics[width=\linewidth]{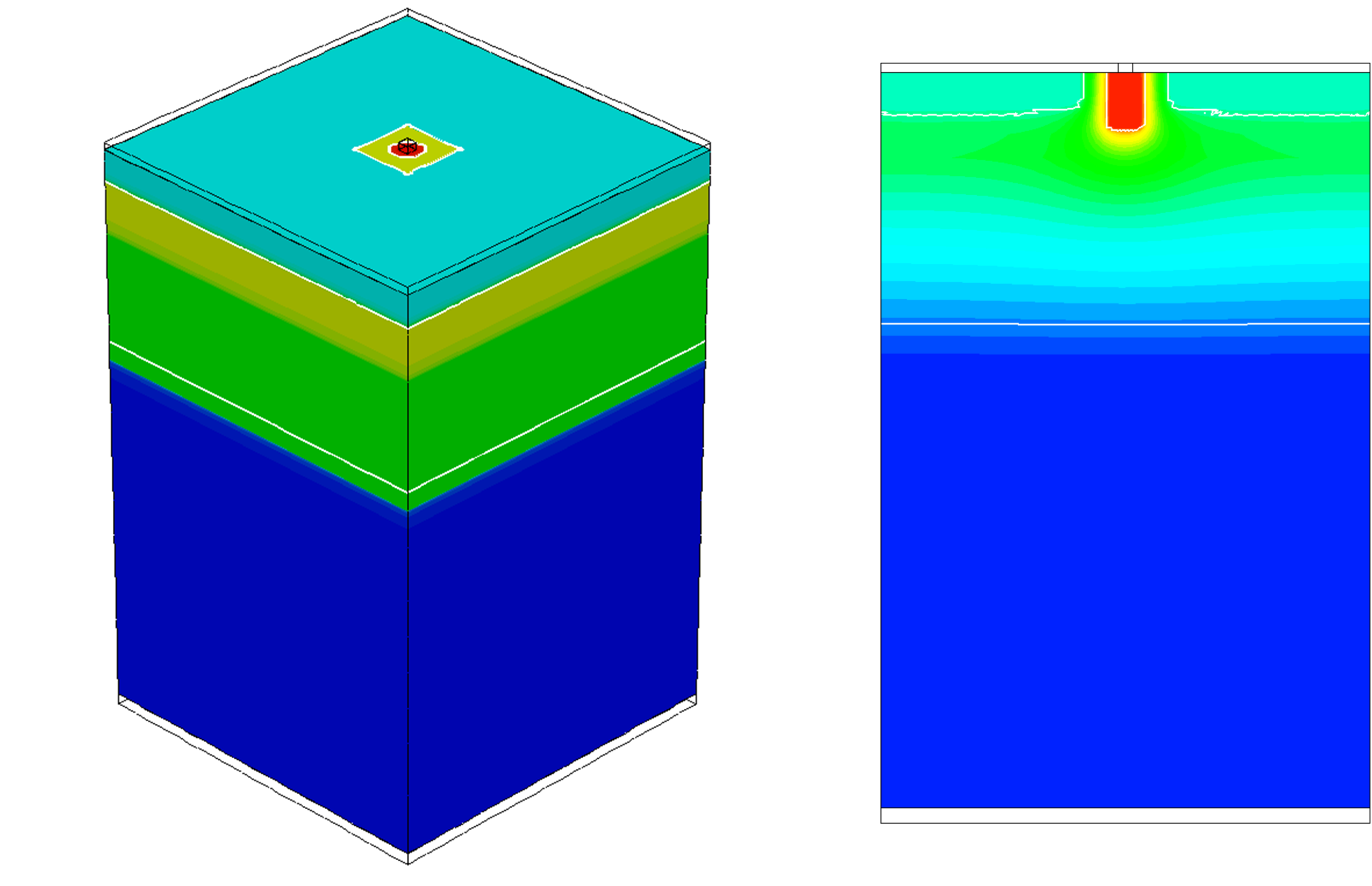}
        \caption{}
    \end{subfigure}
    \hfill
    \begin{subfigure}{0.45\textwidth}
        \centering
        \includegraphics[width=\linewidth]{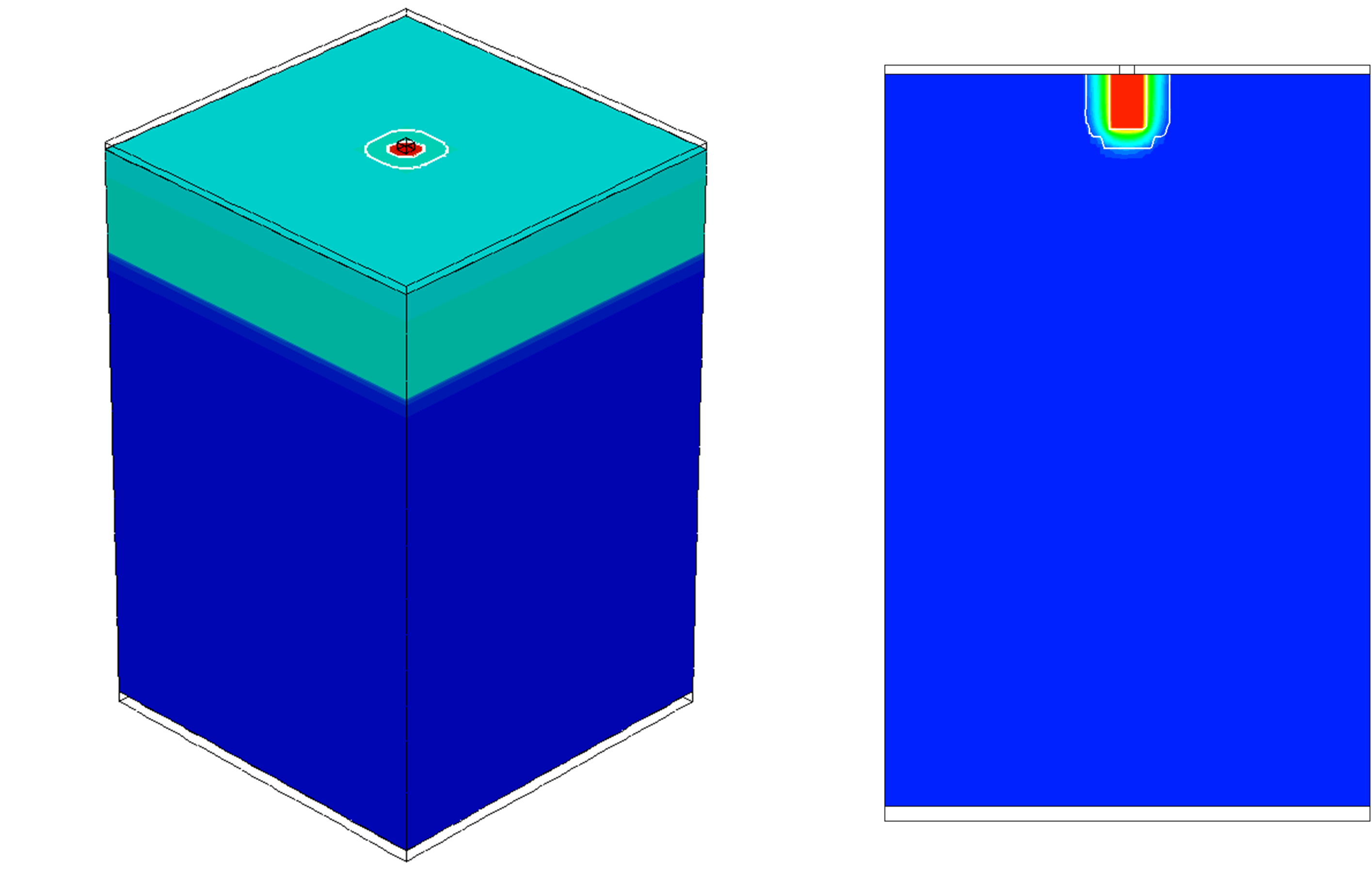}
        \caption{}
    \end{subfigure}
    \hfill
    \begin{subfigure}{0.45\textwidth}
        \centering
        \includegraphics[width=\linewidth]{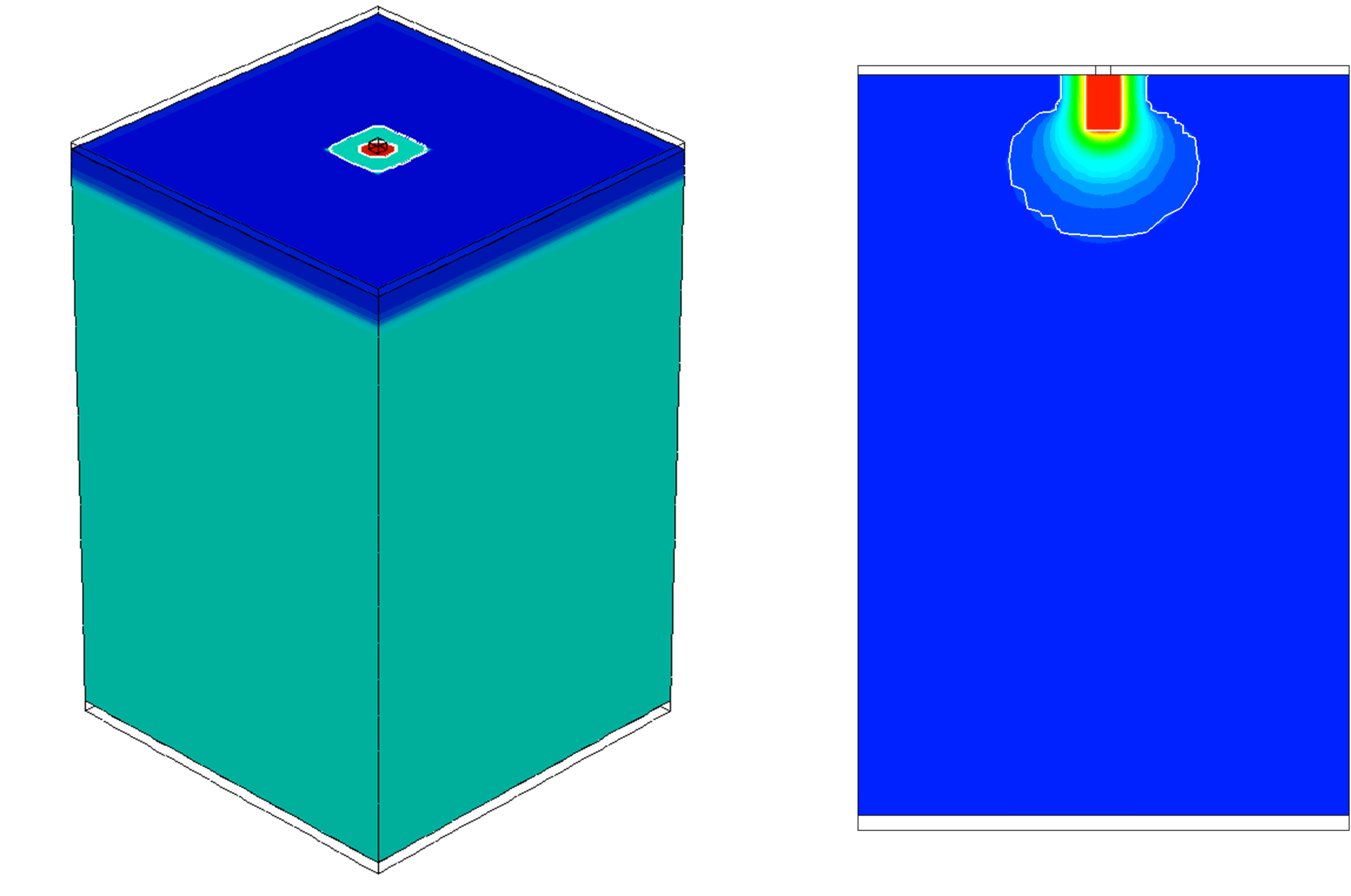}
        \caption{}
    \end{subfigure}
    
\caption{Single-pixel TCAD models for four process variants under an operational substrate bias of $V_{sub} = \SI{-6}{V}$: (a) HR epi, (b) N blanket, (c) LR epi, and (d) HR substrate. For each process, a 3D map of the doping concentration plot and a 2D cross-sectional view of the electrostatic potential through the pixel center are shown. White contour lines in the 2D plots represent the boundaries of the depletion region.}\label{process_tcad}
\end{figure}

For each of the four process variants described above, a 3D model of the pixel is constructed in TCAD based on the given doping profiles (cf. Fig.~\ref{process_tcad}). A two-dimensional cross-sectional view of the electric potential across the central plane of the pixel is shown for each variant, under a substrate bias voltage of \SI{-6}{V}. The \SI{-6}{V} bias is chosen as a safe upper limit, balancing breakdown risk of the circuit and collection performance of the sensor.

% \begin{figure}[htbp]
% \centering
% \includegraphics[width=\textwidth]{process variants TCAD.png}
% \caption{Single pixel models for four process variants in TCAD: HR epi (a), N blanket (b), LR epi (c), and HR substrate (d). The color scale represents the doping concentration.}\label{process_tcad}
% \end{figure}

\begin{figure}[htbp]
\centering
\includegraphics[width=0.8\textwidth]{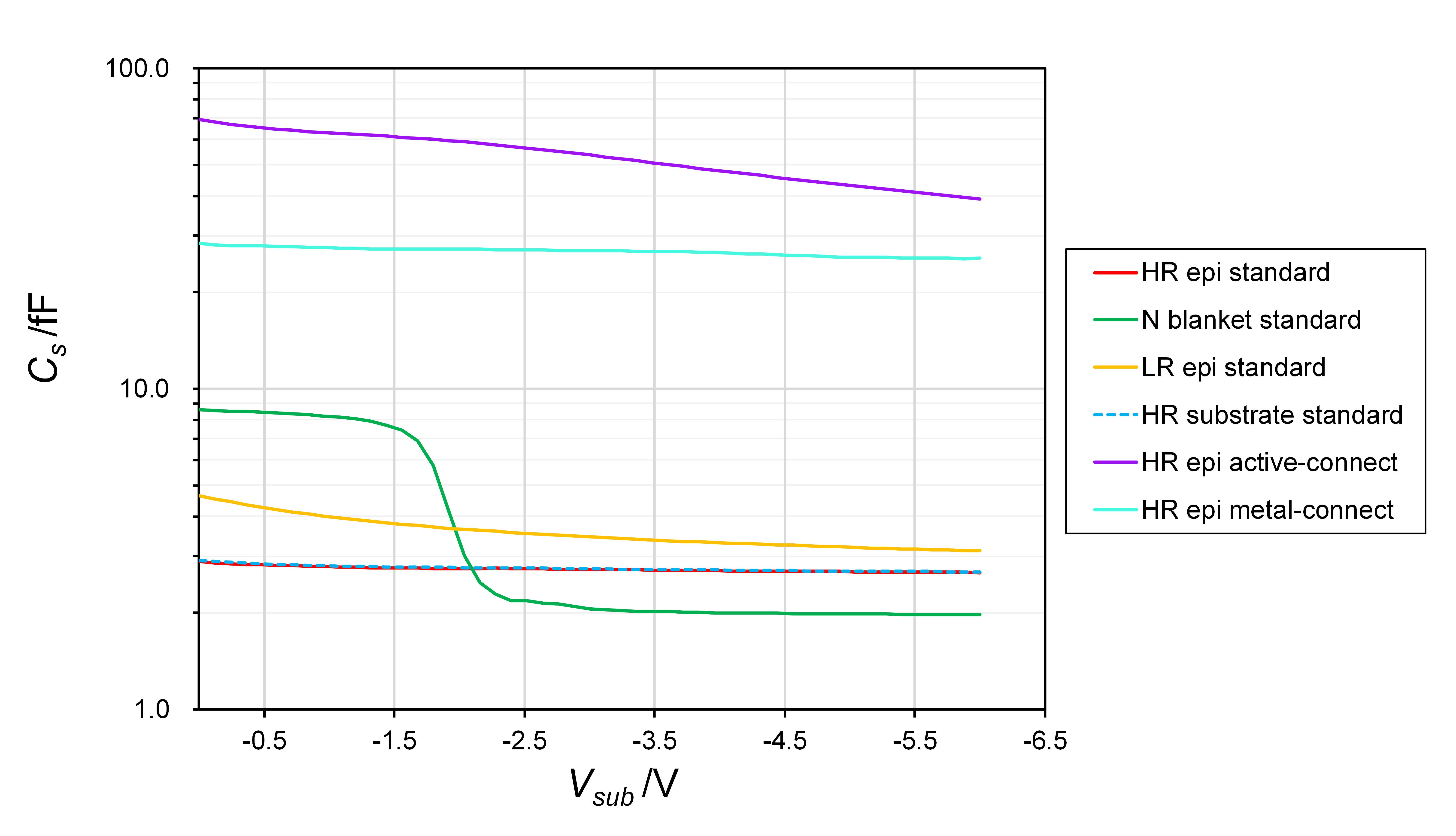}
\caption{Sensor capacitance $C_s$ as function of substrate voltage $V_{sub}$ (with n-well voltage $V_{nw}$ fixed at \SI{0.8}{V}) for different sensor variants.}\label{capacitance}
\end{figure}

Fig.~\ref{capacitance} shows the resulting C-V curve for different processes. The n-well voltage $V_{nw}$ is fixed at \SI{0.8}{V}, while the substrate bias varies from 0 to \SI{-6}{V}.
The simulation reveals a general trend of decreasing capacitance with increasing bias voltage for all processes, attributed to the expansion of the depletion region. Notably, for the ``N blanket'' variant, an abrupt drop in $C_s$ is observed at approximately $V_{sub}=\SI{-2}{V}$. This phenomenon arises due to the depletion boundaries, which initially expand from the deep p-well/n-implant interface and the epitaxial layer/n-implant interface, gradually converge as the bias increases. Only when these boundaries are fully merged and the entire n-implant layer is depleted does the capacitance exhibit this sharp change. At $V_{sub}=\SI{-6}{V}$, the capacitance for all processes stabilize. Among the processes, the ``N blanket'' variant demonstrates the lowest sensor capacitance at \SI{2.0}{fF} while the ``LR epi'' variant exhibits a highest value at \SI{3.1}{fF}, reflecting significant differences in their depletion volumes.

\begin{figure}[htbp]
\centering
    \begin{subfigure}{0.45\textwidth}
        \centering
        \includegraphics[width=\linewidth]{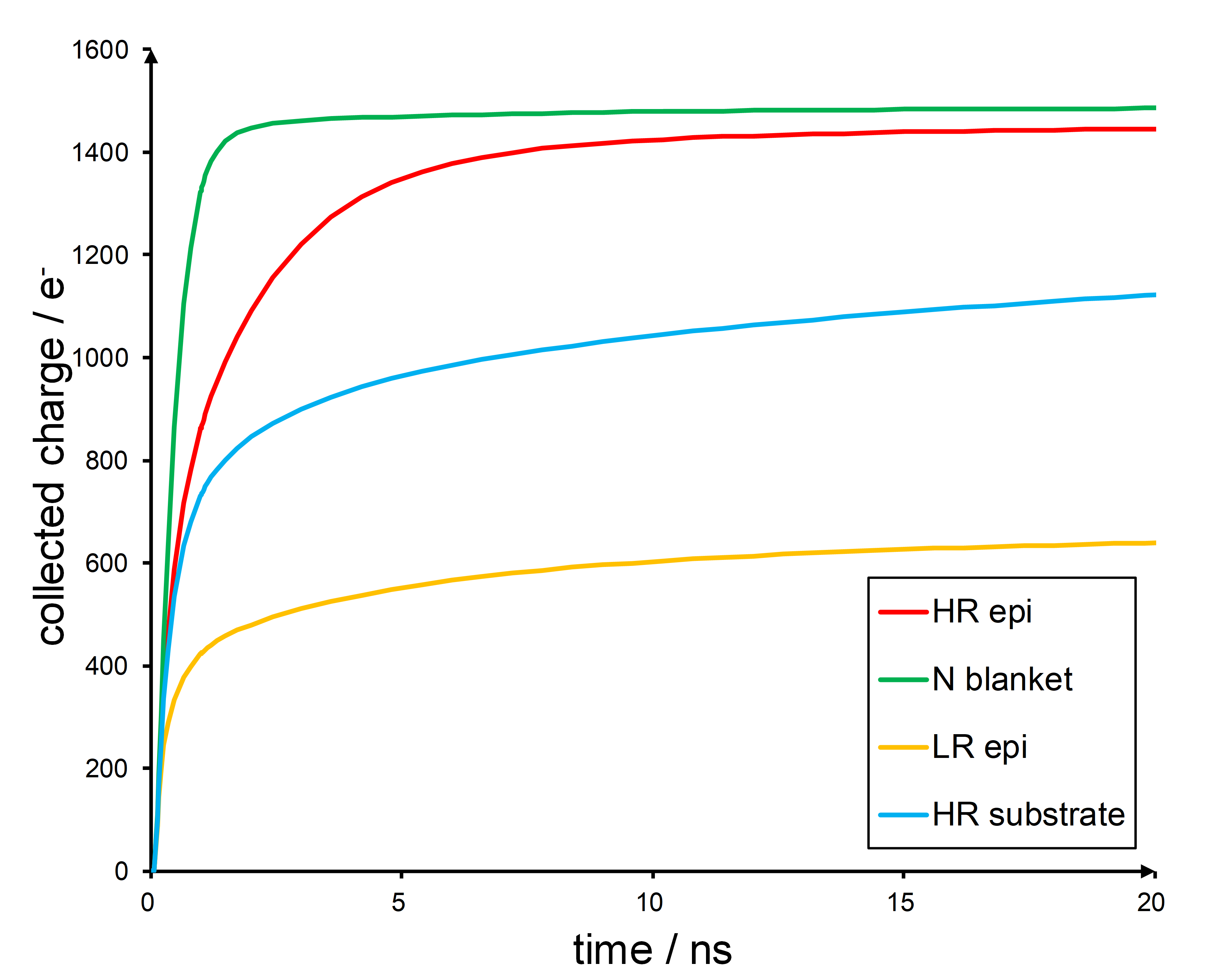}
        \caption{}
    \end{subfigure}
    \hfill
    \begin{subfigure}{0.45\textwidth}
        \centering
        \includegraphics[width=\linewidth]{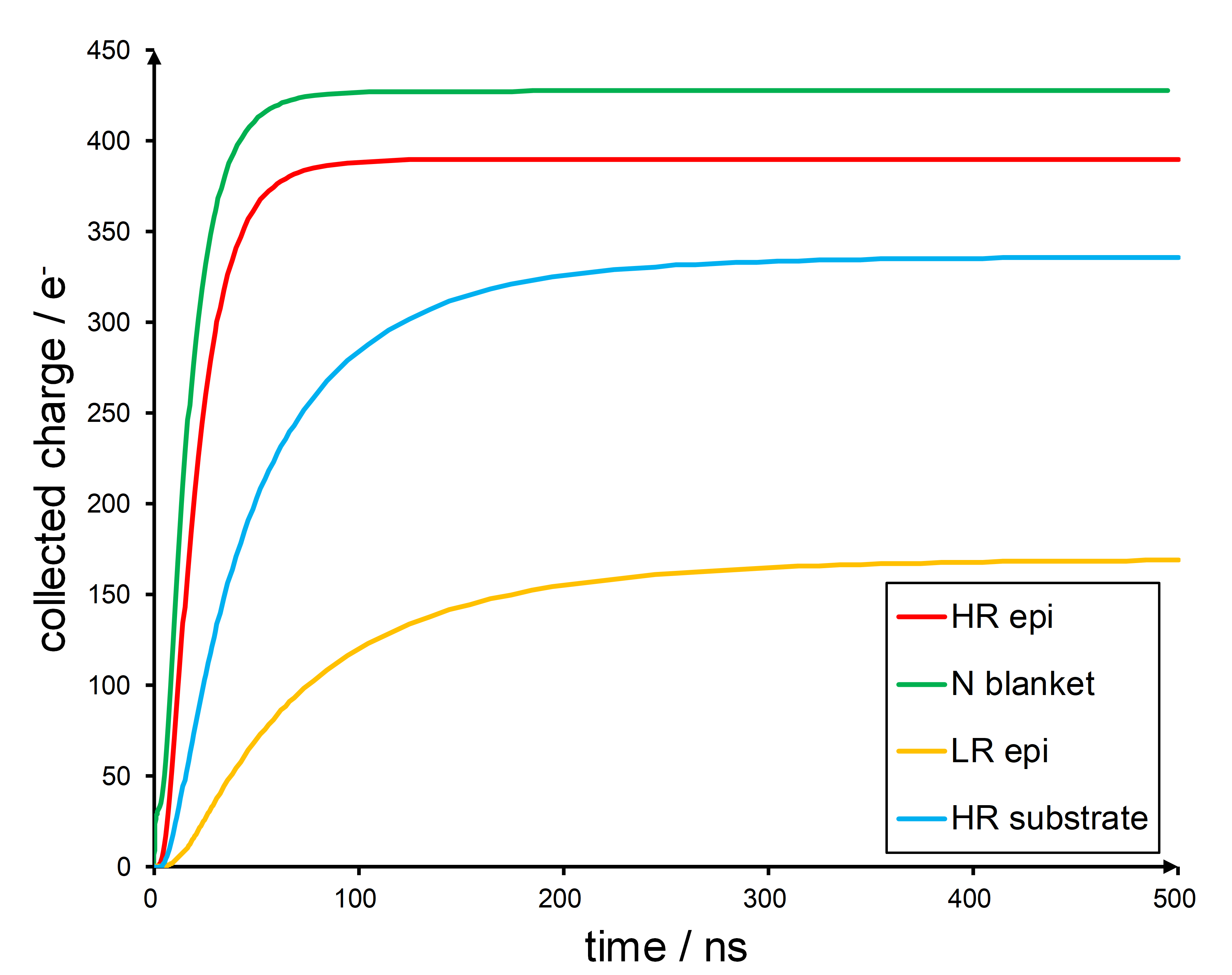}
        \caption{}
    \end{subfigure}
\caption{The collected charge signals for different process variants when the ionization track crosses the pixel center (a) and the pixel corner (b). Note that the time scales are different.} \label{signal}
\end{figure}

The pixel's response to ionizing particles is evaluated through transient simulation performed at $V_{sub}=\SI{-6}{V}$. Simulations are conducted under two extreme conditions: the incident track located at the pixel center and at the pixel corner. In order to avoid edge effects, a $3\times3$ pixel array consisting of identical pixels is modeled, and only the signal from the central pixel is extracted. Collected charge signals of different processes are shown in Fig.~\ref{signal}. Several key observations are summarized below: 

% \begin{figure}[htbp]
% \centering
% \includegraphics[width=0.6\textwidth]{charge signal.png}
% \caption{The collected charge signals for different process variants when the ionization track crosses the pixel center (a) and the pixel corner (b). Note that the time scales are different.} \label{signal}
% \end{figure}

\begin{itemize}
    \item The total charge collected in the corner-incidence scenario is approximately one-quarter of that in the center-incidence scenario. This discrepancy is due to charge-sharing effects when a particle strikes near the corner, with the originally deposited charge being nearly equally distributed among the four adjacent pixels.
    \item For the ``N blanket'' and ``HR epi'' processes, the collected charge exceeds the expectation based on a \SI{20}{\micro\meter} epitaxial layer alone (cf. Table~\ref{tab1})---\SI{1200}{e^-} and \SI{300}{e^-} in the center and corner scenarios, respectively. This result suggests a non-negligible contribution from the substrate to the overall collected charge.
    \item Both the total collected charge and the collection speed (defined as the time difference between reaching 10\% and 90\% of the maximum collected charge) follow the order: ``N blanket''\textgreater``HR epi''\textgreater``HR substrate''\textgreater``LR epi''. This result can serve as a practical criterion for evaluating the charge collection efficiency of different fabrication processes.
    % \textcolor{red}{This result gives a criterion on the charge collection efficiency of these processes}.
    \item The ``LR epi'' process exhibits a significantly lower collected charge, primarily due to its thinner epitaxial layer and lower resistivity (cf. Table~\ref{tab1}).
    \item Assuming a pixel threshold of \SI{150}{e^-}, the time-of-arrival for center- and corner-incidences in the ``N blanket'' process is determined to be \SI{40}{ps} and \SI{11}{ns}, respectively. These results indicate that an intrinsic time resolution of \SI{10}{ns} is achievable.
\end{itemize}

Additionally, the results of three geometrical variants in Fig.~\ref{geometry} are compared. The ``HR epi'' process, as a representative, is used in this simulation as consistent trends have been observed across different processes. The most notable result is the larger sensor capacitance value in the two strip-like pixel designs (cf. Fig.~\ref{capacitance}). At $V_{sub}=\SI{-6}{V}$, the sensor capacitance is \SI{2.7}{fF} for the small pixel, compared to \SI{39}{fF} for the active-connect pixel and \SI{26}{fF} for the metal-connect pixel. This increase in capacitance is expected to reduce the signal amplitude. However, the strip-like pixels demonstrate faster charge collection and higher amount of total charge, particularly at the pixel corners. A comprehensive evaluation of these variants will only be feasible after considering the Monte Carlo results discussed in the following.

\subsection{Sensor response to minimum ionizing particles}
\label{MIP}

The response of MAPS to MIP is investigated with \SI{1}{GeV/c} $\mu^-$ incident on the sensor. Unlike the TCAD simulation discussed in the previous section, which provides a deterministic approximation of charge deposition, here a Geant4-based Monte Carlo simulation is employed to account for the stochastic nature of particle energy loss and secondary particle generation. In order to determine the detection efficiency, a square region of $\pm 3$ times the pixel pitch around the true hit position is searched for any clusters exceeding threshold. The efficiency is consequently defined as the probability of finding an associated cluster with the hit.

\begin{figure}[htbp]
\centering
    \begin{subfigure}{0.45\textwidth}
        \centering
        \includegraphics[width=\linewidth]{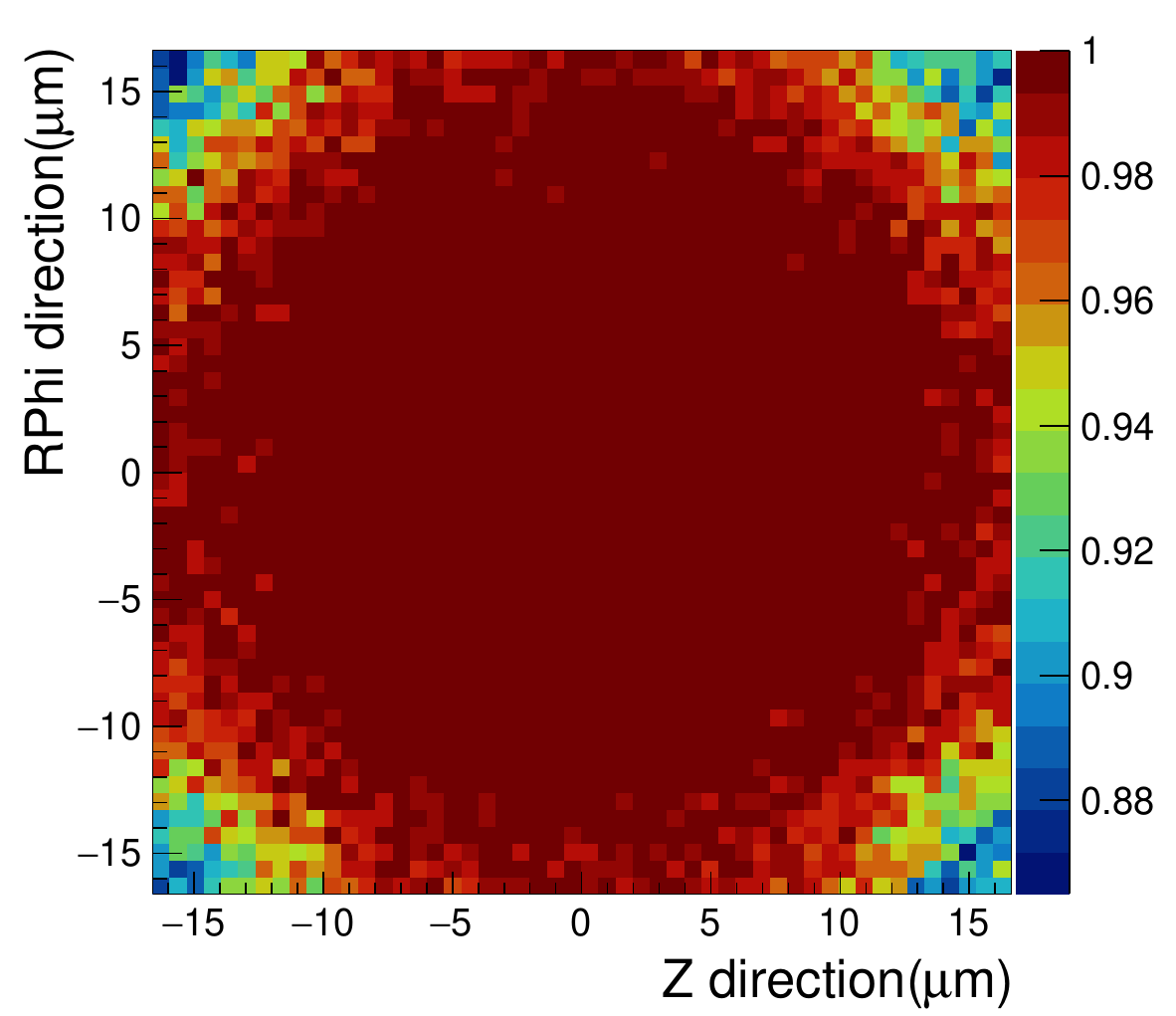}
        \caption{HR epi, average efficiency 98.9\%}
    \end{subfigure}
    \hfill
    \begin{subfigure}{0.45\textwidth}
        \centering
        \includegraphics[width=\linewidth]{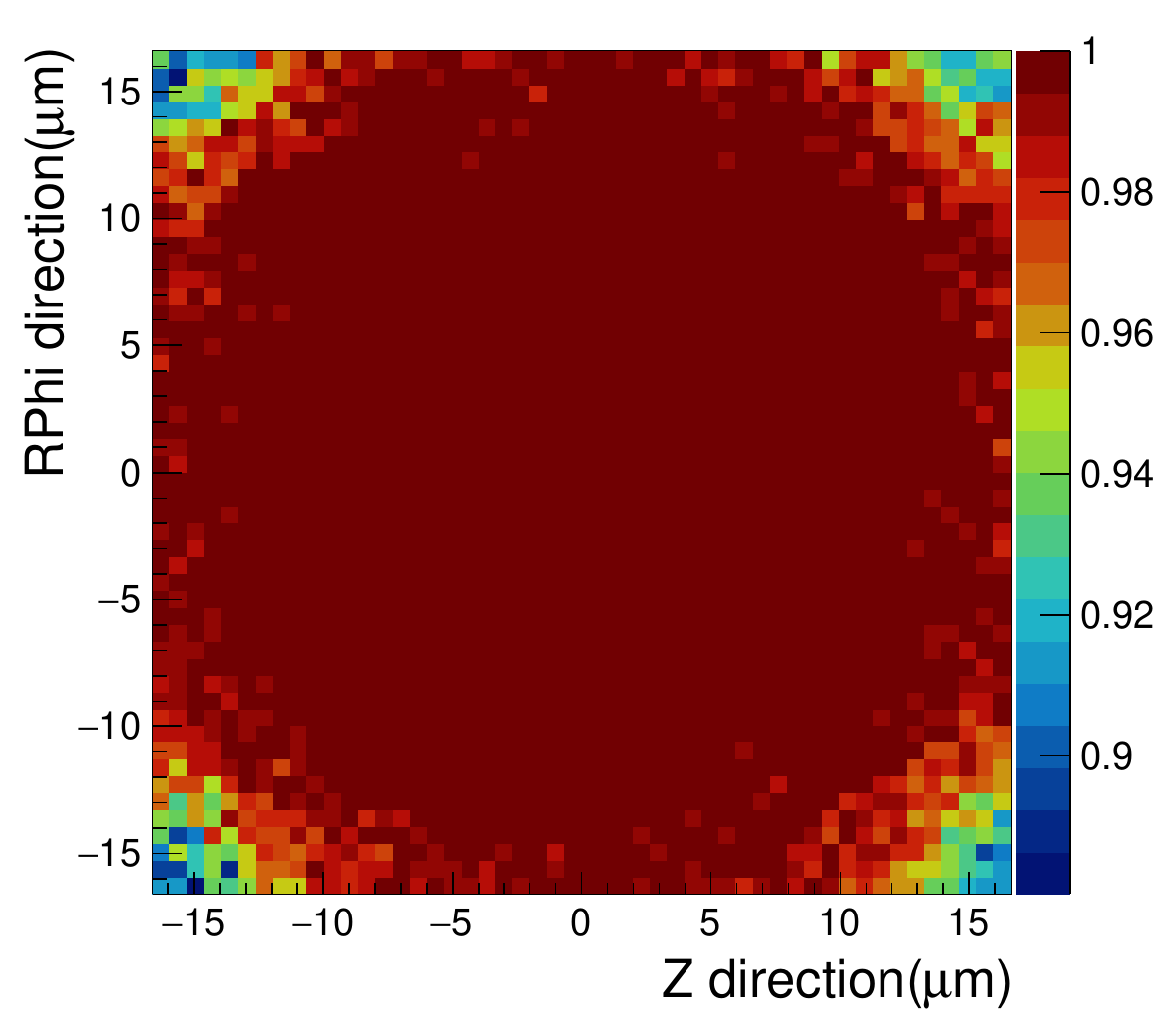}
        \caption{N blanket, average efficiency 99.2\%}
    \end{subfigure}
    \hfill
    \begin{subfigure}{0.45\textwidth}
        \centering
        \includegraphics[width=\linewidth]{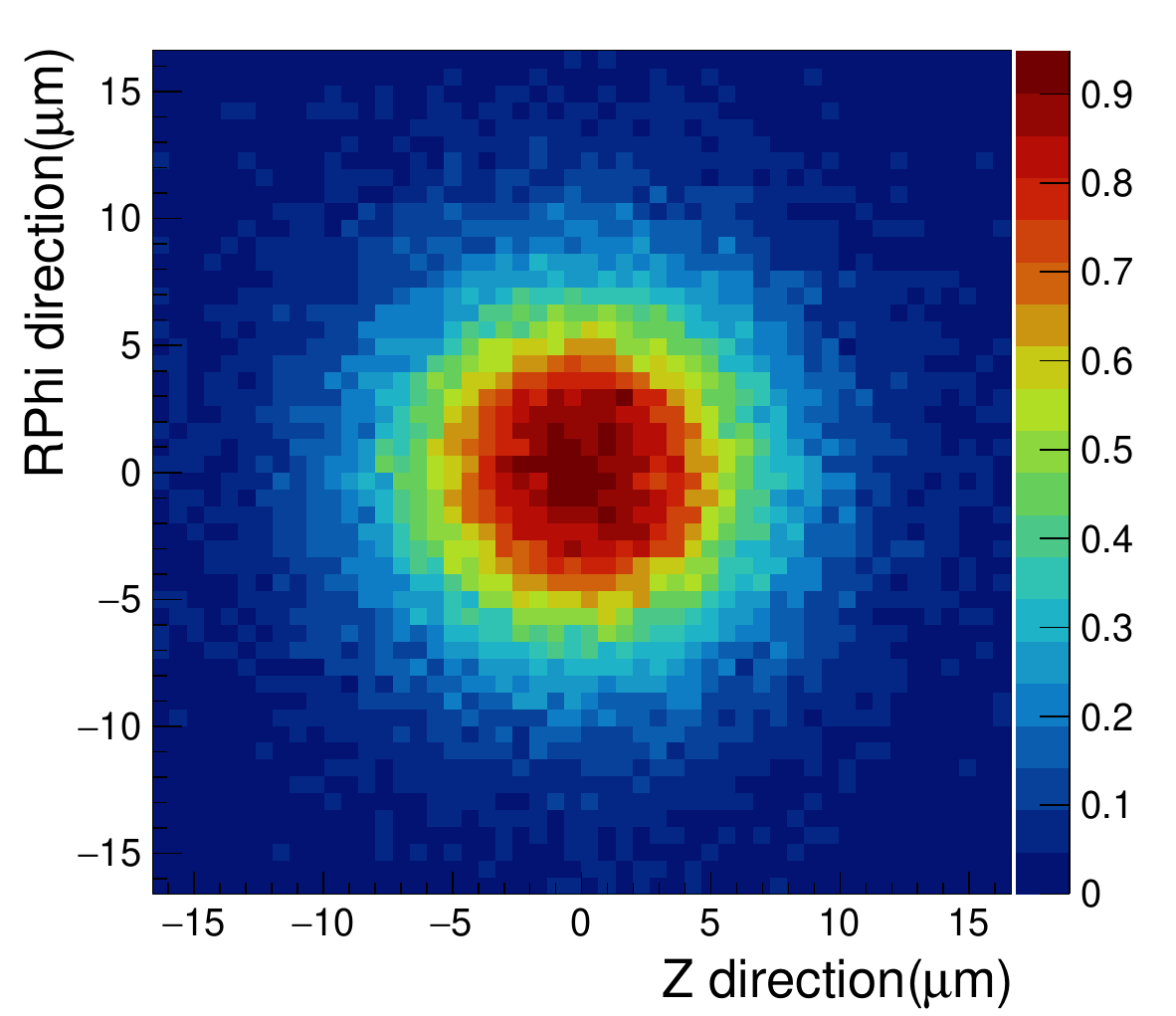}
        \caption{LR epi, average efficiency 16.8\%}
    \end{subfigure}
    \hfill
    \begin{subfigure}{0.45\textwidth}
        \centering
        \includegraphics[width=\linewidth]{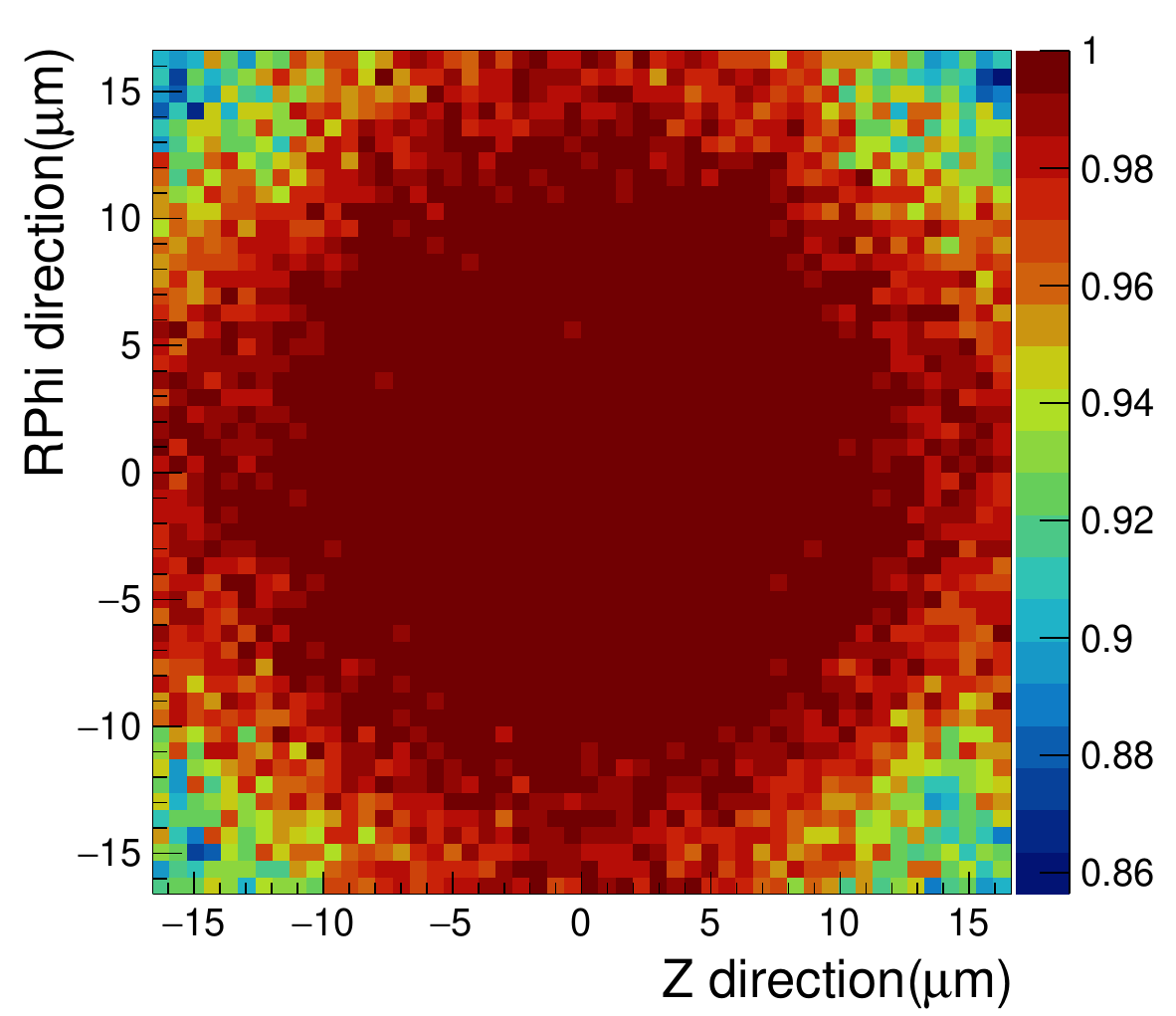}
        \caption{HR substrate, average efficiency 98.2\%}
    \end{subfigure}

\caption{In-pixel detection efficiency maps for four different process variants, at a pixel threshold of \SI{300}{e^-}, $V_{sub}=\SI{-6}{V}$, $V_{nw}=\SI{0.8}{V}$. The average detection efficiencies are also displayed.} \label{efficiency}
\end{figure}

\begin{figure}[htbp]
\centering
    \begin{subfigure}{0.9\textwidth}
        \centering
        \includegraphics[width=\linewidth]{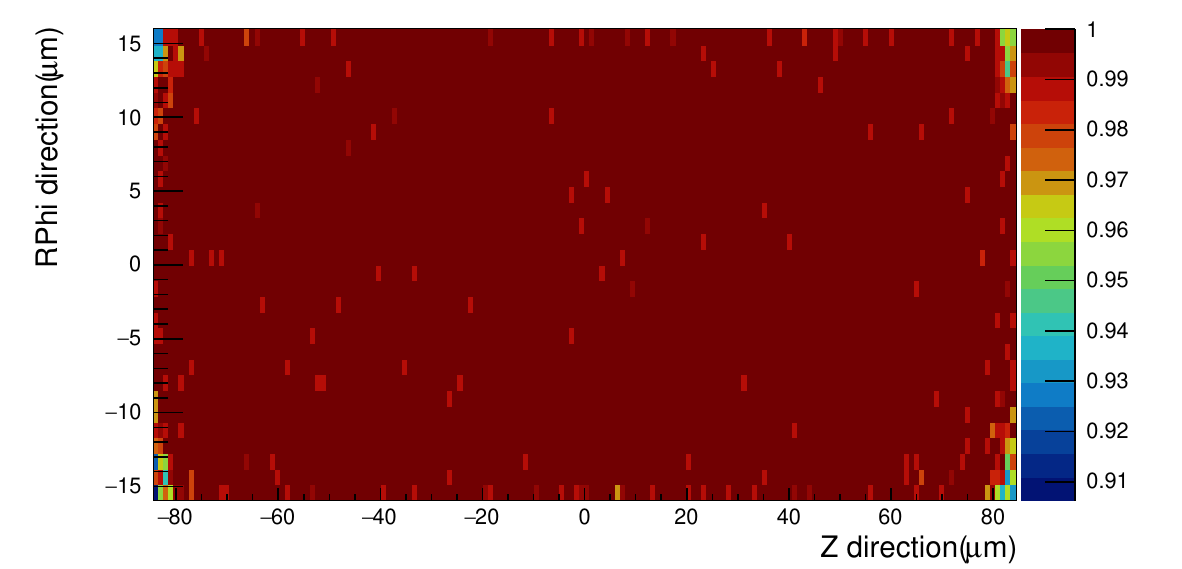}
        \caption{HR epi active-connect, average efficiency 99.7\%}
    \end{subfigure}
    \hfill
    \begin{subfigure}{0.9\textwidth}
        \centering
        \includegraphics[width=\linewidth]{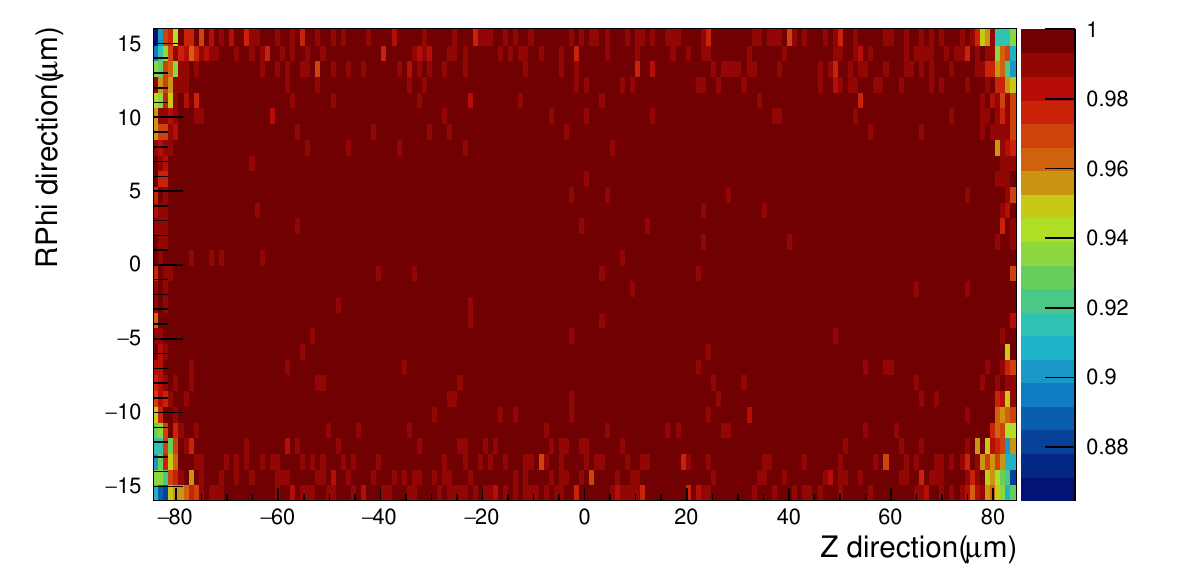}
        \caption{HR epi metal-connect, average efficiency 99.2\%}
    \end{subfigure}

\caption{In-pixel detection efficiency maps for two strip-like pixel designs, at a pixel threshold of \SI{300}{e^-}, $V_{sub}=\SI{-6}{V}$, $V_{nw}=\SI{0.8}{V}$.} \label{efficiency2}
\end{figure}

Fig.~\ref{efficiency} and \ref{efficiency2} present the detection efficiency as a function of in-pixel hit position at a threshold of \SI{300}{e^-}, which is chosen as a common reference in this study~\footnote{This threshold is motivated by the following consideration: for a \SI{20}{\micro\meter} epitaxial layer, a MIP deposits approximately \SI{1400}{e^-}, and even in the corner-sharing scenario each pixel is expected to collect about \SI{400}{e^-} (cf. Fig.~\ref{signal}), which ensures robust detection efficiency.}. Results for the standard pixel geometry at a lower threshold are provided in~\ref{app0} for comparison with other experiments. The ``LR epi'' process exhibits the lowest overall efficiency of 17\% while the efficiency for other processes remains above 98\%. This deterioration arises from the combined effects of a reduced depletion region and a short carrier lifetime, which together severely limit charge collection several micrometers away from the pixel center~\cite{STAR_HR_epi}. In addition, for the two strip-like sensor designs, the efficiency distribution is notably more uniform across the pixel area, a feature that is critical for their practical application in MAPS.
% \begin{figure}[htbp]
% \centering
% \includegraphics[width=0.6\textwidth]{efficiency_map.png}
% \caption{In-pixel detection efficiency maps for four different process variants, at a pixel threshold of \SI{300}{e^-}, $V_{sub}=\SI{-6}{V}$, $V_{nw}=\SI{0.8}{V}$. The average detection efficiencies are also displayed in the plot.} \label{efficiency}
% \end{figure}

\begin{figure}[htbp]
\centering
    \begin{subfigure}{0.45\textwidth}
        \centering
        \includegraphics[width=\linewidth]{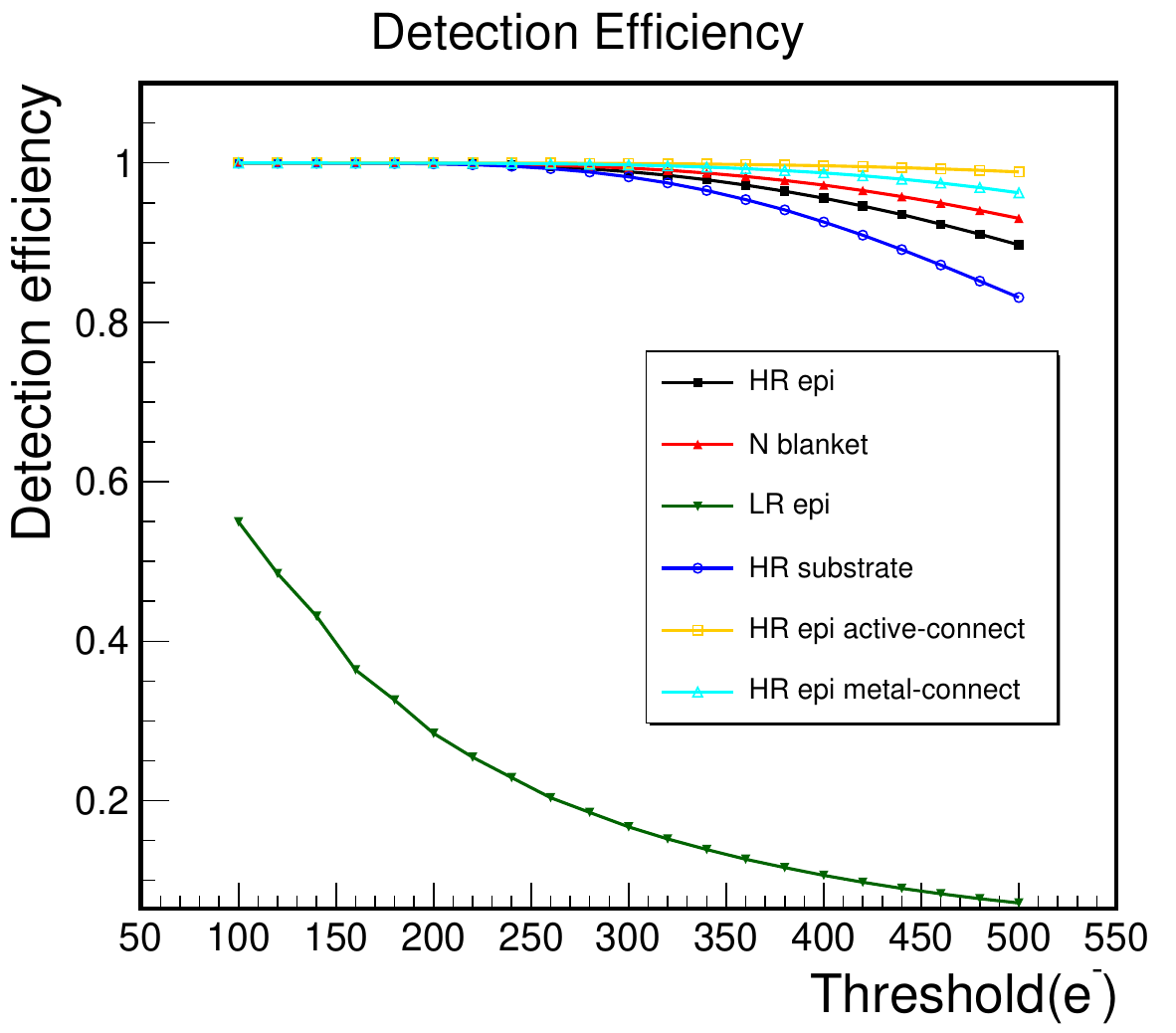}
        \caption{}
    \end{subfigure}
    \hfill
    \begin{subfigure}{0.45\textwidth}
        \centering
        \includegraphics[width=\linewidth]{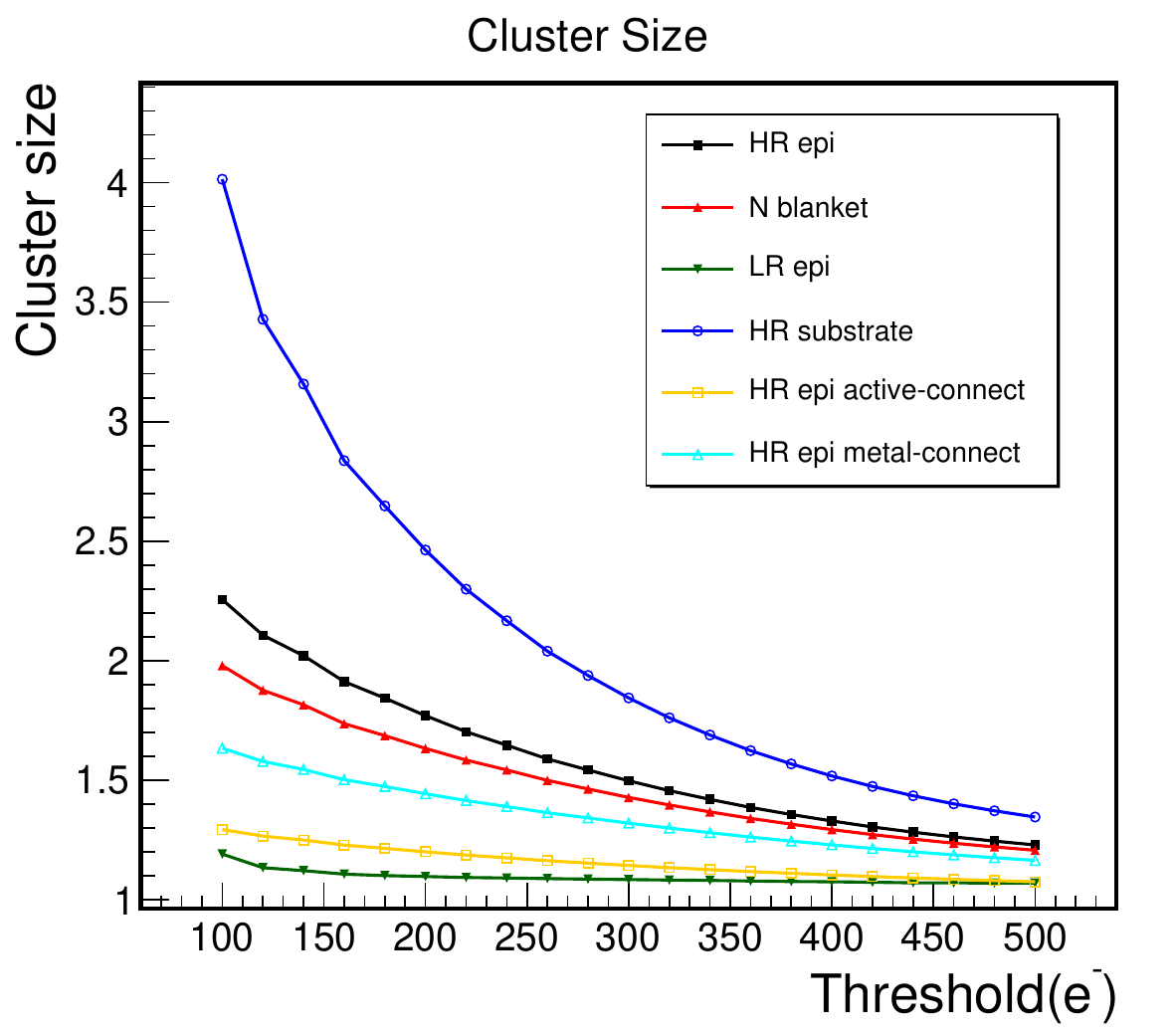}
        \caption{}
    \end{subfigure}
    \hfill
    \begin{subfigure}{0.45\textwidth}
        \centering
        \includegraphics[width=\linewidth]{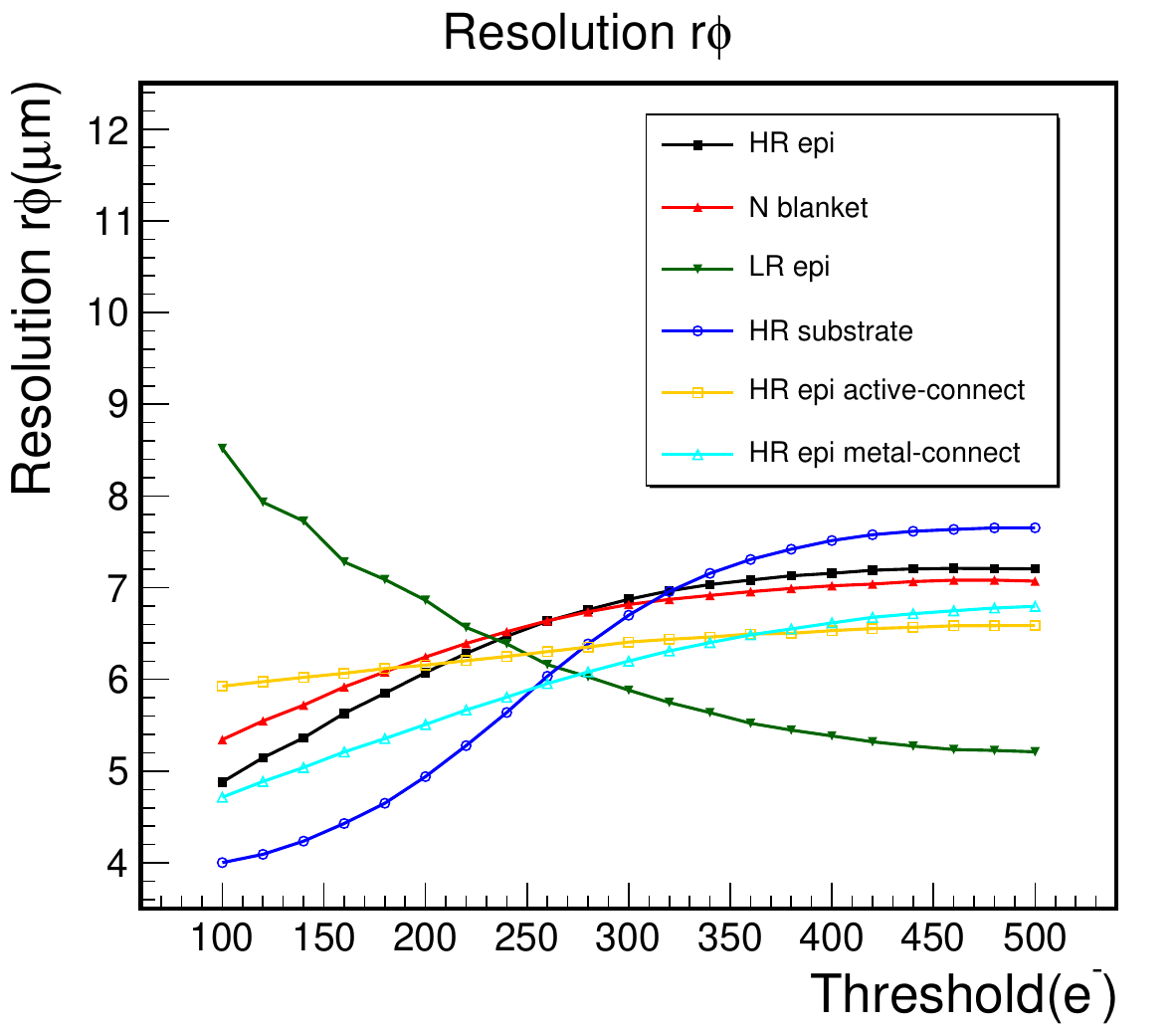}
        \caption{}
    \end{subfigure}
    \hfill
    \begin{subfigure}{0.45\textwidth}
        \centering
        \includegraphics[width=\linewidth]{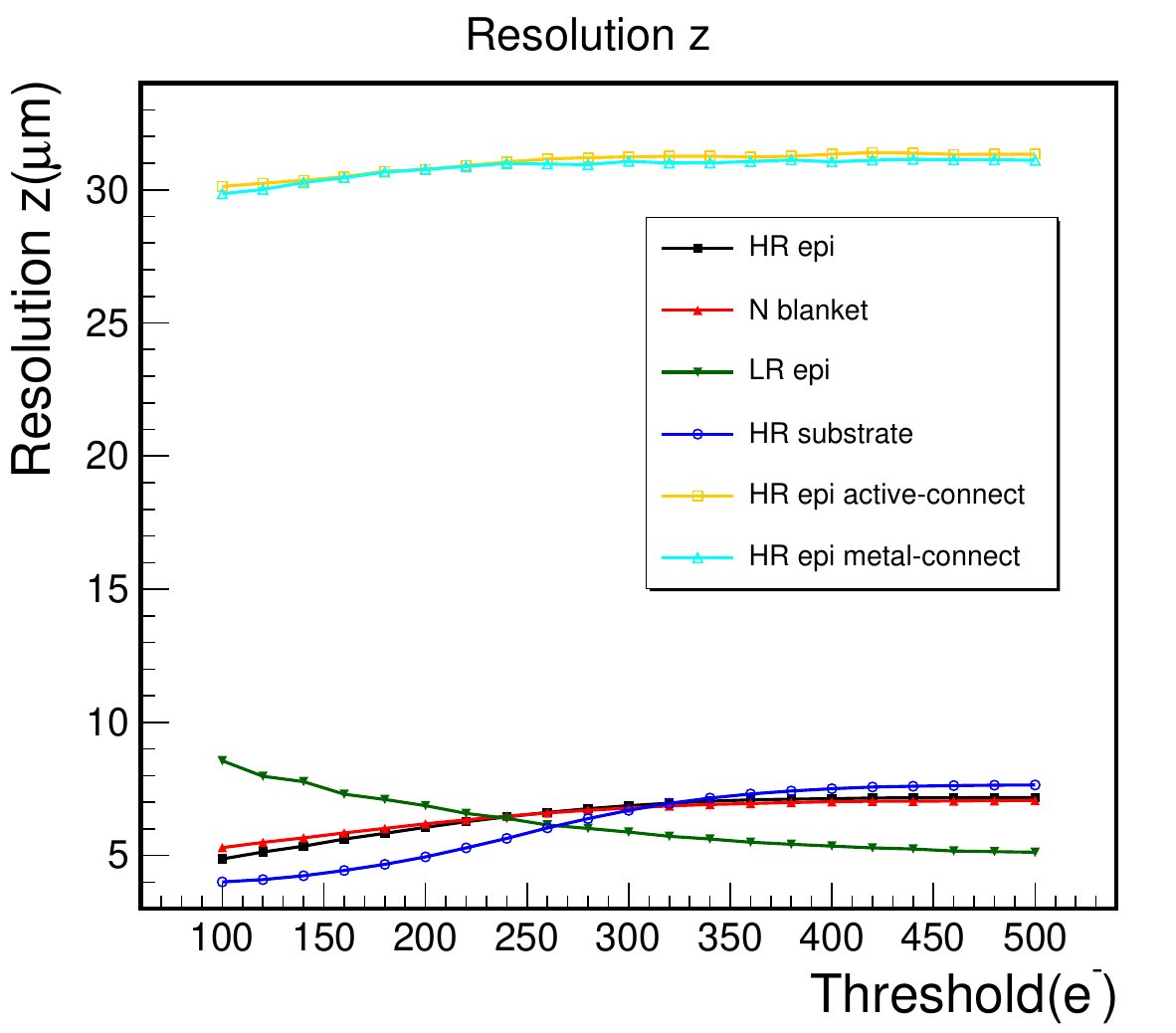}
        \caption{}
    \end{subfigure}
\caption{MIP response performance of different MAPS variants with varying pixel threshold, at $V_{sub}=\SI{-6}{V}$, $V_{nw}=\SI{0.8}{V}$: detection efficiency (a), cluster size (b), spatial resolution in $r-\phi$ direction (c) and z direction (d).} \label{scan}
\end{figure}

The main performance parameters---spatial resolution, detection efficiency, and cluster size---are shown in Fig.~\ref{scan} as functions of the pixel threshold, where threshold dispersion and electronic noise are not considered. Results for two strip-like pixel designs in ``HR epi'' process are also presented for comparison. Hits are reconstructed using charge-weighted center-of-gravity method to calculate the spatial resolution. The resolutions in the $r-\phi$ and $z$ directions correspond to the short and long sides of the pixel pitch, respectively. The results lead to several noteworthy conclusions:

% \begin{figure}[htbp]
% \centering
% \includegraphics[width=0.6\textwidth]{threshold_scan_refined.png}
% \caption{MIP response performance of different MAPS variants with varying pixel threshold, at $V_{sub}=\SI{-6}{V}$, $V_{nw}=\SI{0.8}{V}$.} \label{scan}
% \end{figure}

\begin{itemize}
    \item For the ``LR  epi'' variant, the maximum efficiency is only about 55\% even at a threshold of \SI{100}{e^-}, indicating that a high-resistivity layer is essential for enhancing the charge collection performance of small collection electrode MAPS.
    \item Among the other three processes, the detection efficiency decreases gradually as the threshold increases. The ranking of detection efficiency aligns with the charge collection capability observed in TCAD simulations.
    \item The ``HR substrate'' variant features a considerably larger cluster size compared to the two variants with an epitaxial layer, particularly at low thresholds. This implies a stronger charge sharing effect, which arises not only from the larger effective sensitive thickness but also from the absence of an epitaxial layer/substrate boundary. In epitaxial processes, this HR-LR boundary creates a built-in electric field that helps confine carriers within the epitaxial layer, thereby limiting lateral diffusion and reducing cluster size. While stronger charge sharing improves spatial resolution at low thresholds, it degrades efficiency at high thresholds. Detection performance at an increased bias voltage for this process can be found in~\ref{app1}.
    \item The two strip-like pixel designs outperform all small pixel designs in terms of detection efficiency. Specifically, the ``active-connect'' design maintains an efficiency greater than 98\% at a threshold of \SI{500}{e^-}. Also, the cluster sizes for these two designs are among the smallest, indicating their superior charge collection ability.
\end{itemize}

\subsection{Sensor response to X-ray}
\label{xray}

Since a MIP causes continuous ionization along its trajectory, it is not an ideal measure for investigating charge collection at specific spatial points within the pixel. In contrast, X-rays deposit a well-defined amount of energy at a localized point, making them more suitable for studying spatial charge collection characteristics. For this reason, X-rays are widely used in MAPS performance studies, and simulating their response is particularly valuable as it facilitates direct comparison with experimental measurements. In this work, we therefore complement the MIP-based simulations with a dedicated study using an $^{55}$Fe source. To simplify the simulation, $^{55}$Mn-$K_{\alpha}$ and $^{55}$Mn-$K_{\beta}$ X-rays are generated directly using a particle gun, with their relative intensities taken into account. The simulation sample consists of 30.1\% of \SI{5.888}{keV} $^{55}$Mn-$K_{\alpha2}$ X-rays, 59.4\% of \SI{5.899}{keV} $^{55}$Mn-$K_{\alpha1}$ X-rays and 10.5\% of \SI{6.49}{keV} $^{55}$Mn-$K_{\beta}$ X-rays~\cite{nudat3}.

\begin{figure}[htbp]
\centering
    \begin{subfigure}{0.45\textwidth}
        \centering
        \includegraphics[width=\linewidth]{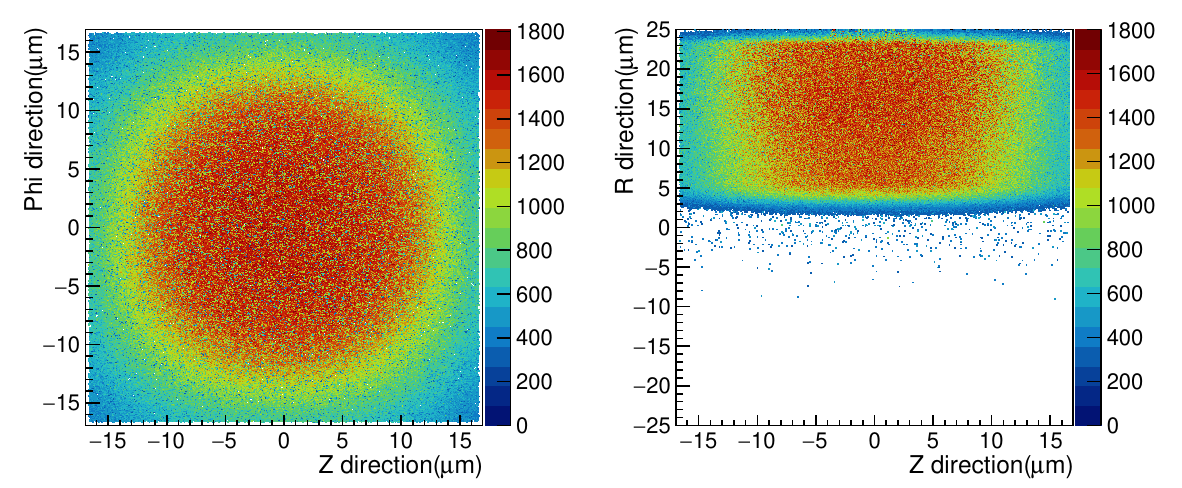}
        \caption{HR epi}
    \end{subfigure}
    \hfill
    \begin{subfigure}{0.45\textwidth}
        \centering
        \includegraphics[width=\linewidth]{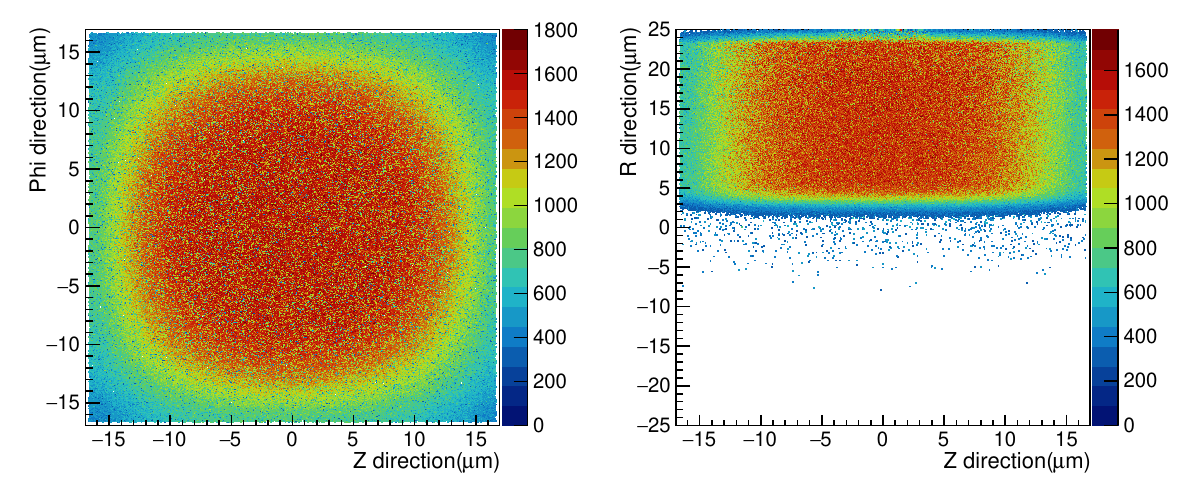}
        \caption{N blanket}
    \end{subfigure}
    \hfill
    \begin{subfigure}{0.45\textwidth}
        \centering
        \includegraphics[width=\linewidth]{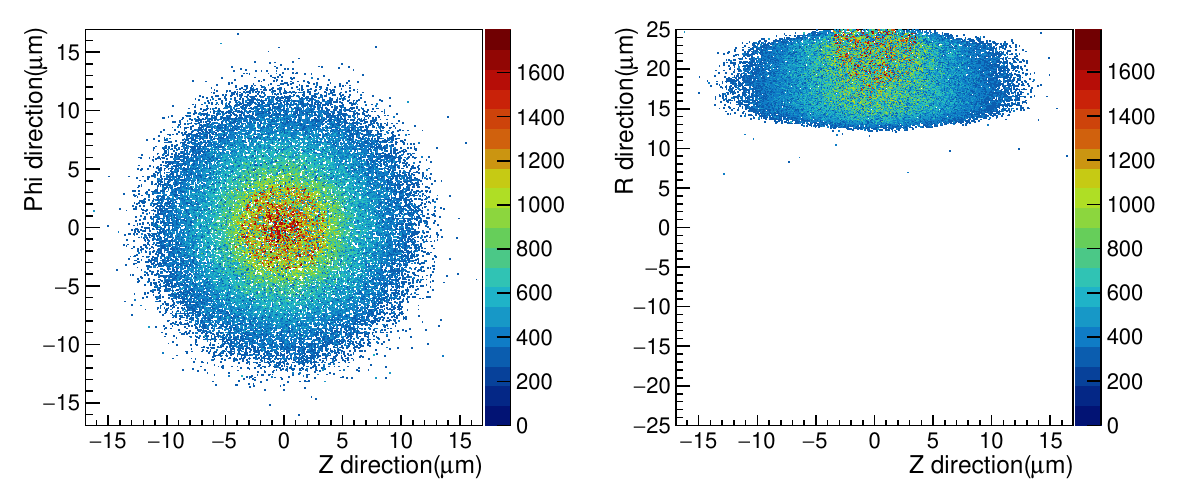}
        \caption{LR epi}
    \end{subfigure}
    \hfill
    \begin{subfigure}{0.45\textwidth}
        \centering
        \includegraphics[width=\linewidth]{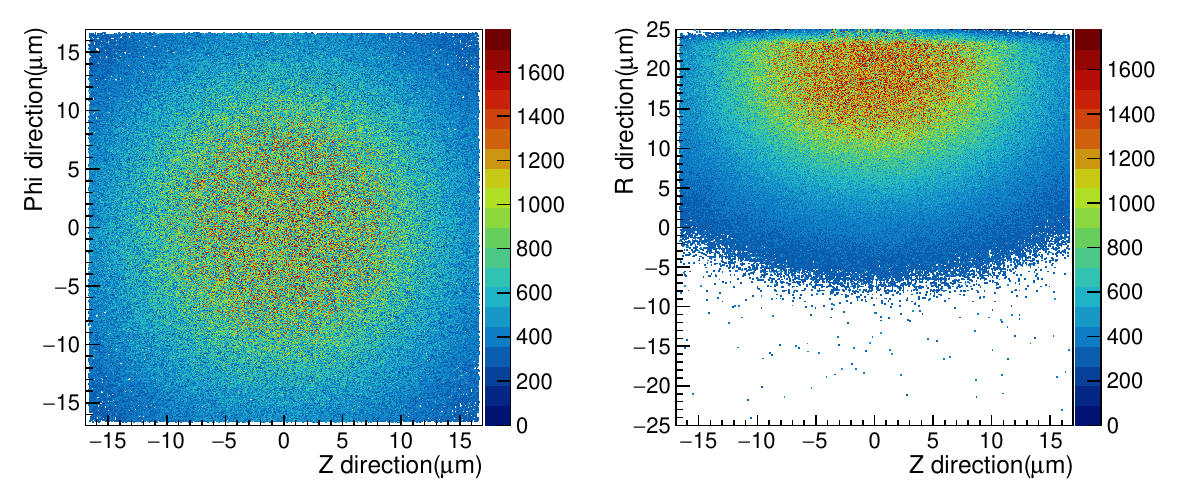}
        \caption{HR substrate}
    \end{subfigure}

\caption{Charge collection map for four different process variants at \SI{300}{e^-} threshold, $V_{sub}=\SI{-6}{V}$, $V_{nw}=\SI{0.8}{V}$: HR epi (a), N blanket (b), LR epi (c), and HR substrate (d). Color bar stands for the collected seed charge in terms of e\textsuperscript{-}.} \label{charge_collection1}
\end{figure}

\begin{figure}[htbp]
\centering
    \begin{subfigure}{\textwidth}
        \centering
        \includegraphics[width=\linewidth]{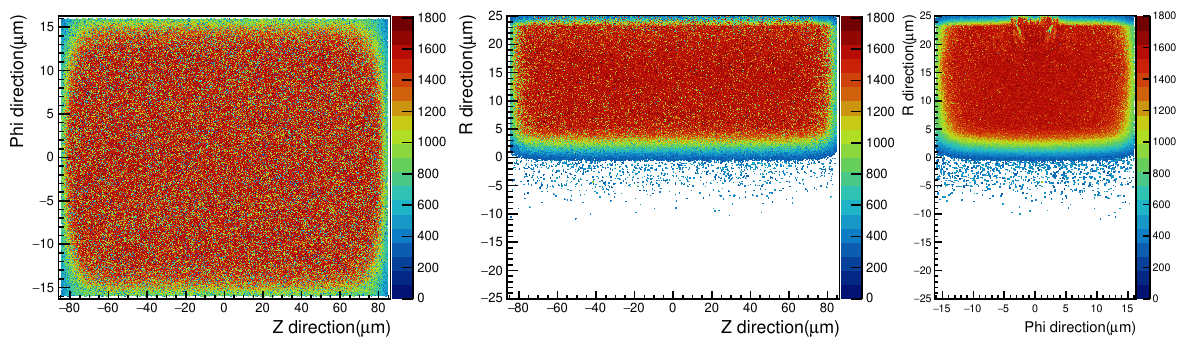}
        \caption{Active connect}
    \end{subfigure}
    \hfill
    \begin{subfigure}{\textwidth}
        \centering
        \includegraphics[width=\linewidth]{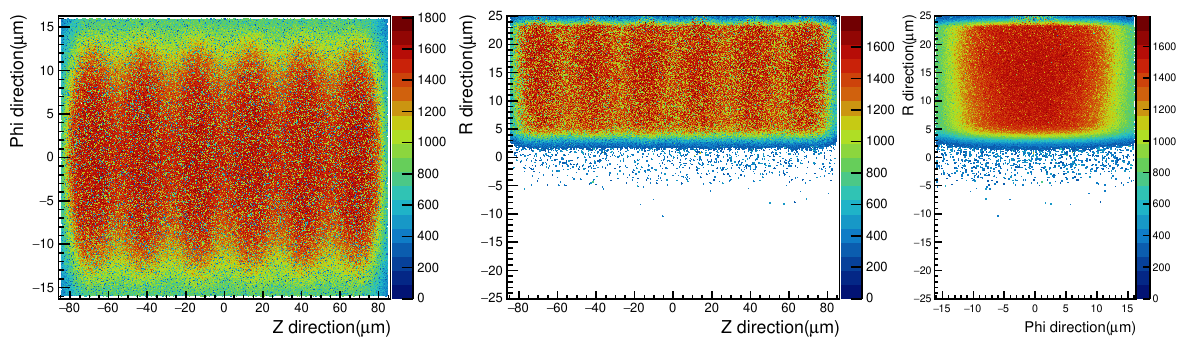}
        \caption{Metal connect}
    \end{subfigure}

\caption{Charge collection map for two strip-like pixel designs at \SI{300}{e^-} threshold, $V_{sub}=\SI{-6}{V}$, $V_{nw}=\SI{0.8}{V}$: active-connect (a), and metal-connect (b). Color bar stands for the collected seed charge in terms of e\textsuperscript{-}.} \label{charge_collection2}
\end{figure}

Charge collection maps are obtained for different variants (cf. Fig.~\ref{charge_collection1} and Fig.~\ref{charge_collection2}) with the X-ray being absorbed at different 3D position inside the pixel. The charge collection ability is represented by the collected charge of the seed pixel, which is the pixel within the cluster that collects the largest amount of charge. The plots are depicted as projections onto two dimensional planes for visualization purposes. In these plots, the $r$ direction corresponds to the pixel depth, the $\phi$ direction corresponds to the shorter pixel pitch, and the $z$ direction corresponds to the longer pixel pitch, which are in accordance with the pixel arrangements in the inner tracker of STCF. For the four process variants, only two projections are presented since the pixel pitches are identical in both directions.

% \begin{figure}[htbp]
% \centering
% \includegraphics[width=\textwidth]{charge_collection_4processes.png}
% \caption{Charge collection map for four different process variants at \SI{300}{e^-} threshold, $V_{sub}=\SI{-6}{V}$, $V_{nw}=\SI{0.8}{V}$: HR epi (a), N blanket (b), LR epi (c), and HR substrate (d). Color bar stands for the collected seed charge in terms of e\textsuperscript{-}.} \label{charge_collection1}
% \end{figure}

% \begin{figure}[htbp]
% \centering
% \includegraphics[width=\textwidth]{charge_collection_2geometries.png}
% \caption{Charge collection map for two strip-like pixel designs at \SI{300}{e^-} threshold, $V_{sub}=\SI{-6}{V}$, $V_{nw}=\SI{0.8}{V}$: active-connect (a), and metal-connect (b). Color bar stands for the collected seed charge in terms of e\textsuperscript{-}.} \label{charge_collection2}
% \end{figure}

The distinctive behaviours in charge collection are clearly illustrated in Fig.~\ref{charge_collection1} and Fig.~\ref{charge_collection2}:

\begin{itemize}
    \item Taking the ``HR epi'' small pixel design as an example, charge is fully collected by one pixel in its central regions. In the pixel corners, the seed charge amount drops due to both sharing with adjacent pixels and recombination loss. Additionally, two distinct boundaries in the $r$ direction are visible at \SI{2}{\micro\meter} and \SI{20}{\micro\meter} from the sensor surface. These boundaries are attributed to the low electric field regions inside the deep p-well and the substrate. Energy deposition occurring several micrometers below the substrate/epitaxial layer barrier is no longer detectable due to the high recombination rate.
    \item For the ``N blanket'' process, charge collection is enhanced, but the pixel corners still exhibit charge sharing effects due to the low lateral electric field. This has also been identified as the motivation for further process modification in~\cite{modified2}.
    \item The charge collection profile of the ``LR epi'' process provides a reasonable explanation for its poor detection performance in Section~\ref{MIP}. The full-charge-collection region is restricted to the small volume around the n-well, while a large portion of the pixel volume collects almost no charge.
    \item For the ``HR substrate'' process, charge can be collected within a much larger region, owing to the entire high-resistivity substrate. However, this process exhibits a faster decrease in charge collection ability as ionization position moves away from the center, compared to the ``HR epi'' process. One possible explanation is the absence of a doping concentration barrier between epitaxial layer and substrate, which partially acts as a reflecting boundary and improves the charge collection efficiency~\cite{simulation_ALPIDE}.
    \item The strip-like pixels demonstrate better charge collection performance, particularly in pixel corners, compared to the small pixel designs. This result is in accordance with the in-pixel efficiency maps shown in Fig.~\ref{efficiency2}. The ``active-connect'' type achieves full charge collection, i.e., the deposited charge is entirely collected by the seed pixel, across a large portion of the epitaxial layer. For the ``metal-connect'' type, the depletion boundary formed by the six individual n-well electrodes is visible in the $\phi-z$ plane.
\end{itemize}

\begin{figure}[htbp]
\centering
    \begin{subfigure}{0.45\textwidth}
    \centering
        \includegraphics[width=\linewidth]{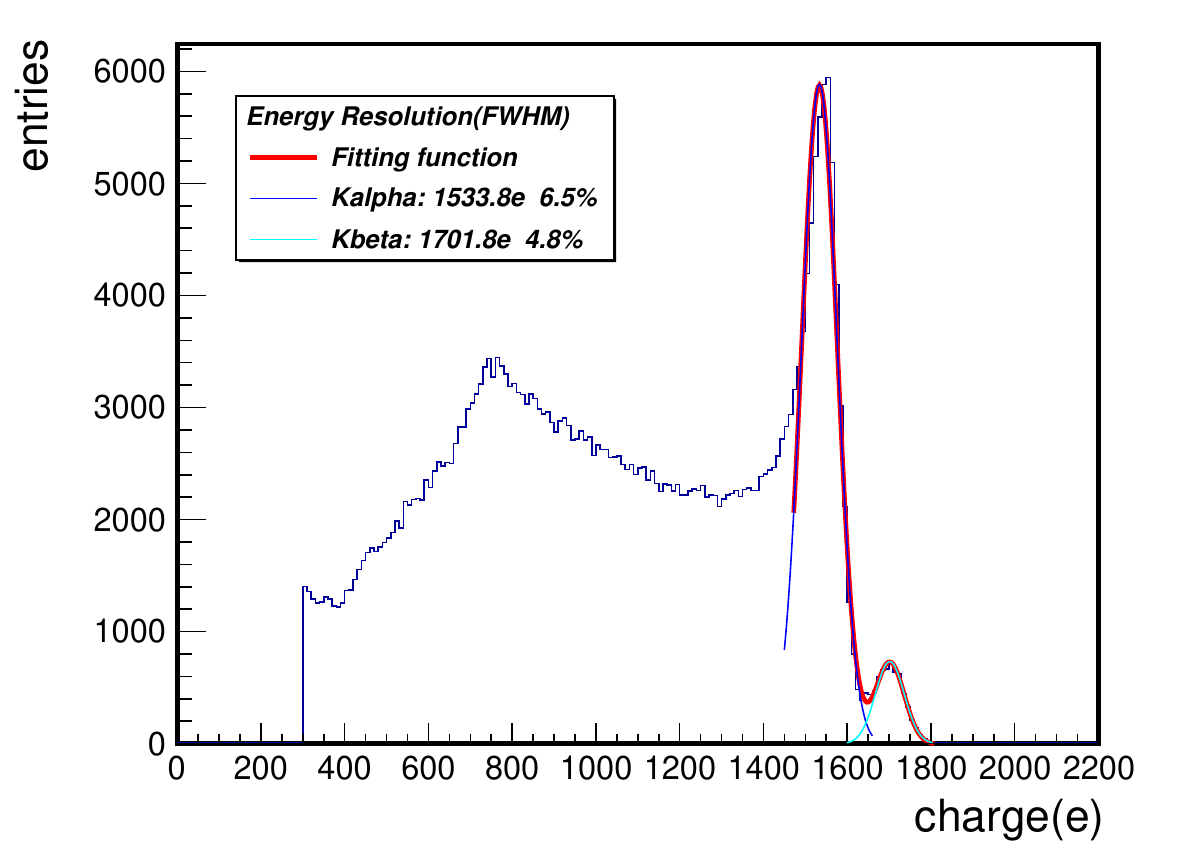}
        \caption{HR epi}
    \end{subfigure}
    \hfill
    \begin{subfigure}{0.45\textwidth}
        \centering
        \includegraphics[width=\linewidth]{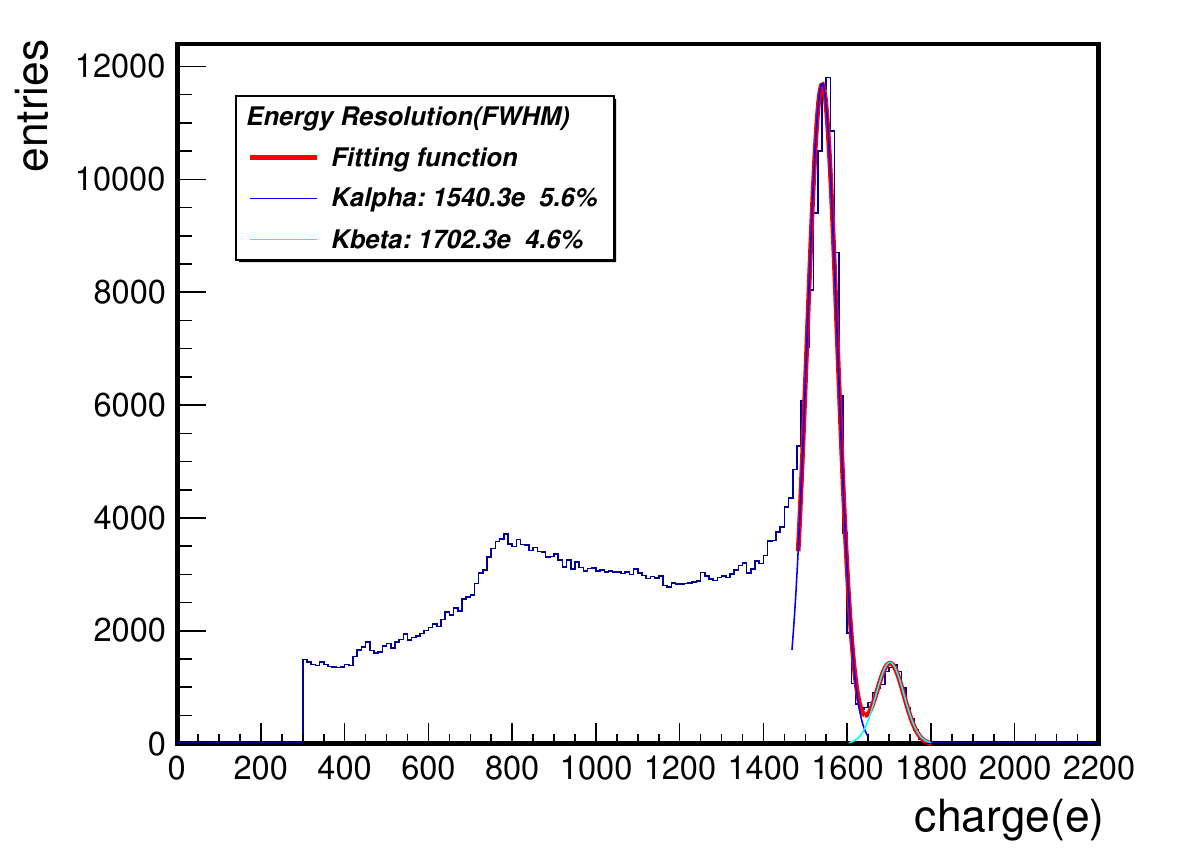}
        \caption{N blanket}
    \end{subfigure}
    \hfill
    \begin{subfigure}{0.45\textwidth}
        \centering
        \includegraphics[width=\linewidth]{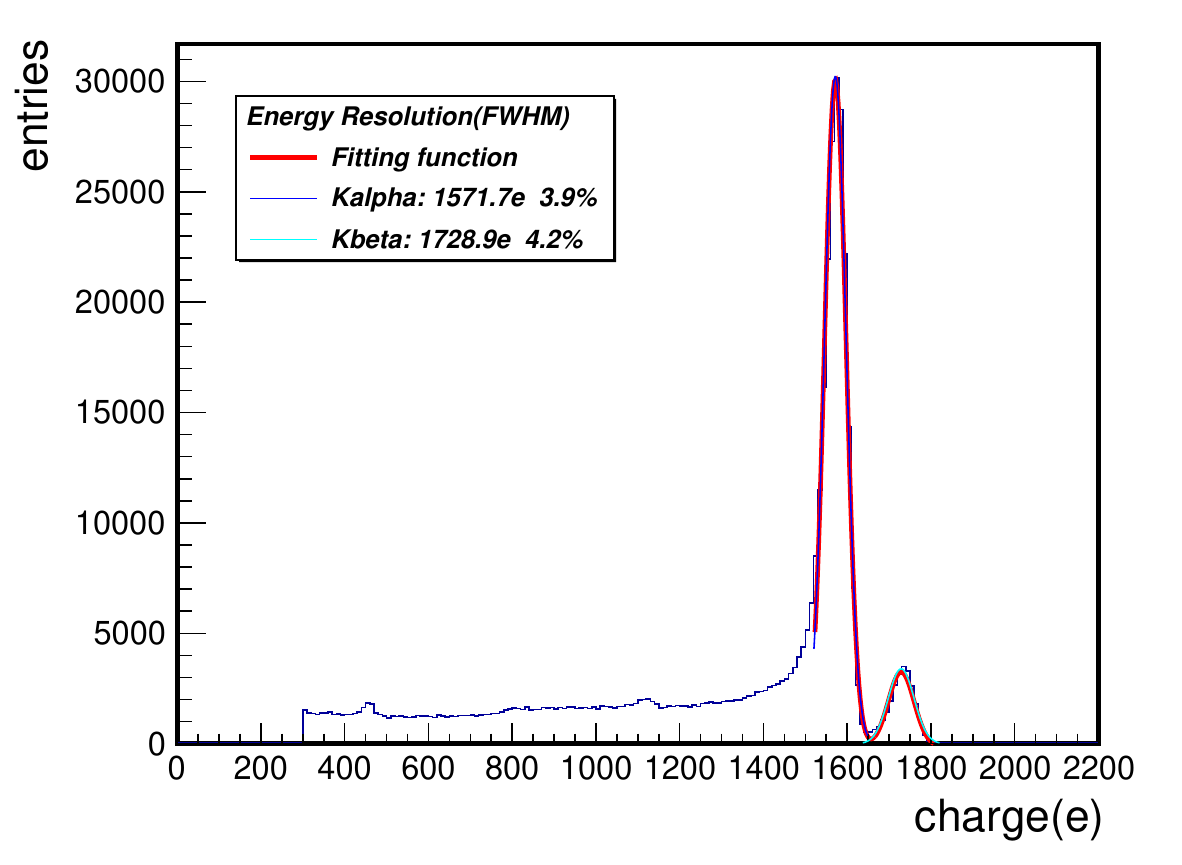}
        \caption{HR epi active connect}
    \end{subfigure}
    \hfill
    \begin{subfigure}{0.45\textwidth}
        \centering
        \includegraphics[width=\linewidth]{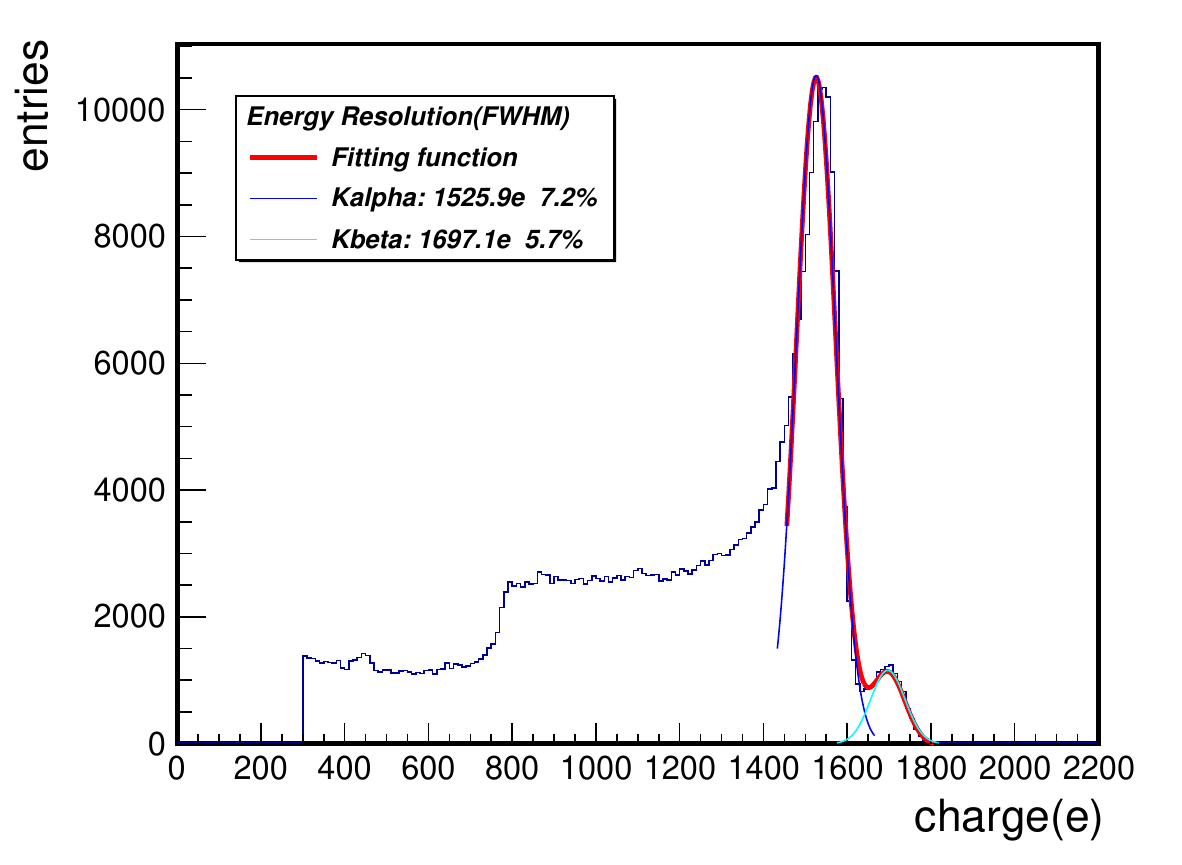}
        \caption{HR epi metal connect}
    \end{subfigure}

\caption{Seed pixel charge spectrum for different process and geometry variants at \SI{300}{e^-}, $V_{sub}=\SI{-6}{V}$, $V_{nw}=\SI{0.8}{V}$: HR epi small pixel (a), N blanket small pixel (b), HR epi active-connect strip-like pixel (c), and HR epi metal-connect strip-like pixel (d). The $K_\alpha$ and $K_\beta$ peaks are fitted with Gaussian functions.} \label{spectrum}
\end{figure}

The seed pixel charge spectrum is depicted in Fig.~\ref{spectrum}. The $K_\alpha$ and $K_\beta$ peaks are clearly distinguishable in all four variants shown. Among these, the ``HR epi active-connect'' pixel shows the best performance in terms of energy resolution and peak-to-total ratio, which is attributed to its superior charge collection efficiency. Additionally, it is observed that the fitted peak positions in all four variants are slightly lower than the expected peak positions (\SI{1620}{e^-} and \SI{1780}{e^-}), suggesting incomplete charge collection caused by recombination. While this is a clear outcome of the simulation, experimental studies typically assume the full-energy peak corresponds to complete collection, and further measurements with precise calibration would be required to validate this effect. The results for ``LR epi'' and ``HR substrate'' processes are not presented, as no distinct full energy peak is observed.
% \begin{figure}[htbp]
% \centering
% \includegraphics[width=\textwidth]{Fe55_spectrum_refined.png}
% \caption{Seed pixel charge spectrum for different process and geometry variants at \SI{300}{e^-}, $V_{sub}=\SI{-6}{V}$, $V_{nw}=\SI{0.8}{V}$: HR epi small pixel (a), N blanket small pixel (b), HR epi active-connect strip-like pixel (c), and HR epi metal-connect strip-like pixel (d). The $K_\alpha$ and $K_\beta$ peaks are fitted with Gaussian functions.} \label{spectrum}
% \end{figure}

\subsection{Discussion}
\label{MAPS_discussion}

Based on the simulation results presented in the previous sections, both the ``HR epi'' and ``N blanket'' processes demonstrate good detection performance. However, after considering process robustness and cost, the ``HR epi'' process is selected as the baseline for the STCF application.

Within this process, an enlarged pixel size with active n-well connection design further improves the charge collection efficiency inside the pixel with an anticipated spatial resolution meeting the requirements. Since the dominant source of power consumption arises from timestamp distribution, this geometry helps reduce the number of readout channels and thus maintain low power consumption. At the same time, it provides sufficient area for integrating more complex in-pixel circuitry for timing measurements, aligning well with the requirements of the STCF MAPS.

One potential concern with this design is the large sensor capacitance (refer to Section~\ref{TCAD}). To address this, special attention should be given to the analog front-end circuitry to ensure stable performance despite the increased input capacitance. Spectre-based simulation of the circuitry shows that a nominal pixel threshold of \SI{310}{e^-}, with an equivalent noise charge (ENC) of \SI{11}{e^-} and a threshold dispersion of \SI{5.7}{e^-} is achievable under a sensor capacitance of \SI{40}{fF} assuming TowerJazz \SI{180}{nm} CMOS technology, together with a power consumption of \SI{55.7}{mW/cm^2}. Notably, even though the strip-like pixels exhibit more than an order of magnitude higher capacitance compared to standard small-pitch pixels, the achievable threshold remains moderate---for instance, comparing to the ALPIDE sensor operating at 100-\SI{150}{e^-}. This confirms the effectiveness of a dedicated front-end circuit design tailored to this sensor geometry. For a more detailed description of the circuit design and its performance, readers are referred to Ref.~\cite{XUAN2025170725}.

These results demonstrate that, by jointly optimizing the sensor and the readout electronics, both high detection efficiency and low power consumption can be achieved simultaneously. Consequently, the ``HR epi'' process with the active n-well connect strip-like pixel has been chosen as the preferred design for the STCF inner tracker performance study in the following section.
% Nevertheless, it is important to note that prototype chips have been designed for both the active-connect and metal-connect pixel types. A final decision will be made following dedicated tests and detailed comparisons of their performance.

\section{Inner tracker performance simulation}
\label{itkm_sim}

\subsection{Detector design and implementation}
\label{detector_setup}
\begin{figure}[htbp]
\centering
\begin{subfigure}{0.4\textwidth}
    \centering
    \includegraphics[width=\linewidth]{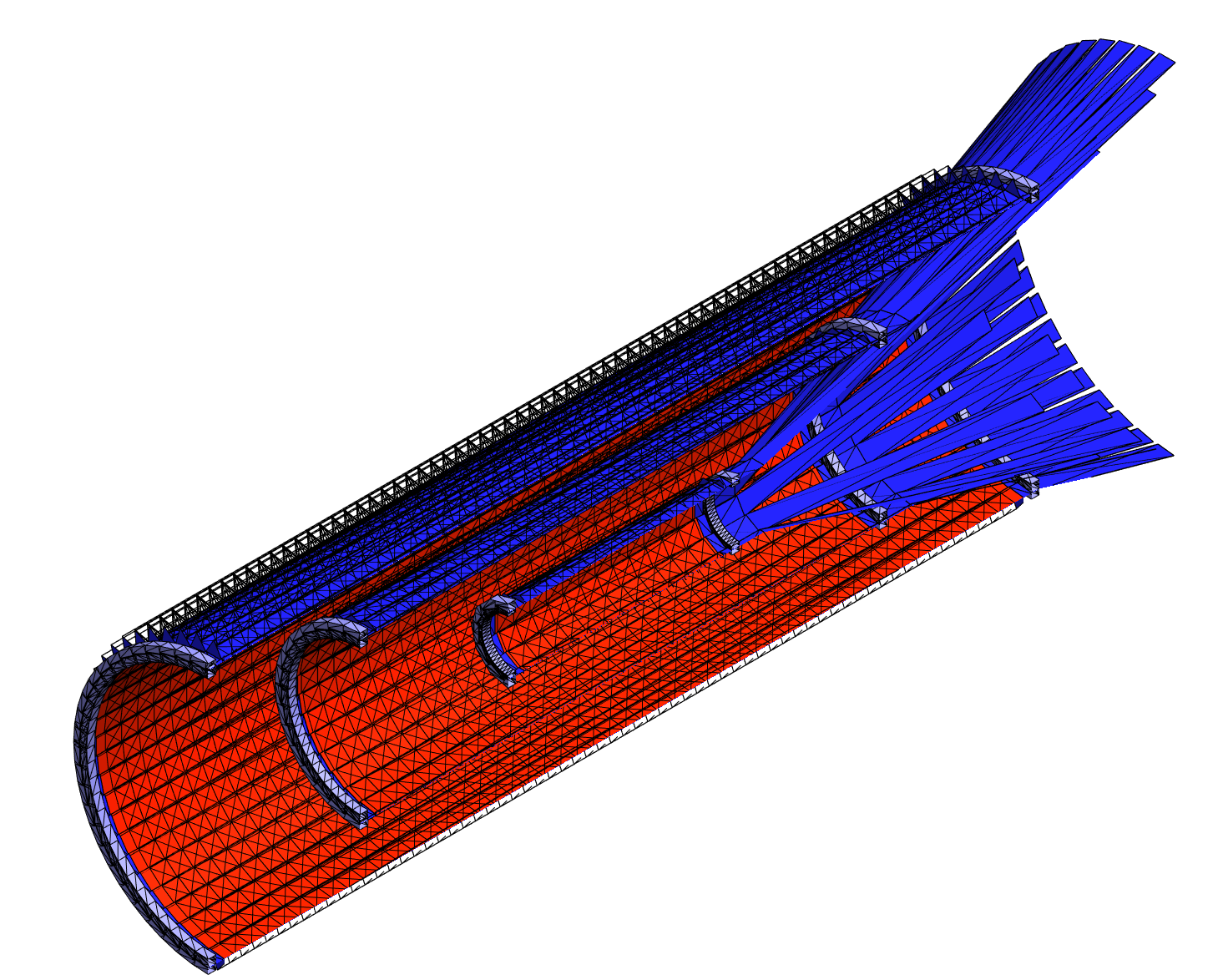}
    \caption{}
    \label{ITKM_a}
\end{subfigure}
\hfill
\begin{subfigure}{0.55\textwidth}
    \centering
    \includegraphics[width=\linewidth]{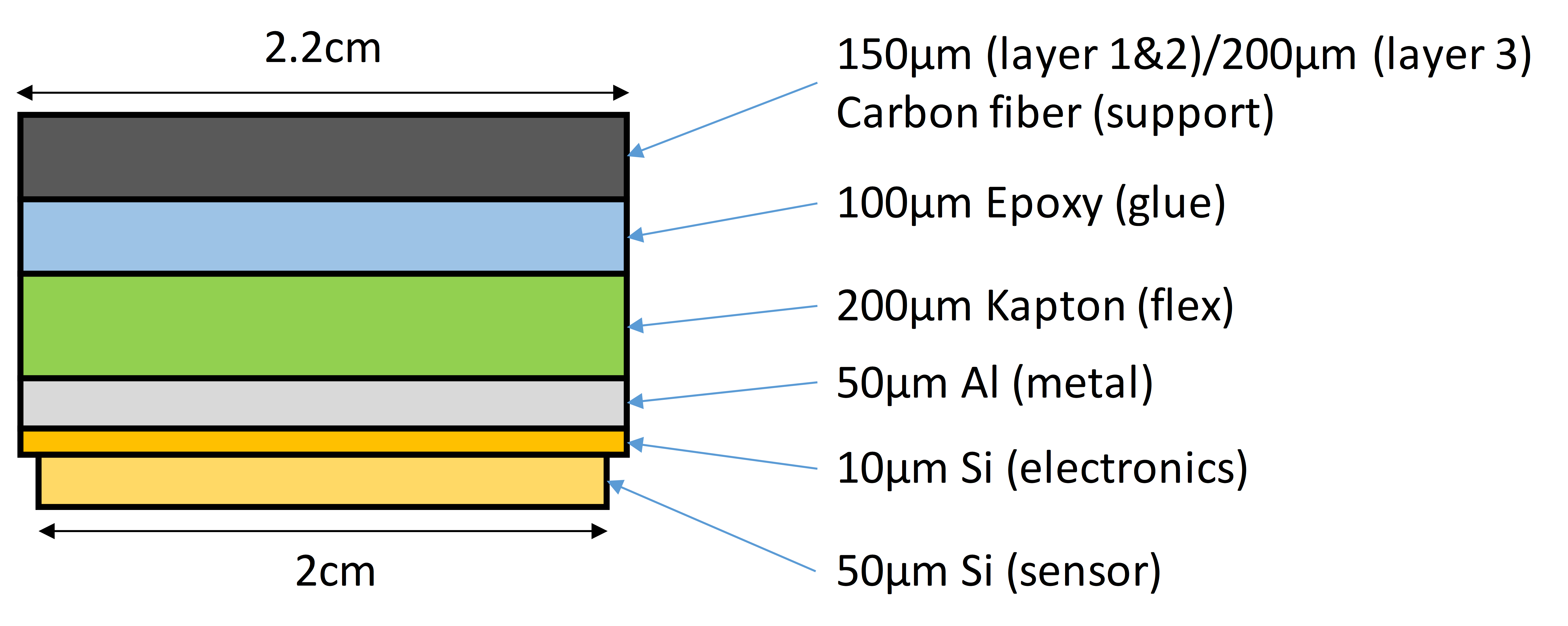}
    \caption{}
    \label{ITKM_b}
\end{subfigure}
\caption{(a). Schematic cross section of the ITKM designed for the STCF experiment. (b). Stave material components and their thickness.} \label{ITKM}
\end{figure}

A conceptual model of a MAPS-based inner tracker for STCF is designed and constructed (cf. Fig.~\ref{ITKM_a}) based on the chosen sensor design. The detector consists of three cylindrical layers of MAPS chips covering the polar angle range between $20^\circ$ and $160^\circ$. The MAPS chips, each with an effective area of $\SI{2}{cm}\times\SI{2}{cm}$, are arranged along the beam direction with a \SI{100}{\micro\meter} gap to form a detector stave. To achieve an extremely low material budget of \textless0.3\%~X$_{0}$ per layer, the thickness of the chips on each stave is reduced to \SI{50}{\micro\meter}. These chips are then mounted on a flexible printed circuit and supported by a carbon fiber frame. Based on the design of ALICE ITS2 inner barrel~\cite{materialbudgetcalculationnew}, a reasonable material thickness is estimated for each ITKM layer (cf. Fig.~\ref{ITKM_b}). A slightly thicker carbon fiber support is employed for the outermost layer considering its mechanical strength. The calculated material budget is approximately 0.27\%~X$_{0}$ for the inner two layers and 0.29\%~X$_{0}$ for the outermost layer~\footnote{These values are obtained along the radial direction ($\theta=90^\circ$) by summing the individual contributions of the stave materials, without accounting for overlaps, edge regions, or non-uniformities.}. The staves are arranged in a pinwheel-like pattern to ensure full coverage of particles from the interaction point. Each layer’s staves are mounted on two end-cap flanges, with the flex cables connected from one side for data transmission. The designed parameters for the three ITKM layers are summarized in Table~\ref{tab2}.

% \begin{figure}[htbp]
% \centering
% \includegraphics[width=\textwidth]{ITKM_and_material.png}
% \caption{(a). Schematic cross section of the ITKM designed for STCF experiment. (b). Stave material components and their thickness.} \label{ITKM}
% \end{figure}

\begin{table}[htbp]
\centering
\caption{Parameters for three layers of ITKM.\label{tab2}}
\begin{tabular}{lllll}
\hline
Layer & Radius/\si{mm} & Stave no. & Chip no. per stave & Area/\si{cm^2} \\
\hline
1 & 36.0 & 12 & 12 & 660.0 \\
2 & 98.0 & 32 & 30 & 4364.8 \\
3 & 160.0 & 52 & 48 & 11440.0 \\
\hline
\end{tabular}
\end{table}

\subsection{Digitization}
\label{digitization}
\begin{figure}[htbp]
\centering
\begin{subfigure}{0.8\textwidth}
    \centering
    \includegraphics[width=\linewidth]{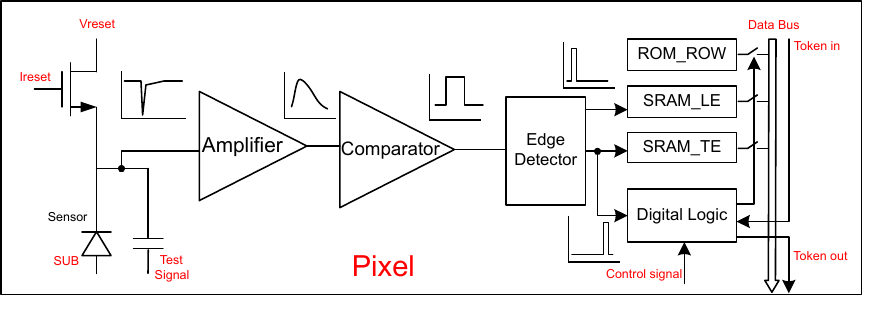}
    \caption{}
    \label{front_end_a}
\end{subfigure}
\hfill
\begin{subfigure}{0.6\textwidth}
    \centering
    \includegraphics[width=\linewidth]{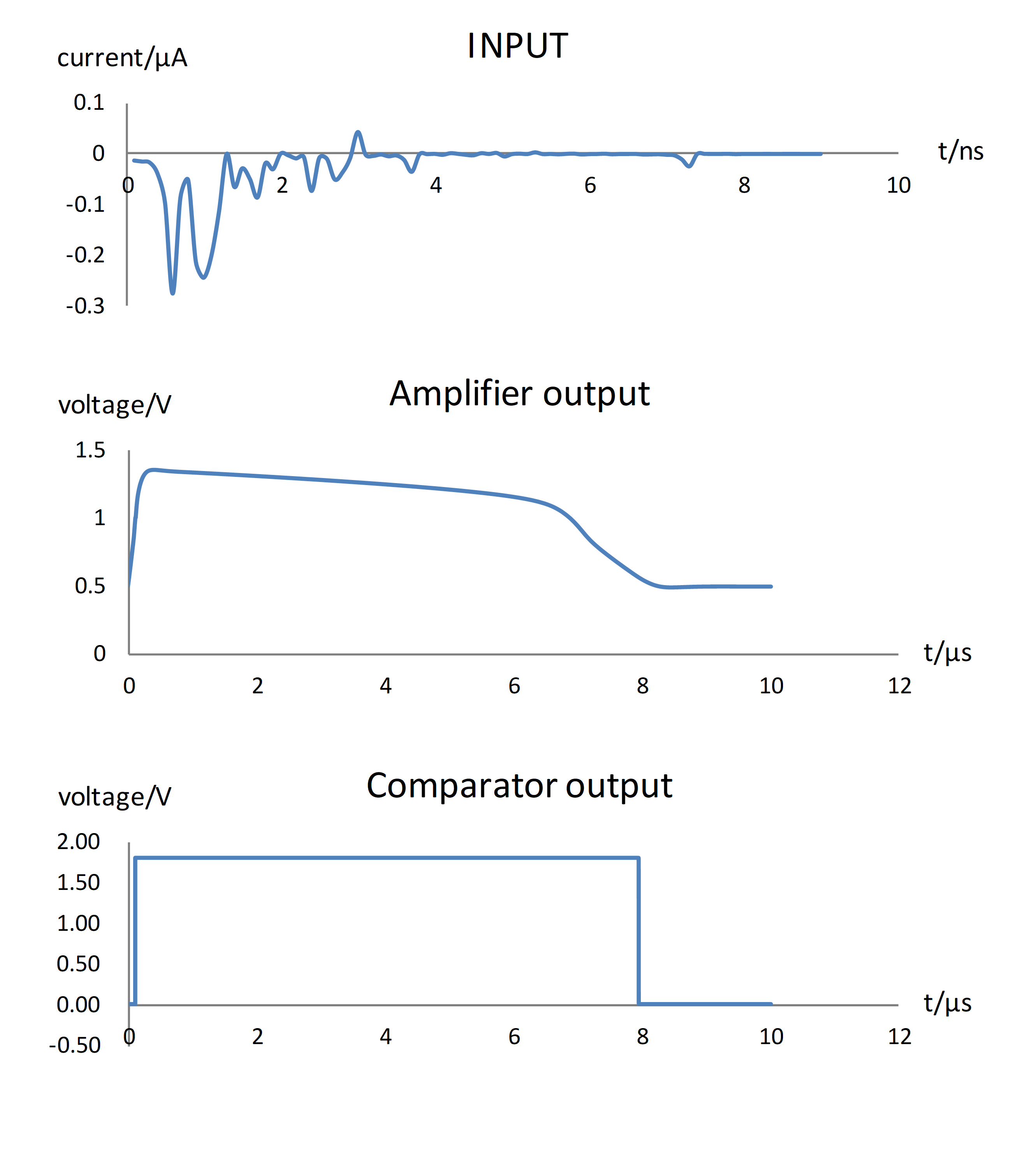}
    \caption{}
    \label{front_end_b}
\end{subfigure}
\caption{(a): Schematic diagram of the in-pixel circuit for the STCF MAPS. (b): A typical input current signal and the corresponding outputs after the amplifier and comparator. Note that the input and output waveforms have different time scale.} \label{front_end}
\end{figure}

Detector digitization is a crucial part of simulation in order to produce results that closely reflects the actual data. Accurate digitization requires a faithful reproduction of the electronics response. For the STCF MAPS, the pixel front-end consists of an open-loop cascode amplifier stage followed by a current comparator (cf. Fig.~\ref{front_end_a})~\cite{alpide_front_end}. When a pixel is hit by a charged particle, current is induced on the collection n-well and a negative voltage signal proportional to the integrated charge is generated on the front-end input node. The signal is then amplified and discriminated, generating a digital pulse with its width proportional to the induced charge of the signal. The leading edge (LE) and trailing edge (TE) of the signal are sampled at a clock frequency of \SI{20}{MHz} and stored to in-pixel SRAMs, waiting for readout~\cite{FEI3}. The final output information of a fired pixel includes the pixel position, LE timestamp and TE timestamp. The LE time is also referred to as the signal ToA, while the time difference between LE and TE corresponds to the signal ToT. The ToA and ToT are then calibrated to provide the time and charge information of the hit.

% \begin{figure}[htbp]
% \centering
% \includegraphics[width=0.5\textwidth]{front-end_and_waveforms.png}
% \caption{(a): In-pixel front-end circuit for the STCF MAPS. (b): A typical input current signal and the simulated output on two nodes -- OUT\_A and OUT\_C. Note that the input and output waveforms have different time scale.} \label{front_end}
% \end{figure}

\begin{figure}[htbp]
\centering
\includegraphics[width=\textwidth]{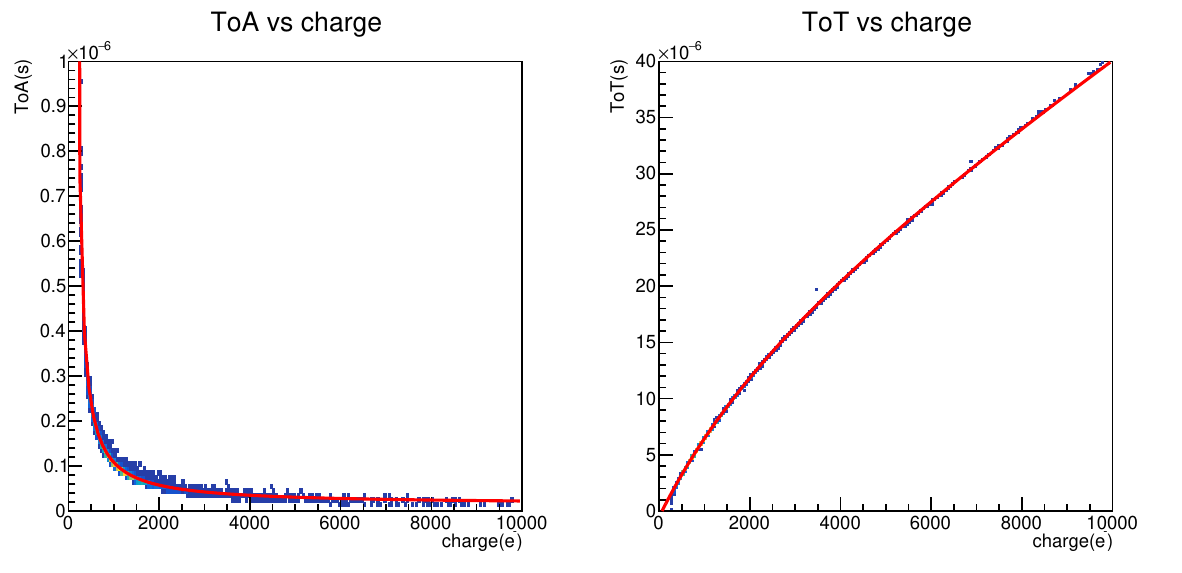}
\caption{Correlation between ToA, ToT and the pixel collected charge.} \label{TQ}
\end{figure}

One straightforward approach for digitization is to convolve the input signal with a front-end transfer function and extract the useful information from the result. However, unlike a Current Sense Amplifier (CSA), a common transfer function is not available for different signal shapes and amplitudes with this front-end design. Therefore, a sampling method based on the collected charge is implemented to achieve efficient and practical digitization. Firstly, a large number of incident MIPs from the interaction point are simulated to generate the induced current signals on the ITKM pixels. The signals are then passed to the front-end, and the corresponding discrimination output signals are simulated with Cadence Spectre. The output waveforms of a typical input signal are shown in Fig.~\ref{front_end_b}. A one-to-one relation of signal charge to ToA (ToT) can therefore be established (the ToA and ToT values at this stage are ideal, without considering the limited clock frequency). As shown in Fig.~\ref{TQ}, there is a strong correlation between ToA (ToT) and charge, which forms the basis for this sampling method. Empirical functions are applied to fit the ToA–charge and ToT–charge relationships~\cite{timepix3}:
\begin{gather}
    ToA = a_0 + \frac{a_1}{(Q + a_2)}, \\
    ToT = b_0 + b_1 \cdot Q + \frac{b_2}{(Q + b_3)},
\end{gather}
where $a_0, a_1, a_2, b_0, b_1, b_2, b_3$ are the fitting parameters.

% \begin{figure}[htbp]
% \centering
% \includegraphics[width=\textwidth]{TQ_relationship.png}
% \caption{Correlation between ToA, ToT and the pixel collected charge.} \label{TQ}
% \end{figure}

To digitize a signal on a pixel, ToA and ToT values are first calculated according to the collected charge of this pixel using the above fitted functions. Gaussian fluctuations are then introduced to account for variations in the front-end response to signals with the same input charge. Finally, the continuous ToA and ToT values are discretized using a \SI{20}{MHz} synchronization clock. It is worth emphasizing that, while this process currently relies solely on simulation results, it is fully calibratable through future experimental measurements. This calibration is essential to ensure accurate MC results for physics analysis.

The digitization serves as a optional procedure in the simulation workflow discussed in Section~\ref{sim_workflow}. The following reconstruction process remains largely unchanged, except that the collected charge used for cluster reconstruction is not the ideal value but rather a converted value derived from the ToT.

\subsection{Results and discussion}

MC simulations of ITKM are performed based on Geant4, with 1\,\si{GeV/c} $\mu^-$ particles generated from the interaction point via a particle gun. The particles are emitted isotropically within a polar angle range of $20^\circ$ to $160^\circ$, matching the detector acceptance. The interaction vertex is approximated as point-like, without accounting for the spatial extent of the interaction diamond. The pixel operates under nominal bias conditions of $V_{sub}=\SI{-6}{V}$ and $V_{nw}=\SI{0.8}{V}$, with a uniform threshold of \SI{300}{e^-} applied to all pixels. Under this threshold, the average ToA is \SI{518}{ns} according to Fig.~\ref{TQ}. A conservative event time window of \SI{1}{\micro\second} is thus applied to ensure most of the over-threshold pixels being properly recorded---only pixel hits occurring within this time window relative to the event start time are assigned a valid timestamp and are treated as part of the same event.

\begin{figure}[htbp]
\centering
    \begin{subfigure}{0.45\textwidth}
        \centering
        \includegraphics[width=\linewidth]{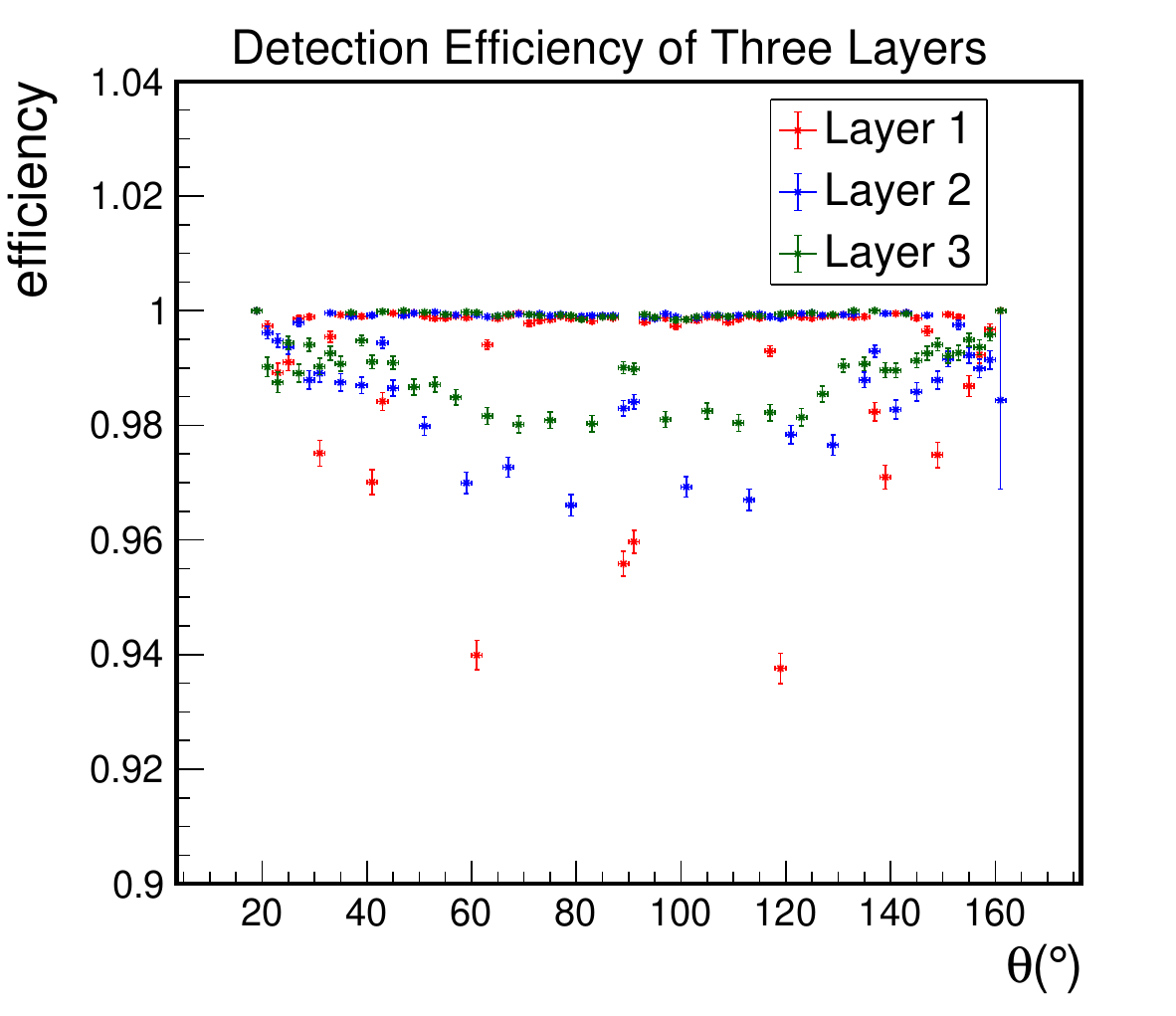}
        \caption{}
        \label{efficiency_clustersize_a}
    \end{subfigure}
    \hfill
    \begin{subfigure}{0.45\textwidth}
        \centering
        \includegraphics[width=\linewidth]{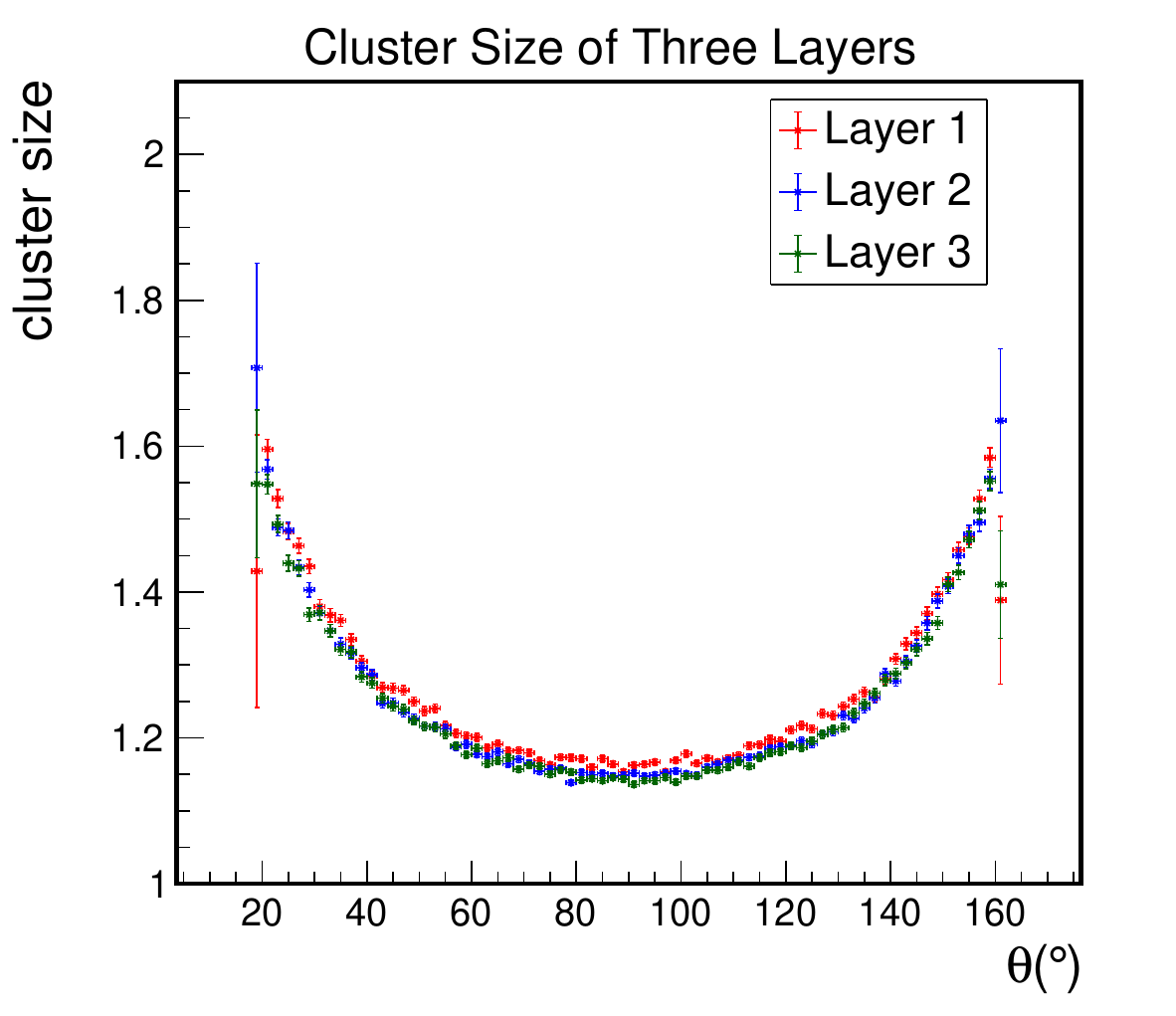}
        \caption{}
        \label{efficiency_clustersize_b}
    \end{subfigure}

    \caption{Detection efficiency (a) and cluster size (b) of three ITKM layers at various polar angles.}
    \label{efficiency_clustersize}
\end{figure}

The detection efficiencies of three ITKM layers as a function of the polar angle $\theta$ are illustrated in Fig.~\ref{efficiency_clustersize_a}. Abrupt drops in efficiency are observed at certain $\theta$ values, which is attributed to the dead areas between adjacent chips along the beam pipe direction. Despite this, average detection efficiencies of 99.3\%, 99.3\% and 99.4\% are achieved for layer 1, 2 and 3, respectively. Additionally, the efficiency tends to increase at the forward and backward regions of the detector, due to the more inclined incident angle resulting in a longer particle trajectory and, consequently, greater energy deposition within the sensitive volume. This effect is further evidenced by the cluster size distribution, as shown in Fig.~\ref{efficiency_clustersize_b}. At the edges of the angular coverage, the simulated cluster sizes ($\sim$1.6) are slightly larger than the geometric expectation ($\sim$1.3), which can be naturally attributed to additional charge sharing arising from diffusion.

% \begin{figure}[htbp]
% \centering
% \includegraphics[width=\textwidth]{ITKM_efficiency_clustersize.png}
% \caption{Detection efficiency (a) and cluster size (b) of three ITKM layers at various polar angles.} \label{efficiency_clustersize}
% \end{figure}

\begin{figure}[htbp]
\centering
\begin{subfigure}{0.45\textwidth}
    \centering
        \includegraphics[width=\linewidth]{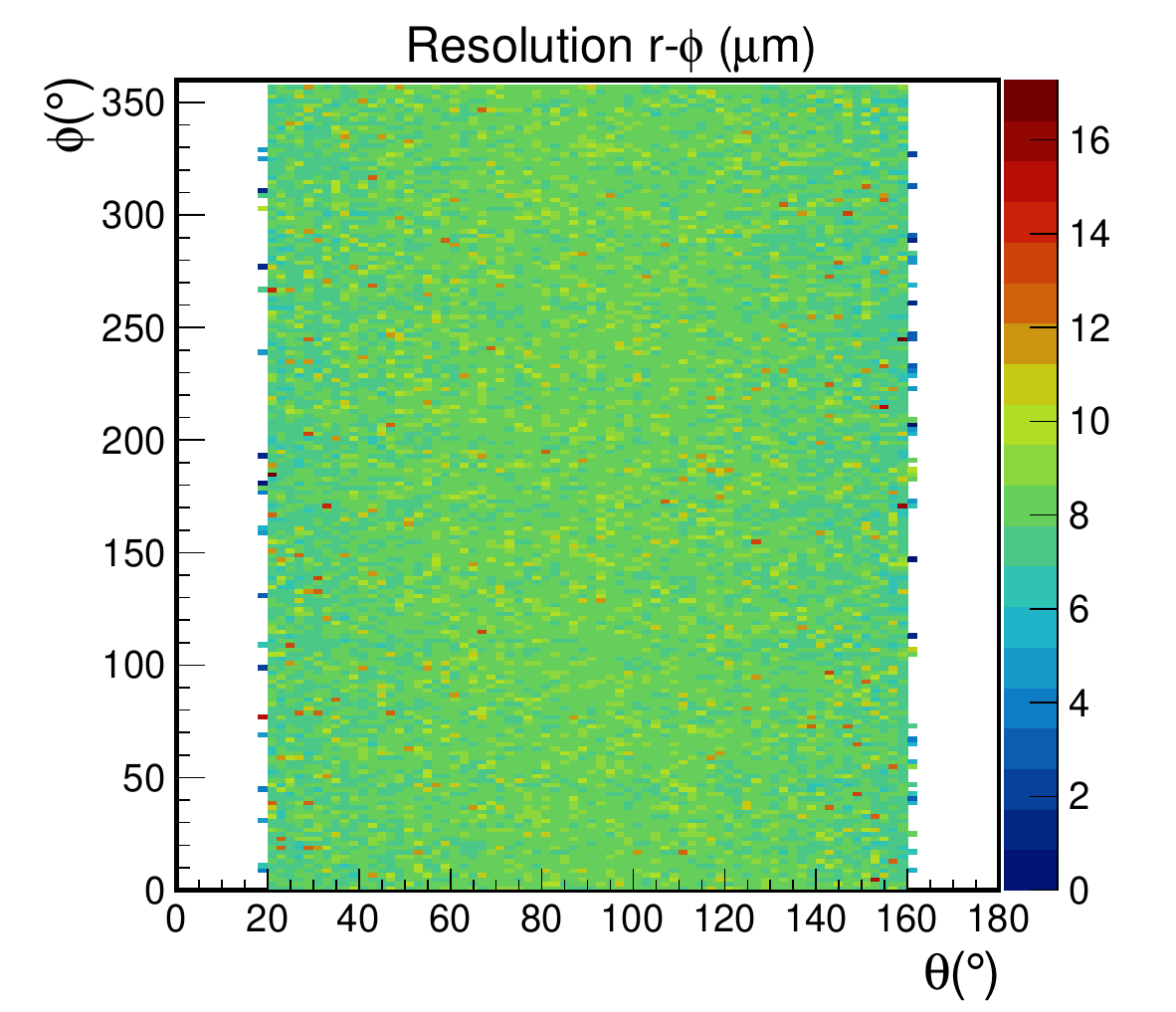}
        \caption{}
    \end{subfigure}
    \hfill
    \begin{subfigure}{0.45\textwidth}
        \centering
        \includegraphics[width=\linewidth]{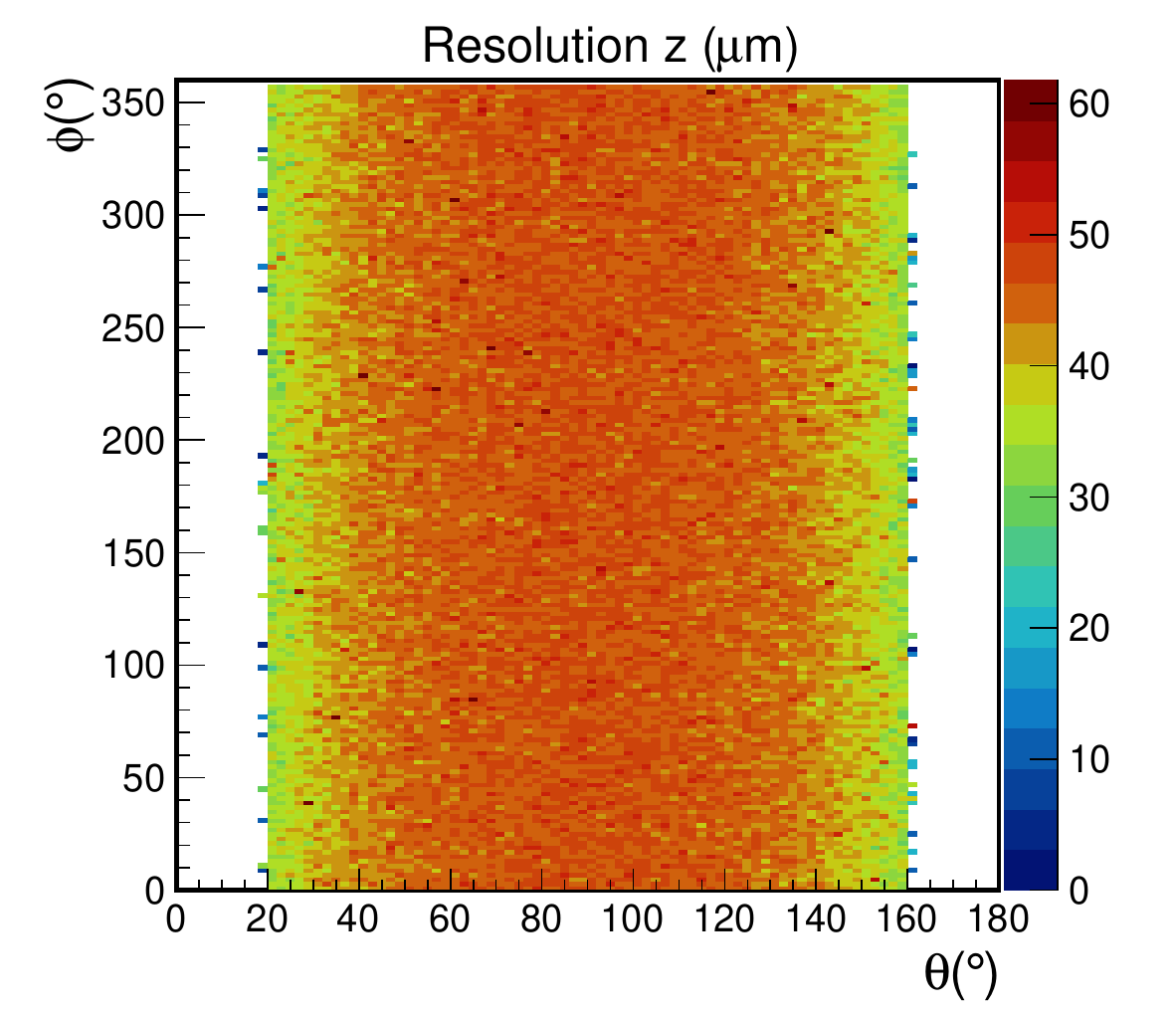}
        \caption{}
    \end{subfigure}

\caption{Three layer averaged $r-\phi$ dimensional (a) and $z$ dimensional (b) spatial resolution at different polar and azimuthal angles.} \label{position_resolution}
\end{figure}

For the detected hits, the spatial resolution is calculated as the Root Mean Square (RMS) of the residual, which is the distance between reconstructed position and the MC truth position. A 2D map illustrating the resolution across various spatial locations is presented in Fig.~\ref{position_resolution}. The resolution in the $r-\phi$ direction is almost uniform throughout the entire solid angle coverage, except for a few regions where delta-ray events cause slightly higher values. In contrast, the $z$ direction resolution exhibits a dependence on the $\theta$ angle, with better resolution observed towards the ends of detector barrel. This trend is related to the charge-centering method used for cluster reconstruction, which achieves higher accuracy when more pixel hits provide additional information. The overall average spatial resolution is \SI{8.2}{\micro\meter} in the $r-\phi$ direction and 44.8\,\si{\micro\meter} in the $z$ direction.

% \begin{figure}[htbp]
% \centering
% \includegraphics[width=\textwidth]{ITKM_position_resolution.png}
% \caption{Three layer averaged $r-\phi$ dimensional (a) and $z$ dimensional (b) spatial resolution at different polar and azimuthal angles.} \label{position_resolution}
% \end{figure}

\begin{figure}[htbp]
\centering
\includegraphics[width=\textwidth]{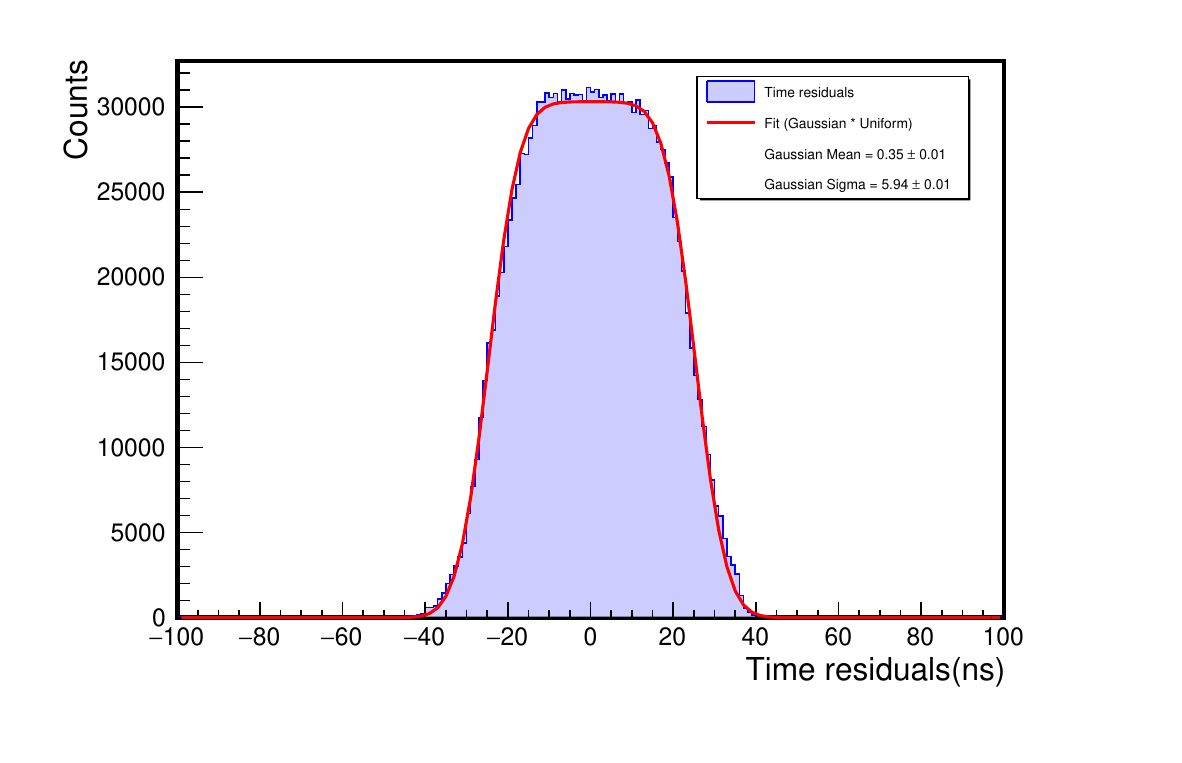}
\caption{Distribution of cluster time residuals for all three layers. A convolution of a Gaussian function with a uniform distribution function is used to fit the distribution.} \label{ITKM_time}
\end{figure}

Additionally, the timing performance of ITKM is investigated. For each cluster, the timing information of the seed pixel is used to calculate the cluster time---the ToA is corrected using the T-A relationship derived from Fig.~\ref{TQ}, and the corrected value serves as an estimate of the particle hit time. The corrected ToA is then compared with the MC truth event time to determine the time residual for each cluster. The distribution of the time residuals for all detected clusters is presented in Fig.~\ref{ITKM_time}. The residual distribution comprises two distinct components: one resulting from the discrete clock period and the other reflecting the intrinsic time resolution of the sensor. To accurately model this distribution, a Gaussian-Uniform convolution function is used:
\begin{equation}
    F(t) = \int_a^b G(t - t^{\prime}; A, \mu, \sigma) \, dt^{\prime},
\end{equation}
where $a, b$ represents the lower and upper bounds of the uniform distribution function, and $G(t; A, \mu, \sigma)$ is a Gaussian function with amplitude $A$, mean $\mu$ and standard deviation $\sigma$. For this fit, $a$ and $b$ are fixed to \SI{-25}{ns} and \SI{25}{ns}, respectively, corresponding to the \SI{50}{ns} clock period used in the prototype chip design. The fitted function, shown as the red line in Fig.~\ref{ITKM_time}, provides a good description of the data. From the fit, the intrinsic single-cluster time resolution of ITKM is determined to be \SI{5.9}{ns}. Considering the contribution from the \SI{50}{ns} clock period, the overall time resolution is evaluated to be \SI{16}{ns} in terms of RMS. This result demonstrates that the \SI{50}{ns} resolution requirement for the STCF MAPS can be met even with a moderate clock frequency. Furthermore, it reveals promising potential for achieving sub-\SI{10}{ns} single-cluster timing capability through dedicated optimizations of the MAPS chip design.

\FloatBarrier
\section{Conclusions}

The small collection electrode MAPS has been selected as a candidate technology for the inner tracker of STCF, a proposed next generation high-luminosity $e^+ e^-$ collider experiment operating in the tau-charm energy region, due to its high spatial resolution and low material budget. A dedicated simulation workflow, combining TCAD and MC softwares, has been developed to support the sensor design and evaluate its expected performance. Four process variants for MAPS fabrication and three pixel geometries have been explored. Based on simulation results, an active-connect strip-like pixel design with a high-resistivity epitaxial layer has been chosen as the baseline due to its superior charge collection performance. Using this pixel design, a three-layer cylindrical MAPS-based inner tracker model has been constructed and tested in simulation. The results indicate an average detection efficiency exceeding 99\% for all three layers, with spatial resolutions of \SI{8.2}{\micro\meter} and 44.8\,\si{\micro\meter} in the $r-\phi$ and $z$ directions, respectively, and an intrinsic sensor time resolution of \SI{5.9}{ns}. These performance metrics surpass the STCF experiment's requirements for the inner tracker, demonstrating the promising potential of a MAPS-based design.

The Monte Carlo simulation workflow developed in this study has been integrated as a key component of the OSCAR framework, the offline data processing software for the STCF experiment. At present, a complete detector simulation chain for all STCF sub-detectors has been implemented within OSCAR. These integrated efforts have enabled further simulation studies, such as evaluating the tracking performance of the STCF tracking system~\cite{stcf_tracking}, which are essential for optimizing detector design and verifying the feasibility of achieving the STCF’s physics goals.

\section*{Declaration of generative AI and AI-assisted technologies in the writing process}

During the preparation of this work the authors used ChatGPT in order to improve the readability and language of the manuscript. After using this tool/service, the authors reviewed and edited the content as needed and take full responsibility for the content of the published article.

\section*{Acknowledgements}
This work was supported by the National Key Research and Development Program of China under Grant 2022YFA1602203, the National Natural Science Foundation of China under Grant 12341502 and 12125505, the Fundamental Research Funds for the Central Universities of China under Grant WK2360000014, and the STCF Key Technology Research and Development Project. We thank the Hefei Comprehensive National Science Center for its strong support on the promotion and development of the STCF project.

\appendix
\section{In-pixel efficiency at \SI{150}{e^-} threshold}
\label{app0}

\begin{figure}[htbp]
\centering
    \begin{subfigure}{0.45\textwidth}
        \centering
        \includegraphics[width=\linewidth]{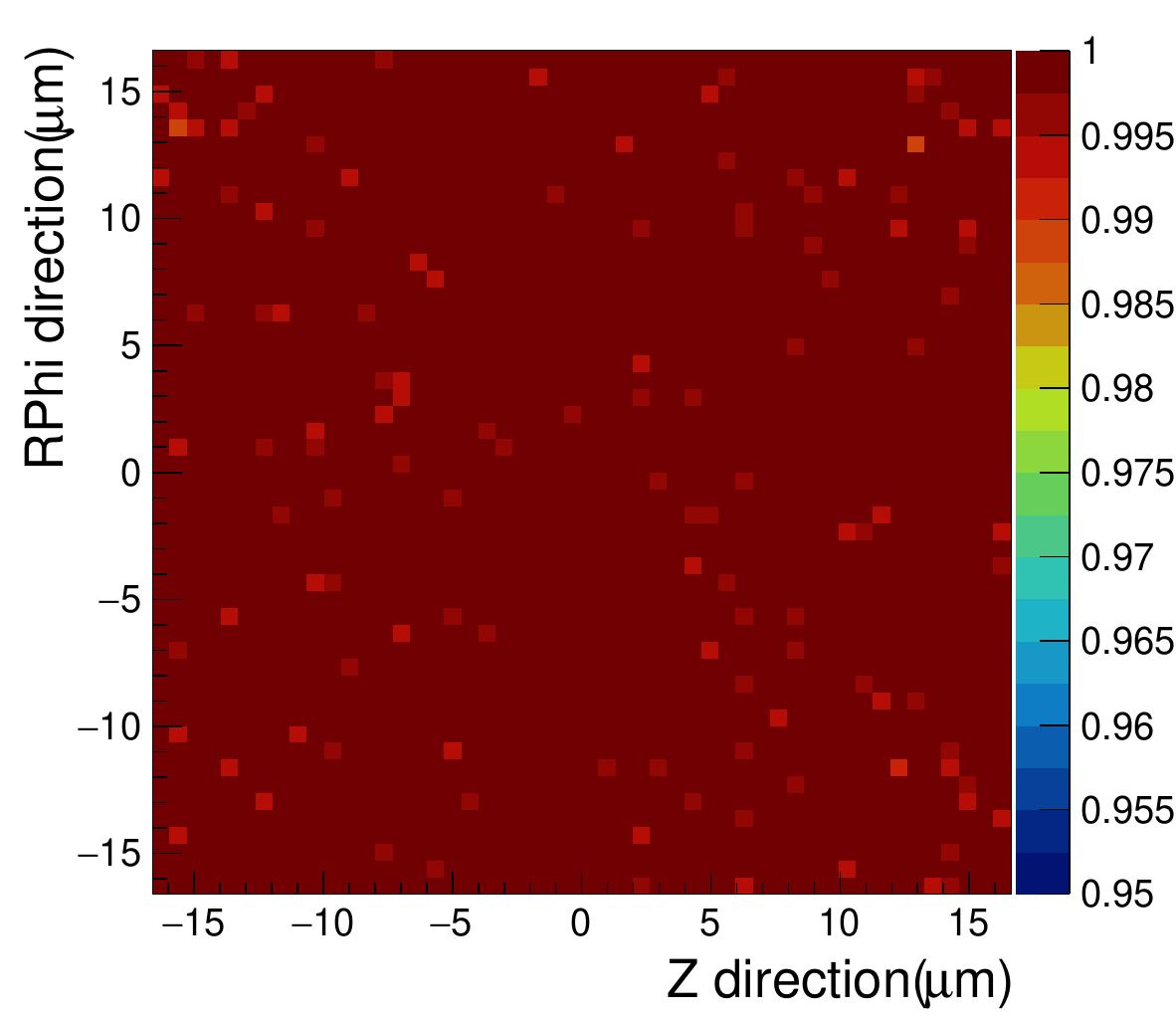}
        \caption{HR epi, average efficiency 99.97\%}
    \end{subfigure}
    \hfill
    \begin{subfigure}{0.45\textwidth}
        \centering
        \includegraphics[width=\linewidth]{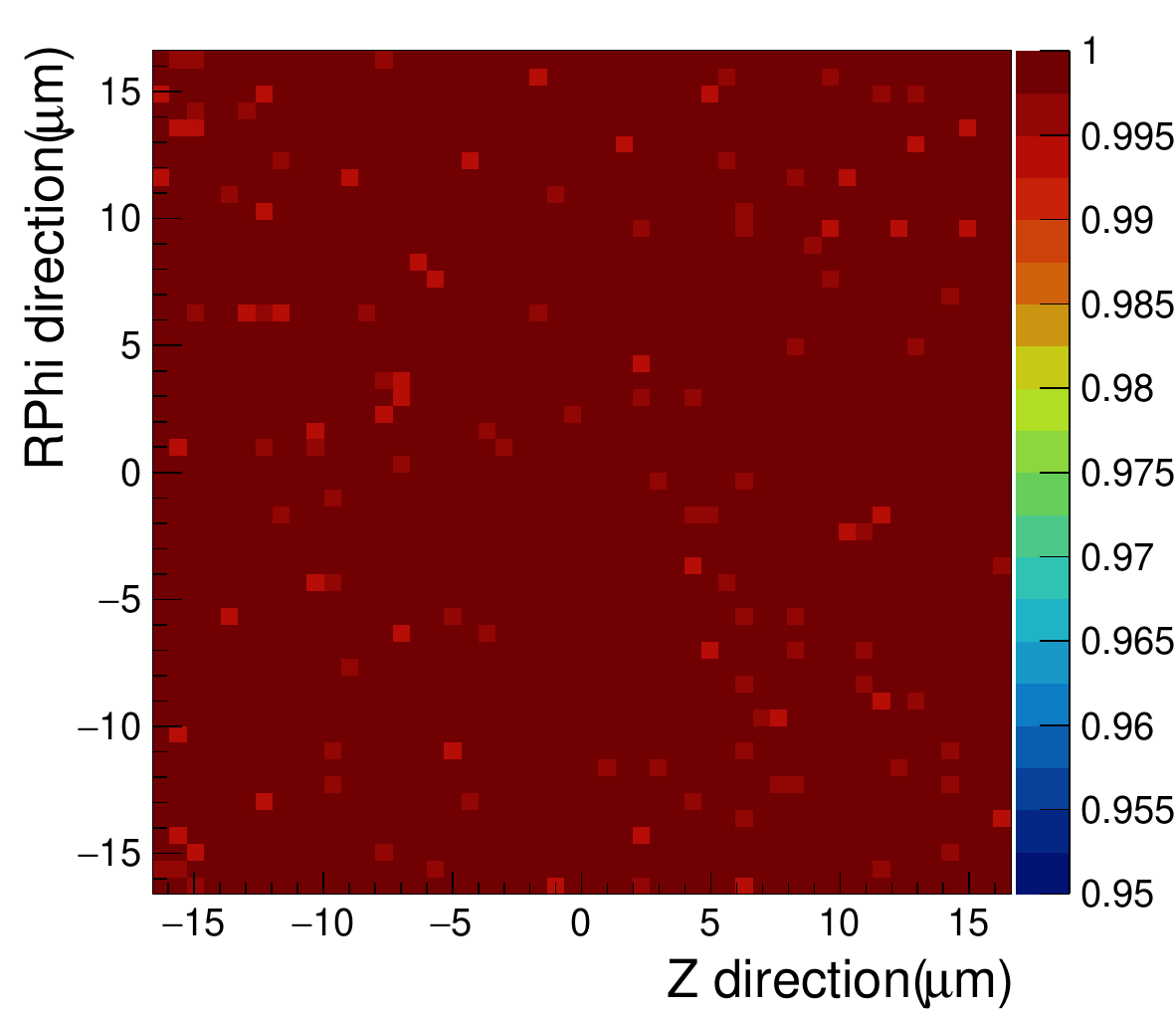}
        \caption{N blanket, average efficiency 99.98\%}
    \end{subfigure}
    \hfill
    \begin{subfigure}{0.45\textwidth}
        \centering
        \includegraphics[width=\linewidth]{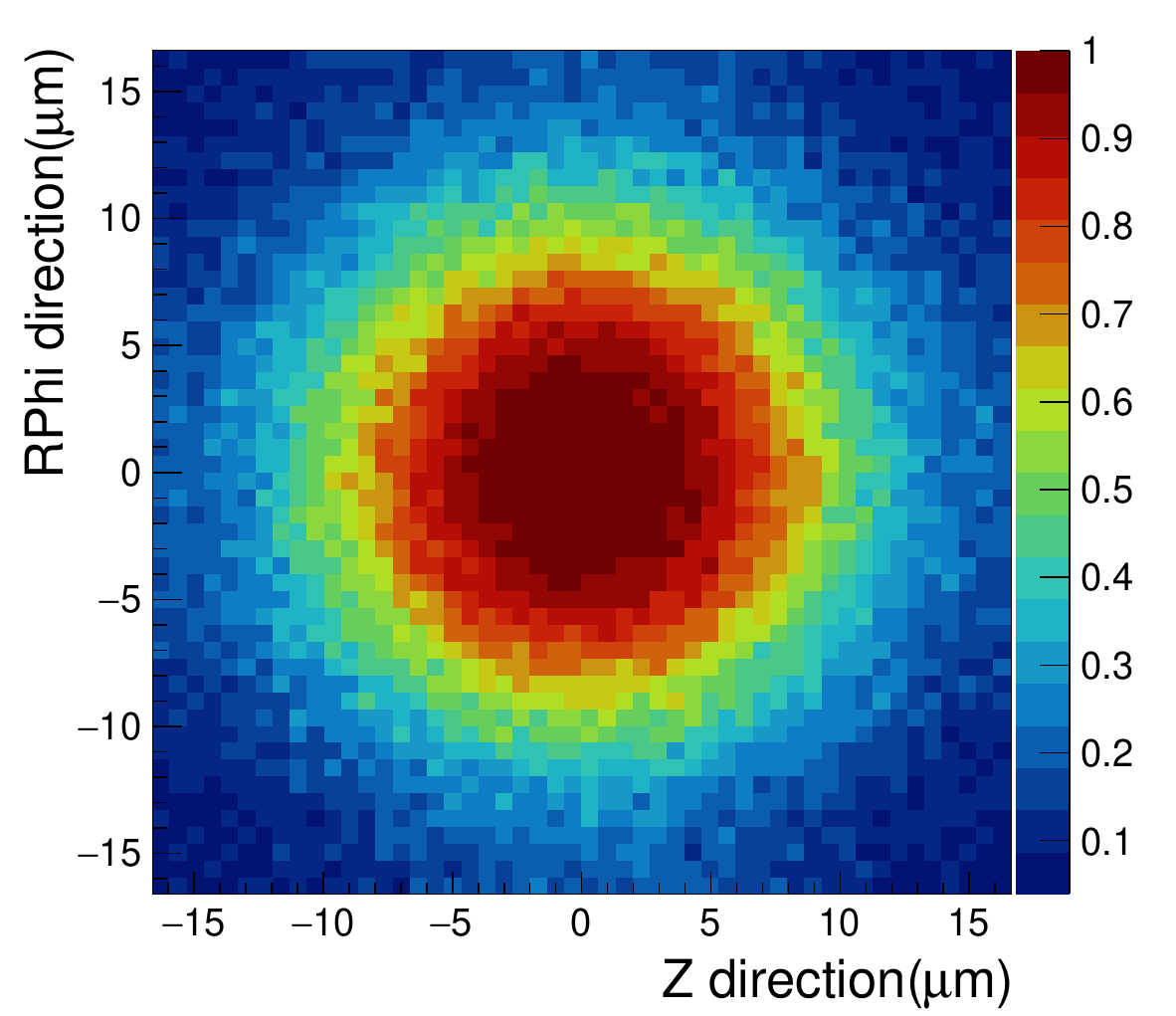}
        \caption{LR epi, average efficiency 38.8\%}
    \end{subfigure}
    \hfill
    \begin{subfigure}{0.45\textwidth}
        \centering
        \includegraphics[width=\linewidth]{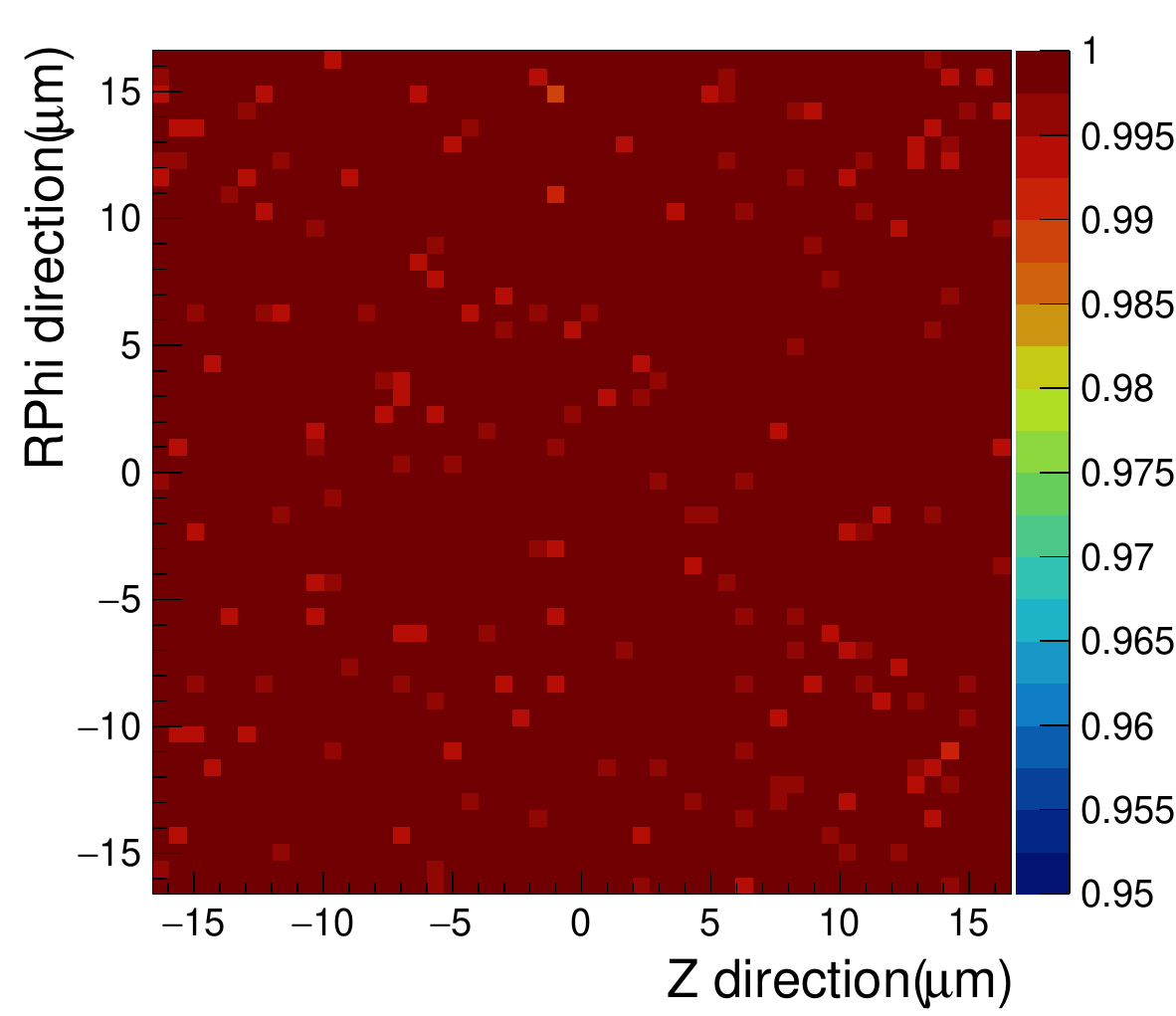}
        \caption{HR substrate, average efficiency 99.97\%}
    \end{subfigure}

\caption{In-pixel detection efficiency maps for four different process variants, at a pixel threshold of \SI{150}{e^-}, $V_{sub}=\SI{-6}{V}$, $V_{nw}=\SI{0.8}{V}$.} \label{efficiency_150e}
\end{figure}

The in-pixel detection efficiency at a threshold of \SI{150}{e^-} for a standard pixel ($\SI{33}{\micro\meter}\times\SI{33}{\micro\meter}$) in the four processes is shown in Fig.~\ref{efficiency_150e}. This threshold is representative of typical operating conditions achievable for sensors with this pixel pitch, and thus provides a useful benchmark for comparison with results from other experimental groups. The simulation conditions are identical to those described in Section~\ref{MIP}.

\section{Effect of substrate bias on ``HR substrate'' pixel}
\label{app1}

Three different substrate voltages are applied to the ``HR substrate'' MAPS with a pixel size of $\SI{33}{\micro\meter}\times\SI{33}{\micro\meter}$: \SI{-6}{V}, \SI{-9}{V}, and \SI{-12}{V}. The variation in sensor capacitance is minimal, decreasing slightly from \SI{2.7}{fF} to \SI{2.6}{fF} as the voltage increases from \SI{-6}{V} to \SI{-12}{V}. The transient TCAD simulation results under these bias settings, including the collected charge and collection time for an \SI{60}{e^-/\micro\meter} ionization trajectory, are summarized in Table~\ref{tab3}. The charge collection time is defined as the time interval between collecting \SI{10}{\percent} and \SI{90}{\percent} of the total charge.

\begin{table}[htbp]
\centering
\caption{TCAD simulated collected charge and collection time for ``HR substrate'' process under different substrate bias.\label{tab3}}
\begin{tabular}{lllll}
\hline
\multirow{2}{*}{Substrate bias/\si{V}} & \multicolumn{2}{c}{Center incident} & \multicolumn{2}{c}{Corner incident} \\
\cline{2-5}
 & \makecell{Collected \\ charge/\si{e^-}} & \makecell{Collection \\ time/\si{ns}} & \makecell{Collected \\ charge/\si{e^-}} & \makecell{Collection \\ time/\si{ns}} \\
\hline
-6 & 1330 & 39 & 340  & 125 \\
-9 & 1410 & 34 & 370 & 111 \\
-12 & 1470 & 32 & 390 & 101 \\
\hline
\end{tabular}
\end{table}

\begin{figure}[htbp]
\centering
    \begin{subfigure}{0.45\textwidth}
    \centering
        \includegraphics[width=\linewidth]{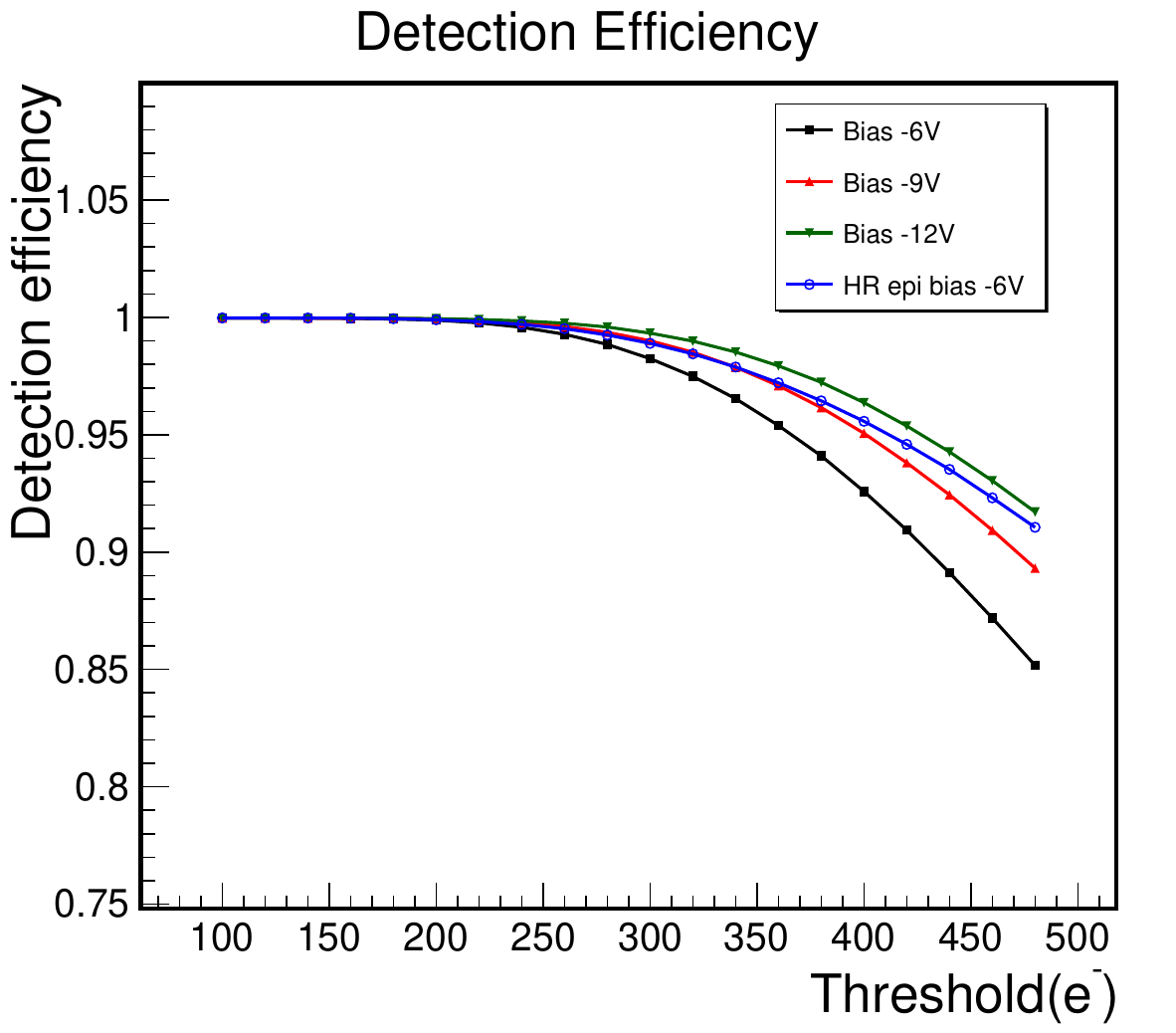}
        \caption{}
    \end{subfigure}
    \hfill
    \begin{subfigure}{0.45\textwidth}
        \centering
        \includegraphics[width=\linewidth]{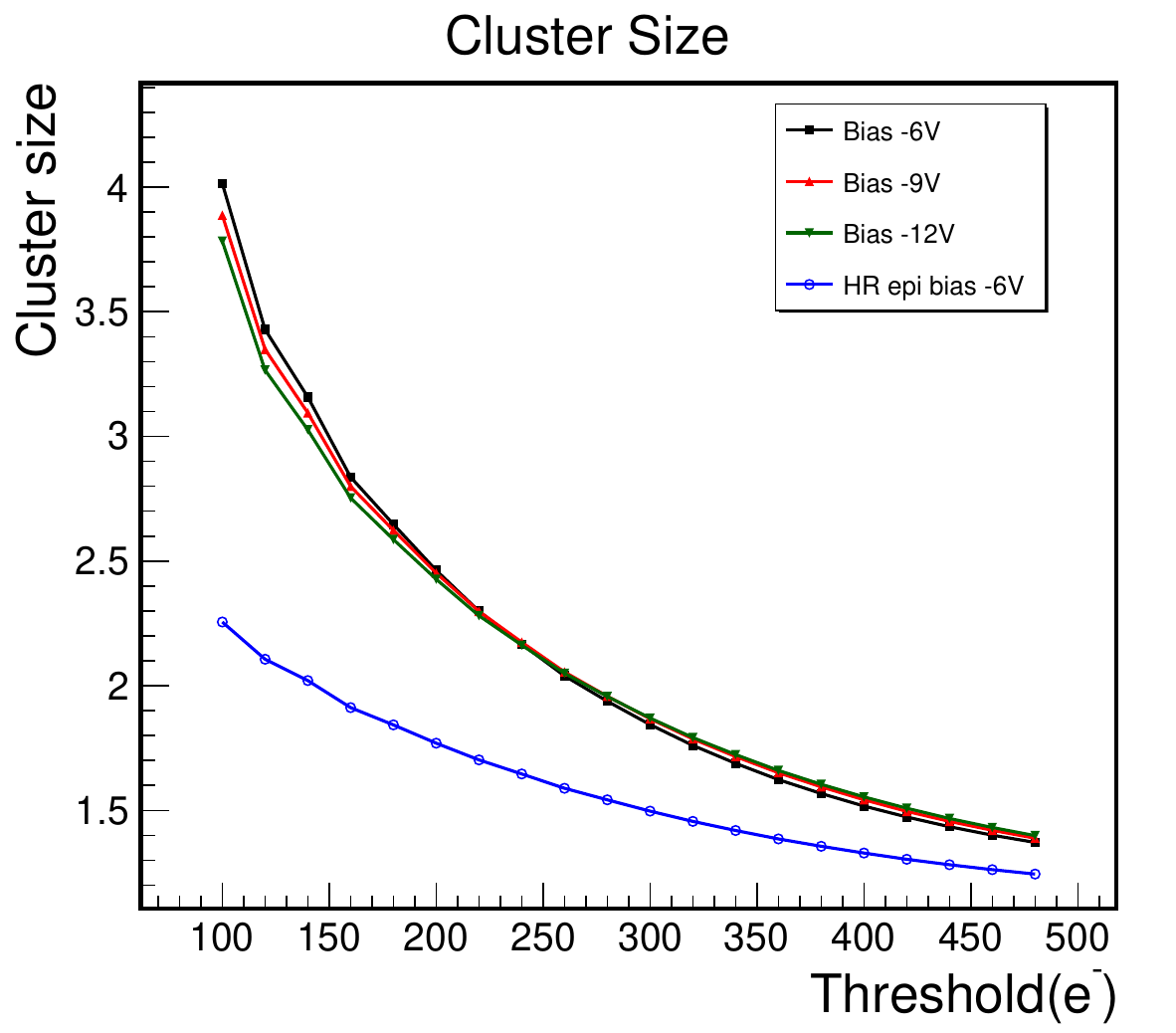}
        \caption{}
    \end{subfigure}
    \hfill
    \begin{subfigure}{0.45\textwidth}
        \centering
        \includegraphics[width=\linewidth]{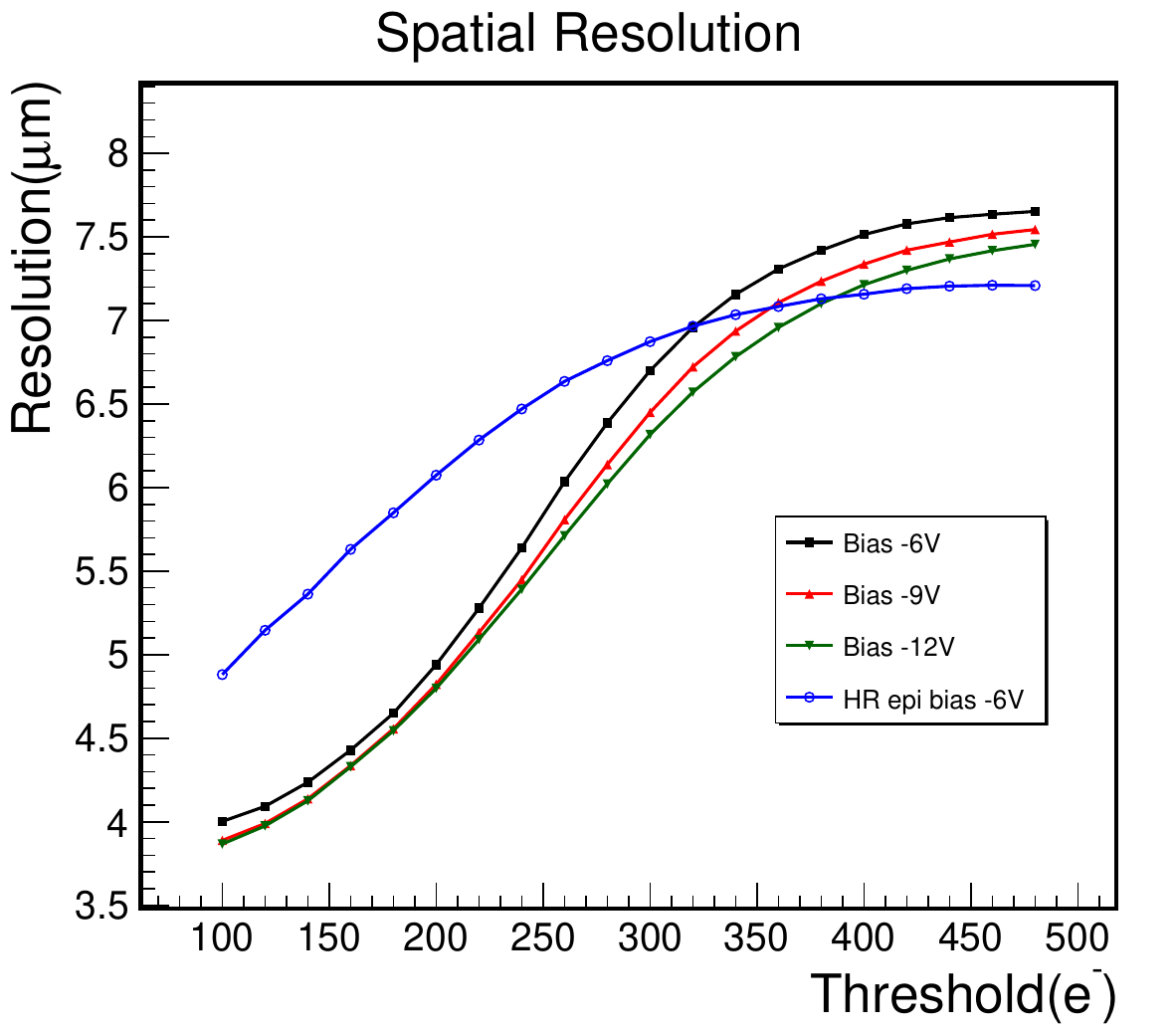}
        \caption{}
    \end{subfigure}

\caption{The ``HR substrate'' pixel's repsonse to MIPs under various substrate settings, with $V_{nw}$ fixed at \SI{0.8}{V}. Results of ``HR epi'' pixel under nominal bias settings are shown for comparison.} \label{HR_bias_scan}
\end{figure}

The response to minimum ionizing particles is simulated based on the settings described in Section~\ref{MIP}. The detection efficiency, cluster size, and spatial resolution are presented in Fig.~\ref{HR_bias_scan}. For comparison, the results of ``HR epi'' pixel at a bias voltage of \SI{-6}{V} are also included. Since the pixel pitches are identical in both directions, the spatial resolution is only shown for one direction. The degradation in detection efficiency is flattened at higher bias voltages, achieving a performance comparable to that of the ``HR epi'' type. Additionally, a slight improvement in spatial resolution is observed, attributed to the larger amount of charge collected.

% \begin{figure}[htbp]
% \centering
% \includegraphics[width=0.5\textwidth]{HR_sub_bias_scan.png}
% \caption{The ``HR substrate'' pixel's repsonse to MIPs under various substrate settings, with $V_{nw}$ fixed at \SI{0.8}{V}. Results of ``HR epi'' pixel under nominal bias settings are shown for comparison.} \label{HR_bias_scan}
% \end{figure}

%%       \cite{<label>} ==> [1]

%%
% Example citation, See \cite{star}.

\bibliographystyle{elsarticle-num} 
\bibliography{biblio.bib}




% %% For numbered reference style

%   \textit{\LaTeX: a document preparation system},
%   Addison Wesley, Massachusetts,

% \end{thebibliography}

\end{document}